\documentclass[11pt,twoside]{book}
\usepackage{a4wide,graphicx,longtable,fancyheadings,array,epsfig}
\usepackage{epstopdf}
\usepackage[round]{natbib}
\usepackage{titlesec}
\usepackage{rotating,setspace,amssymb,pifont}
\usepackage[activeacute,spanish]{babel}
\usepackage[latin1]{inputenc}
\usepackage{isolatin1}
\usepackage[colorlinks=true,linkcolor=black,citecolor=night,breaklinks=true]{hyperref}
\usepackage{color}
\usepackage{pdfpages}
\DeclareGraphicsExtensions{.eps,.ps}
\usepackage{sidecap}

\newcommand{\bigrule}{\titlerule[0.5mm]}
\titleformat{\chapter}[display] 
{\bfseries\Huge} 
{
 \titlerule 
 \filleft 
 \Large\chaptertitlename\ 
 \Large\thechapter} 
{0mm} 
{\filleft} 
[\vspace{0.5mm} \bigrule] 

\newcommand{\OIII}{O{\sc iii}}
\newcommand{\OII}{O{\sc ii}}
\newcommand{\OI}{O{\sc i}}
\newcommand{\SIII}{S{\sc iii}}
\newcommand{\SII}{S{\sc ii}}
\newcommand{\NII}{N{\sc ii}}

\newcommand{\NeIII}{Ne{\sc iii}}
\newcommand{\FeIII}{Fe{\sc iii}}

\newcommand{\ArIII}{Ar{\sc iii}}
\newcommand{\ArIV}{Ar{\sc iv}}

\newcommand{\SIV}{S{\sc iv}}

\newcommand{\clearemptydoublepage}{\newpage{\pagestyle{empty}\cleardoublepage}}

\def\ergs{~erg\,s$^{-1}$}

\def\kms{~km\,s$^{-1}$}

\definecolor{violet}{cmyk}{0.07,0.90,0,0.34}
\definecolor{night}{cmyk}{0.98,0.13,0,0.43}
\begin{document}
\renewcommand{\listtablename}{'Indice de tablas} 
\renewcommand{\tablename}{Tabla} 

\begin{spacing}{1.21}

\graphicspath{{resumen/}{capitulo1/}{capitulo2/}{capitulo3/}{capitulo4/}{capitulo5/}{agradecimientos/}{fin/}
}

\thispagestyle{empty}

\begin{center}

\vskip 2cm

{\Huge \bf 
{\color{night}Regiones H{\sc ii} Gigantes en Galaxias}\\
\vskip 0.2cm
{\color{night}Observables desde el Hemisferio Sur}}

\vskip 3cm

{\LARGE \bf
Ver\'onica Firpo}

\vskip 1cm

{\Large
Director de Tesis: Dr. Guillermo~L. Bosch\\
\vskip 0.2cm
Co-director de Tesis: Dr. Guillermo~F. {H{\"a}gele}}

\vskip 2cm

{\large
Tesis presentada para optar al grado de Doctor en Astronom\'{\i}a}

\vskip 3cm

   \begin{center}
     \includegraphics[width=.15\textwidth]{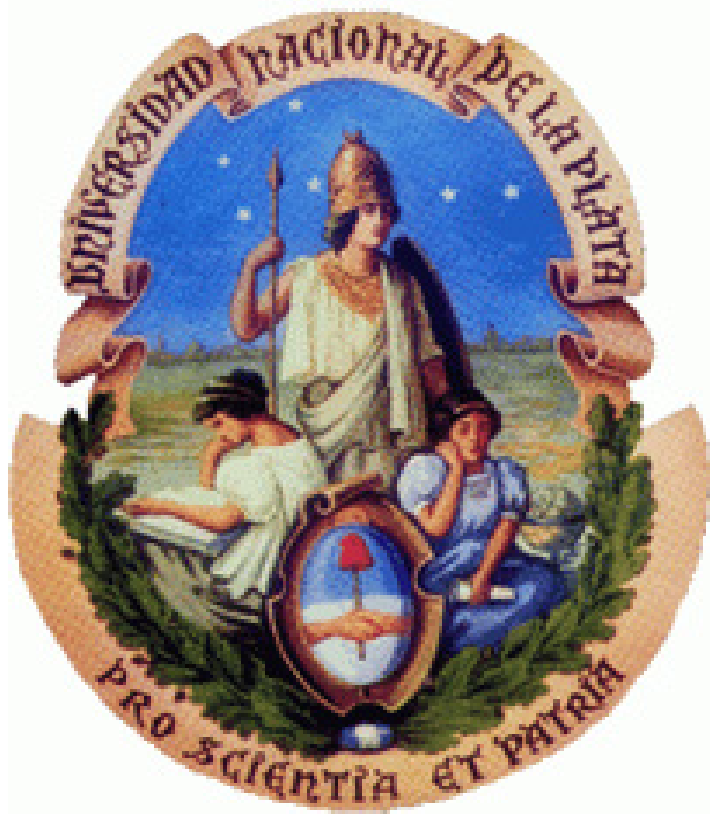}
   \end{center}

\vskip 1cm

{\Large
Facultad de Ciencias Astron\'omicas y Geof\'{\i}sicas\\
Universidad Nacional de La Plata\\
Argentina}

\medskip

{\large
Marzo de 2011}

\end{center}

\clearemptydoublepage
\newpage

 \vskip 3cm
\thispagestyle{empty}
\vspace{15cm}

\begin{figure*}
  \centering
    \vskip 3cm
   \includegraphics[width=1\textwidth]{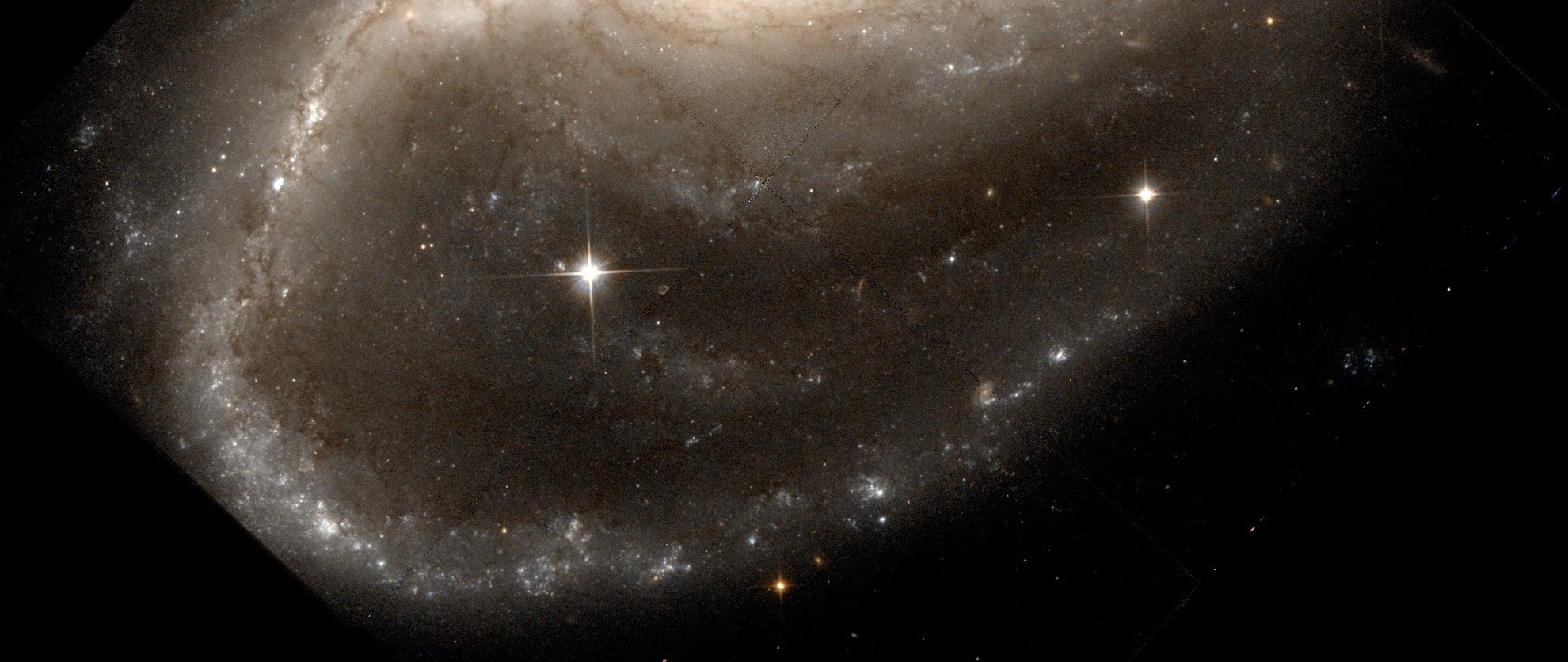}
\end{figure*}
\noindent


\vspace{2em}
\begin{flushright}
\em \Large ``Una mota solitaria \\
en la inmensa oscuridad c'osmica."\\
\hspace{5em}{\em \large  Carl Sagan (referenciando a la Tierra)}
\end{flushright}

\vskip 3cm
\newpage
\vspace{5cm}
\thispagestyle{empty}

\begin{flushleft}
\vspace{25cm}{\em \large La im'agen anterior corresponde a la galaxia NGC\,7479. \\
Basada en observaciones realizadas con el Telescopio Espacial Hubble, NASA/ESA, y obtenida del Hubble Legacy Archive, en colaboraci'on entre  Space Telescope Science Institute (STScI/NASA), Space Telescope European Coordinating Facility (ST-ECF/ESA) y Canadian Astronomy Data Centre (CADC/NRC/CSA).}
\end{flushleft}

\clearemptydoublepage
\newpage
%
%

\thispagestyle{empty}
\begin{picture}(350,630)
\put (250,130){\em \Large A mis padres}\\
\put (150,110){\em \Large por su inagotable paciencia y comprensi'on.} \\
\put (200,90){\em \Large  Y al resto de mi familia}\\
\put (150,70){\em \Large  por el incondicional apoyo en todos estos a'nos.}\\
\end{picture}

\clearemptydoublepage
\pagenumbering{roman}
\chapter*{{\color{night}Resumen}}
\addcontentsline{toc}{chapter}{\numberline{}{Resumen}}
\thispagestyle{empty}

En esta Tesis presento un estudio espectrosc'opico detallado de una muestra de Regiones H{\sc ii} Gigantes en galaxias observables con telescopios del Hemisferio Sur.

Las Regiones H{\sc ii} Gigantes proveen el v\'{\i}nculo entre regiones de formaci\'on estelar de menor escala y los violentos procesos de formaci\'on estelar que ocurren en las galaxias con brotes estelares intensos. Las Regiones H{\sc ii} Gigantes (GH{\sc ii}Rs, de sus siglas en ingl\'es) son objetos extendidos y luminosos, observados en los discos de galaxias espirales, irregulares y galaxias enanas compactas azules (BCDs, de su sigla en ingl'es). Las GH{\sc ii}Rs se forman debido a la presencia de un gran n\'umero de estrellas j\'ovenes y masivas cuyo flujo ultravioleta ioniza el gas que las rodea, y por lo tanto estas regiones indican una presencia de formaci\'on estelar violenta reciente o activa.
 El principal objetivo de este trabajo de Tesis es confirmar la presencia de Regiones H{\sc ii} Gigantes en galaxias del Universo Local y realizar un estudio comparativo de GH{\sc ii}Rs con distintas metalicidades y estados evolutivos. 
 
La identificaci'on de Regiones H{\sc ii} Gigantes fue posible realizarla mediante espectroscop'ia de alta resoluci'on, la cual me ha permitido distinguir la presencia de distintas componentes cinem'aticas en el gas ionizado, con  diferentes velocidades y verificar la naturaleza supers'onica de dichas componentes. 

Definir la naturaleza y condiciones f'isicas de los brotes de formaci'on estelar masiva permite analizar las abundancias qu'imicas que producen el enriquecimiento qu'imico del Universo Local. Para poder caracterizar los procesos f'isicos que gobiernan la formaci'on estelar masiva es necesario estudiar un variado rango de entornos y propiedades. Gracias al amplio rango espectral y la buena calidad de los datos he llevado adelante el estudio de las propiedades f'isicas del gas ionizado. El estudio de los par'ametros f'isicos tales como densidad, temperatura, composici'on qu'imica, etc., comprende el an'alisis de las intensidades relativas de las l'ineas de emisi'on del gas. Este tipo de an'alisis resulta esencial para poder mejorar el estudio de brotes de formaci'on estelar detectados a distancias cosmol'ogicas. 

En esta Tesis se presenta el estudio realizado en seis regiones brillantes de las galaxias espirales, NGC\,6070 y NGC\,7479, y en cinco brotes de formaci'on estelar de la galaxia enana compacta azul Haro\,15.
 
En el Cap'itulo 1 presento una breve introducci'on de las propiedades generales de las Regiones H{\sc ii} Gigantes y del conocimiento que se tiene sobre este tipo de objetos. 
En el Cap'itulo 2 propongo un m'etodo para proceder a un an'alisis detallado de la cinem'atica en aquellas regiones que presentan m'as de una componente revelada cinem'aticamente en los espectros de alta resoluci'on. En ese cap'itulo presento el an'alisis cinem'atico de las seis regiones estudiadas en las dos galaxias espirales y se analiza c'omo la presencia de m'ultiples componentes cinem'aticas influye en la ubicaci'on final en el plano L-$\sigma$.
En el Cap'itulo 3 presento un an'alisis cinem'atico utilizando una metodolog'ia similar que en el cap'itulo anterior, pero ahora enfocado en los brotes de formaci'on estelar violentos de la galaxia BCD Haro\,15. Del  an'alisis cinem'atico detallado en los brotes estudiados con m'as de una componente, se deriva el estudio de las propiedades f'isicas y abundancias qu'imicas presentado en el Cap'itulo 4. En este cap'itulo, sumado a los datos espectrosc'opicos de alta resoluci'on, presento datos de baja resoluci'on espectral observados en la misma galaxia. Aplicando el m'etodo propuesto por el Grupo de Astrof'isica del Departamento de F'isica Te'orica de la Universidad Aut'onoma de Madrid, Espa'na (UAM), para la determinaci'on de los par'ametros f'isicos y abundancias, a partir de un programa generado por dicho grupo, realic'e un cuidadoso y detallado estudio sobre las propiedades f'isicas, y abundancias qu'imicas i'onicas y totales del gas ionizado discriminado para cada componente cinem'atica y para el flujo global de las l'ineas de emisi'on. 
En el Cap'itulo 5 se presentan las conclusiones finales de la Tesis, haciendo un conciso resumen sobre los aportes m'as significativos, y se describen las l'ineas de trabajo a futuro que se pueden plantear como continuaci'on de lo aqu'i expuesto.

\chapter*{{\color{night}Prefacio}}
\addcontentsline{toc}{chapter}{\numberline{}{Prefacio}}

Parte de esta Tesis ha sido publicada:
\begin{itemize}
\item Cap\'{\i}tulo 2 ha sido publicado como\\
{\bf Firpo, V.}, Bosch, G., {H{\"a}gele}, G.~F., Morrell, N., 2010. {\em
Monthly Notices of the Royal Astronomical Society}, Volume 406, Issue 2, pp. 1094-1107.\\

\item Cap\'{\i}tulo 3 ha sido publicado como\\
{\bf Firpo, V.}, Bosch, G., {H{\"a}gele}, G.~F., D\'{\i}az, A. I., Morrell, N., 2011.  {\em
Monthly Notices of the Royal Astronomical Society}, Volume 414, Issue 4, pp. 3288\\

\item Cap\'{\i}tulo 4 ha sido aceptado para su publicaci'on en {\em Monthly Notices of the Royal Astronomical Society}, 2012. {\em ArXiv e-prints:1203.0531v1}\\
{H{\"a}gele}, G.~F., {\bf Firpo, V.}, Bosch, G.,  D\'{\i}az, A. I., Morrell, N.\\
\end{itemize}

Proceedings de conferencias:
\begin{itemize}
\item Cap\'{\i}tulo 2\\
{\bf Giant Extragalactic HII Regions in the Southern Sky}.\\
{\bf Firpo, V.}, Bosch, G., Morrell, N. I., 2006. {\em Revista Mexicana de Astronom'ia y Astrof'isica Conference Series, Vol. 26 de RMxAA, vol. 27, pp 160}\\

{\bf Massive star formation in external galaxies: new giant HII regions}.\\
 {\bf Firpo, V.}, Bosch, G., Morrell, N. I., 2008. {\em Revista Mexicana de Astronom'ia y Astrof'isica Conference Series, Vol. 33 de RMxAA, vol. 27, pp 171-171}\\
 
 \item Cap\'{\i}tulo 3 y Cap\'{\i}tulo 4\\
{\bf Haro15: Is it actually a low metallicity galaxy?}. \\
 {\bf Firpo, V.}, Bosch, G., {H{\"a}gele}, G.~F., D\'{\i}az, A. I., Morrell, N., 2010. {\em K. Cunha, M. Spite, \& B. Barbuy (ed.), IAU Symposium, Vol. 265 of IAU Symposium, pp 243-244}\\
 
{\bf Internal kinematic and physical properties in a BCD galaxy: Haro 15 in detail}.
{\bf Firpo, V.}, Bosch, G., {H{\"a}gele}, G.~F., D\'{\i}az, A. I., Morrell, N., 2010. Highlights of Spanish Astrophysics VI, Proceedings of the IX Scientific Meeting of the Spanish Astronomical Society (SEA), held in Madrid, September 13 - 17, 2010, Eds.: M. R. Zapatero Osorio, J. Gorgas, J. Ma\'{\i}z Apell\'aniz, J. R. Pardo, and A. Gil de Paz., p. 192-197\\
\end{itemize}

\begin{flushright}
Ver\'onica Firpo \\ La Plata, Argentina, Marzo, 2011. 
\end{flushright}
 
\clearemptydoublepage 


\tableofcontents
\listoftables
\clearemptydoublepage
\listoffigures

\clearemptydoublepage

\pagenumbering{arabic}

{\color{night}\chapter[Generalidades]{Generalidades}}
\label{capitulo1}

Las Regiones H\,{\sc ii} Gigantes (GH{\sc ii}Rs, sigla en ingl\'es de Giant H{\sc ii} Regions) son objetos extendidos y luminosos, con luminosidades en la emisi'on de H$\alpha$ de alrededor de $10^{40}$\ergs, observados en los discos de galaxias espirales, irregulares y galaxias enanas compactas azules (BCDs, sigla en ingl'es de Blue Compact Dwarfs). Las GH{\sc ii}Rs se originan debido a la presencia de un gran n\'umero de estrellas j\'ovenes y masivas cuyo flujo ultravioleta ioniza el gas que las rodea y, por lo tanto, estas regiones indican una presencia de formaci\'on estelar violenta reciente o activa \citep{1994ApJ...425..720C}. Esto 'ultimo se ha podido verificar para el caso de la GH\,{\sc ii}R visible m'as cercana, 30 Doradus en la Nube Mayor de Magallanes (ver Figura 1.1), con espectroscop'ia del n'ucleo del c'umulo estelar, conocido como R136, realizada con el Telescopio Espacial Hubble \citep{MH98}. 
Tienen tama'nos t'ipicos de un centenar de parsecs de di'ametro. Existen menciones en la literatura de tama'nos de un orden de magnitud mayor, pero en general se refieren a regiones m'ultiples, las cuales pueden resolverse en varias componentes \citep{B02}. Las GH\,{\sc ii}Rs podr'ian ser clasificadas, de alguna manera, en una escala intermedia entre regiones de formaci\'on estelar de menor escala, como Ori\'on en nuestra galaxia (ver Figura 1.2), y regiones con violentos procesos de formaci\'on estelar que ocurren en las galaxias con brotes estelares intensos (llamadas starburst en ingl\'es) \citep{1972ApJ...173...25S}.

\begin{figure*}
   \begin{center}
     \includegraphics[width=.75\textwidth]{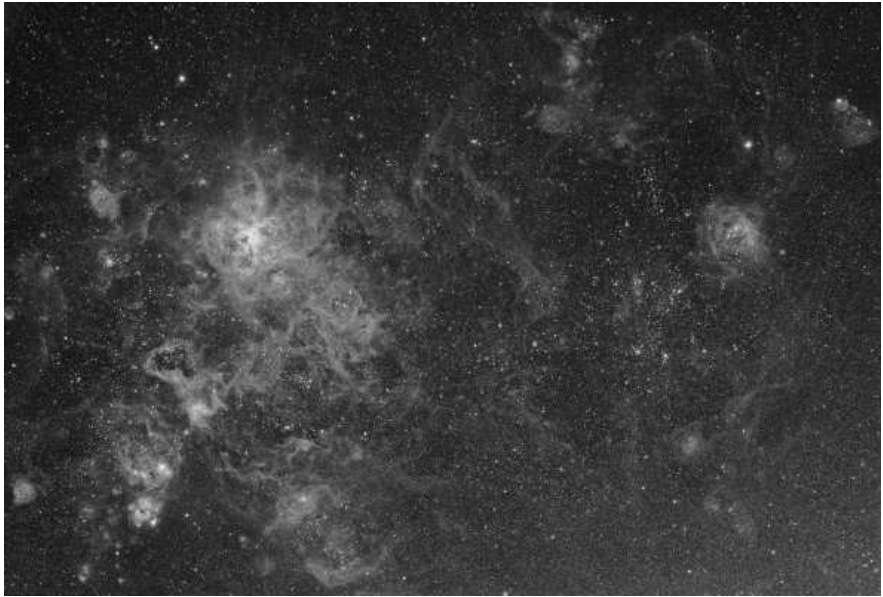}
   \end{center}
   \caption[Imagen H$\alpha$ de la Nebulosa de la Tar'antula 30 Doradus]{Imagen H$\alpha$ de la Nebulosa Tar'antula 30 Doradus, o NGC 2070. Imagen tomada con un telescopio refractor TEC-140 a f/7, en Hakos, Namibia, por Johannes Schedler - PantherObservatory.}
\label{fig:30Dor}
 \end{figure*}

Estas regiones albergan cientos de estrellas masivas que, desde el punto de vista qu'imico y morfol'ogico, son parcialmente responsables de los cambios en la estructura de las galaxias a trav'es de una constante emisi'on de masa a alta velocidad (vientos) o durante uno de los procesos energ'eticos m'as impresionantes del Universo como son las explosiones de supernovas. La gran cantidad de fotones ultravioleta emitida por las estrellas OB j'ovenes y calientes junto con los de las estrellas menos masivas, con menor nivel de emisi'on pero muy numerosas, tiene como consecuencia que en muchas zonas de estas regiones el gas, principalmente hidr'ogeno, est'a altamente ionizado, con temperaturas t'ipicas del orden de 10000 K.

Junto a las zonas de gas caliente ionizado, las GH\,{\sc ii}Rs contienen gas en otros dos estados. Por un lado, se encuentra en un estado molecular fr'io y denso. En este estado el gas puede ser re-generado en zonas m'as densas que pueden formarse por la compresi'on en esas regiones din'amicamente complejas. Es en tales nubes moleculares fr'ias que la actual formaci'on estelar se est'a produciendo. Por otro lado, grandes vol'umenes de gas caliente de baja densidad son producidos por la acci'on combinada de vientos y explosiones de SN que ocurren en la regi'on.
Un entorno tan complejo da lugar a m'as de un evento de formaci'on estelar o a una nueva generaci'on de estrellas. Uno de los principales prop'ositos para estudiar estas regiones radica en ampliar el conocimiento sobre la formaci'on estelar dado que desde muchos aspectos astrof'isicos es esencial para comprender desde la evoluci'on de las galaxias, a trav'es del enriquecimiento qu'imico de las estrellas masivas, hasta la formaci'on de los planetas, con el fin de entender las condiciones iniciales de los discos proto-planetarios. Es muy posible que la mayor'ia de las estrellas que se formaron alguna vez, y la mayor'ia de las que existen en la actualidad, lo hayan hecho en condiciones m'as similares a las de las Regiones H{\sc ii} Gigantes o ``starbursts'', que a las condiciones m'as apacibles en las que se forman hoy en d'ias las estrellas en nuestra vecindad gal'actica.

 \begin{figure*}
   \begin{center}
     \includegraphics[width=.45\textwidth]{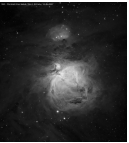}
     \caption[Imagen H$\alpha$ de la Nebulosa de Ori'on (M42)]{Imagen H$\alpha$ de la Nebulosa (M42) de Ori'on. Imagen tomada con un telescopio refractor Takahashi Sky 90 a f/4.5, por Don McCrady}
   \end{center}
\label{fig:Orion}
 \end{figure*}

En los 'ultimos a'nos ha habido grandes progresos para entender la formaci'on estelar. Los modelos te'oricos han tenido 'exito en la reproducci'on de las principales caracter'isticas de las estrellas formadas en las nebulosas, tales como la funci'on inicial de masa (IMF, sigla en ingl'es de Initial Mass Function). Los principales procesos f'isicos que pueden cambiar la masa final de las estrellas y que pueden establecer eficiencia en la formaci'on estelar de las nebulosas son: la fragmentaci'on durante el colapso, la acreci'on de masa, los vientos estelares y la radiaci'on ultravioleta \citep{2001ApJ...556..837K,2003ApJ...585..850M,2005JRASC..99R.132T}. Sin embargo, es muy complicado saber hasta qu'e punto estos procesos son capaces de modificar la funci'on inicial de masa. En la actualidad, hay dos modelos principales que compiten para explicar el origen de la IMF:
i) el modelo que plantea que la IMF estelar se origina en la distribuci'on de masas de los n'ucleos densos de las nubes (CMF, sigla en ingl'es de core mass function) \citep{2007A&A...462L..17A,2008MNRAS.391..205S}. Bajo esta hip'otesis, s'olo se necesita explicar por qu'e la turbulencia en las nubes producir'ia una distribuci'on de masas de los n'ucleos igual a la distribuci'on de masas de las estrellas. Se han construido teor'ias para determinar la CMF a partir de la turbulencia en el medio interestelar \citep{2009arXiv0907.0248P,2008ApJ...684..395H}, y se utilizan simulaciones hidrodin'amicas para comprobar si se verifican las predicciones de las teor'ias (Smith et al.,\citeyear{2009MNRAS.396..830S}a, \citeyear{2009MNRAS.400.1775S}b).
ii) El otro modelo alternativo del origen de la IMF es el de la ``acreci'on competitiva'' de Bate \& Bonnell~(\citeyear{2003MNRAS.339..577B}). En este modelo, las estrellas adquieren su masa compitiendo por el material disponible para ser acretado. En este caso la turbulencia es poco relevante para las masas finales de las estrellas, siendo s'olo el mecanismo que originalmente fragmenta a la nube. Este modelo ha sido criticado por requerir, aparentemente, condiciones demasiado ``restrictivas'' para las estrellas en formaci'on, y que posiblemente s'olo puedan darse en la formaci'on de c'umulos estelares muy poblados. Como se puede ver, hoy en d'ia las discusiones siguen abiertas en este tema.

Por otro lado, el advenimiento de la generaci'on de los grandes telescopios ha permitido estudiar y analizar las regiones de formaci'on estelar no solo del Universo cercano, sino en galaxias distantes, como por ejemplo, las galaxias H{\sc ii} que constituyen un potente indicador de distancia a escalas cosmol\'ogicas. Ya han sido identificadas numerosas galaxias H{\sc ii} en los seguimientos realizados con el Sloan Digital Sky Survey (SDSS) (que en castellano significa ``Relevamiento Digital del Cielo Sloan'').

 \begin{figure*}
   \begin{center}
 \includegraphics[width=.55\textwidth]{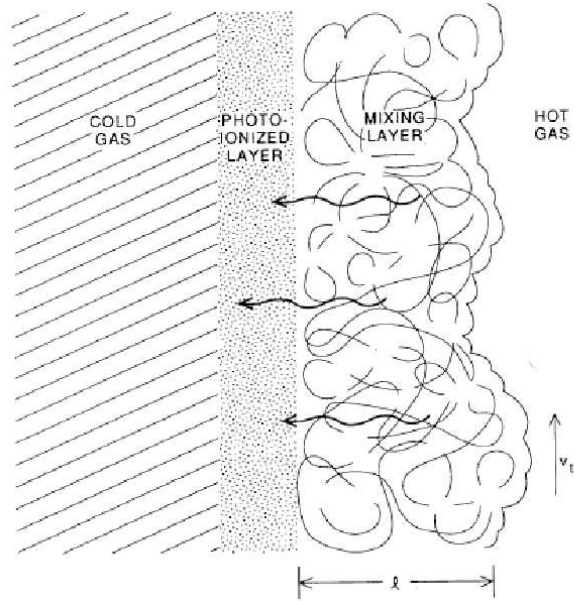}
      \caption[Dibujo esquem'atico que simula la geometr'ia de las capas de mezcla turbulenta]{Dibujo esquem'atico que simula la geometr'ia de las capas de mezcla turbulenta mostrando el gas frio y caliente y la capa fotoionizada lindera a la capa de mezcla. El gas caliente se mueve a una velocidad transversal v$_{t}$ relativa a la capa de mezcla \citep{1993ApJ...407...83S}. La figura pertenece al mismo trabajo.}
   \end{center}
\label{fig:mixinglayer}
 \end{figure*}
 
Las galaxias H{\sc ii} son galaxias con formaci\'on estelar violenta que producen una se\~nal espectral fuerte e inconfundible la cual puede detectarse hasta z$\sim$3 \citep{1997ApJ...481..673L}. Melnick, Terlevich \& Terlevich~(\citeyear{2000MNRAS.311..629M}) propusieron el uso de las galaxias H{\sc ii} como una sonda cosmol\'ogica que resulta \'util para medir la densidad de masa del Universo ($\Omega_M$). Sin embargo este m\'etodo que es potencialmente muy poderoso, se basa en la calibraci\'on del punto cero de la regresi\'on  observada en el plano fundamental para Regiones H{\sc ii} Gigantes Extragal\'acticas (su sigla en ingl'es es GEH\,{\sc ii}Rs).

\section{Caracter\'{\i}sticas de las GH\,{\sc ii}Rs}
Smith \& Weedman ~(\citeyear{SW70}) encontraron que los anchos de los perfiles de las l\'{\i}neas de emisi\'on de las GH{\sc ii}Rs corresponden a velocidades supers\'onicas en el gas. Dicha fuente cinem\'atica observada permanece a\'un sin una clara explicaci\'on, pero permite identificar y distinguir una regi\'on H{\sc ii} gigante de una mera aglomeraci\'on de regiones H{\sc ii} cl\'asicas por medio de espectroscop'ia de alta resoluci\'on.

\begin{figure*}
   \begin{center}
 \includegraphics[angle=270,width=.55\textwidth]{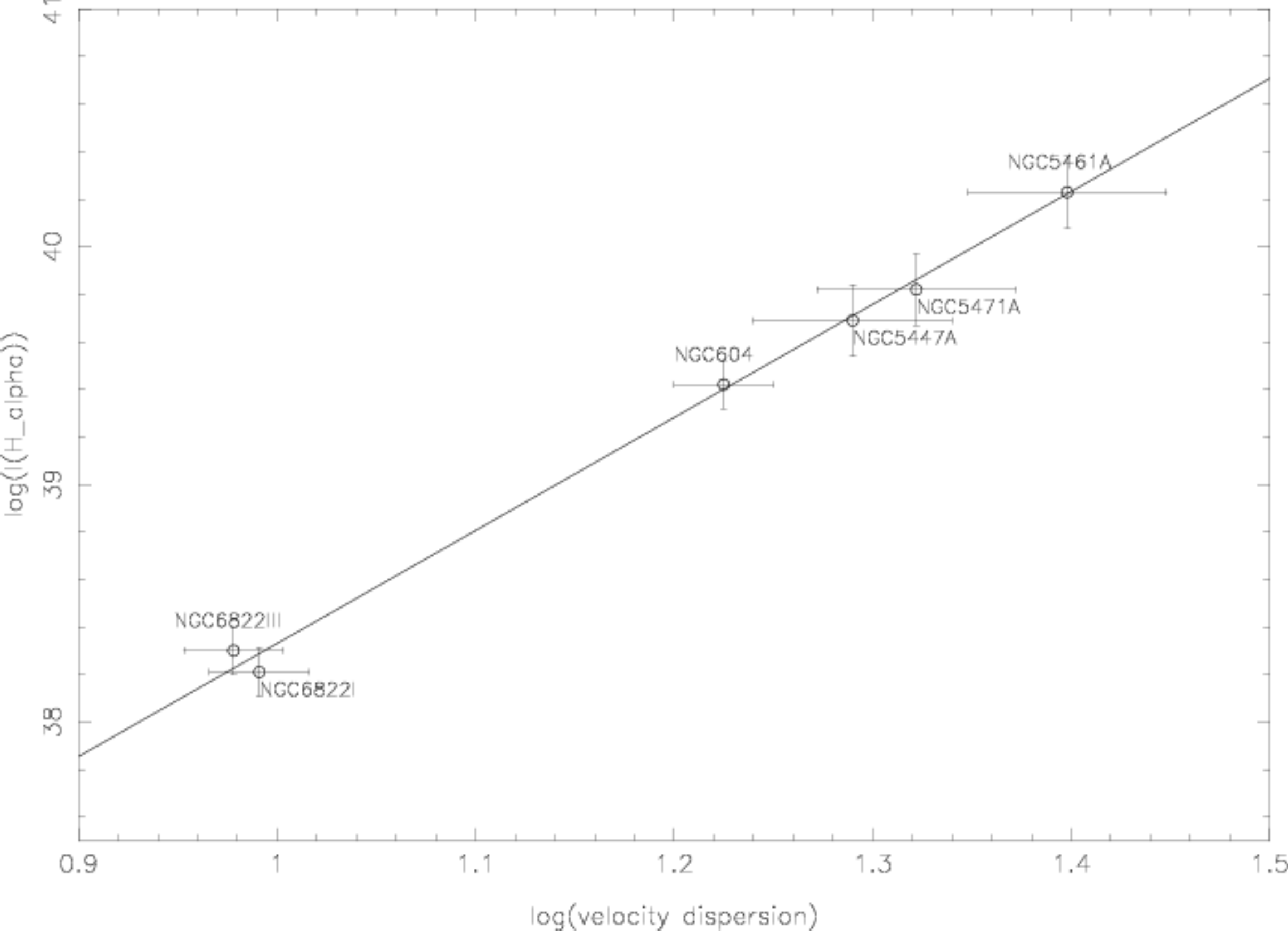}
      \caption[Regresi'on en el plano Luminosidad (L) vs. Dispersi'on de Velocidades ($\sigma$) de Bosch et al.~(\citeyear{B02})]{Regresi'on en el plano Luminosidad (L) vs. Dispersi'on de Velocidades ($\sigma$) de Bosch et al.~(\citeyear{B02}). La figura pertenece al mismo trabajo.}
   \end{center}
\label{fig:regresion}
 \end{figure*}

Melnick~(\citeyear{M77}) encontr\'o una correlaci\'on estrecha entre la luminosidad de las GH{\sc ii}Rs y la dispersi\'on de velocidades observada en el perfil de las l\'{\i}neas de emisi\'on. Diversas investigaciones siguieron a este descubrimiento y confirmaron la existencia de una correlaci\'on, aunque exhib\'{\i}an desacuerdos en las pendientes encontradas para dicha regresi\'on \citep{TM81, R86, B02}. Una gran variedad de mecanismos son los responsables del ensanchamiento que presentan los perfiles de las l'ineas de emisi'on en el espectro, lo que ha originado var'ias teor'ias paralelas para explicar dicho fen'omeno, invocando din'amica de sistemas virializados \citep{TT93}, superposici'on de m'ultiples burbujas de gas en expansi'on \citep{1994ApJ...425..720C}, o turbulencia del mismo gas interestelar \citep{MT97}. Bosch y colaboradores obtuvieron fotometr'ia CCD de banda angosta en l'ineas de emisi'on para un grupo de GH\,{\sc ii}Rs, analizando c'omo incid'ia la incerteza de los flujos publicados sobre las discrepancias de los valores hallados para la regresi'on en el plano Luminosidad (L) vs. Dispersi'on de Velocidades ($\sigma$). Se hall'o una doble lectura, ya que los nuevos datos fotom'etricos no mejoraban la dispersi'on de los puntos en la regresi'on. Sin embargo, los autores realizaron un an'alisis m'as detallado de las regiones, considerando informaci'on sobre el estado evolutivo de algunas GH\,{\sc ii}Rs. Malumuth et al.~(\citeyear{M96}) y Drissen et al.~(\citeyear{D93}) analizaron la componente estelar de NGC 595, una de las GH\,{\sc ii}Rs de la galaxia espiral M33, y encontraron signos de evoluci'on. Con espectroscop'ia Fabry-Perot de alta resoluci'on espacial y espectral de algunas regiones de M33 y M101, Mu\~noz Tu\~n'on et al.~(\citeyear{MT95,1996AJ....112.1636M}) resolvieron el perfil supers'onico global observado en varias componentes discretas, algunas con anchos subs'onicos. De acuerdo al modelo de evoluci'on propuesto por Mu\~n'oz-Tu\~n'on et al.~(\citeyear{1996AJ....112.1636M}), cuando las burbujas de gas sopladas por estrellas masivas dentro de las regiones H\,{\sc ii} gigantes alcanzan el borde de la regi'on, alteran el perfil global observado de la l'inea de recombinaci'on del gas. Al descartar las regiones con evidentes signos de evoluci'on, Bosch et al.~(\citeyear{B02}) obtuvieron una regresi'on muy estrecha en el plano L vs. $\sigma$, aunque basada en una muestra muy reducida (ver Figura 1.4). Si se desea estudiar la relaci\'on entre luminosidad y dispersi\'on de velocidades en una muestra de GH{\sc ii}Rs que tenga sentido estad\'{\i}stico es necesario descubrir nuevas Regiones H{\sc ii} Gigantes. 

\begin{figure*}
   \begin{center}
\includegraphics[width=.70\textwidth]{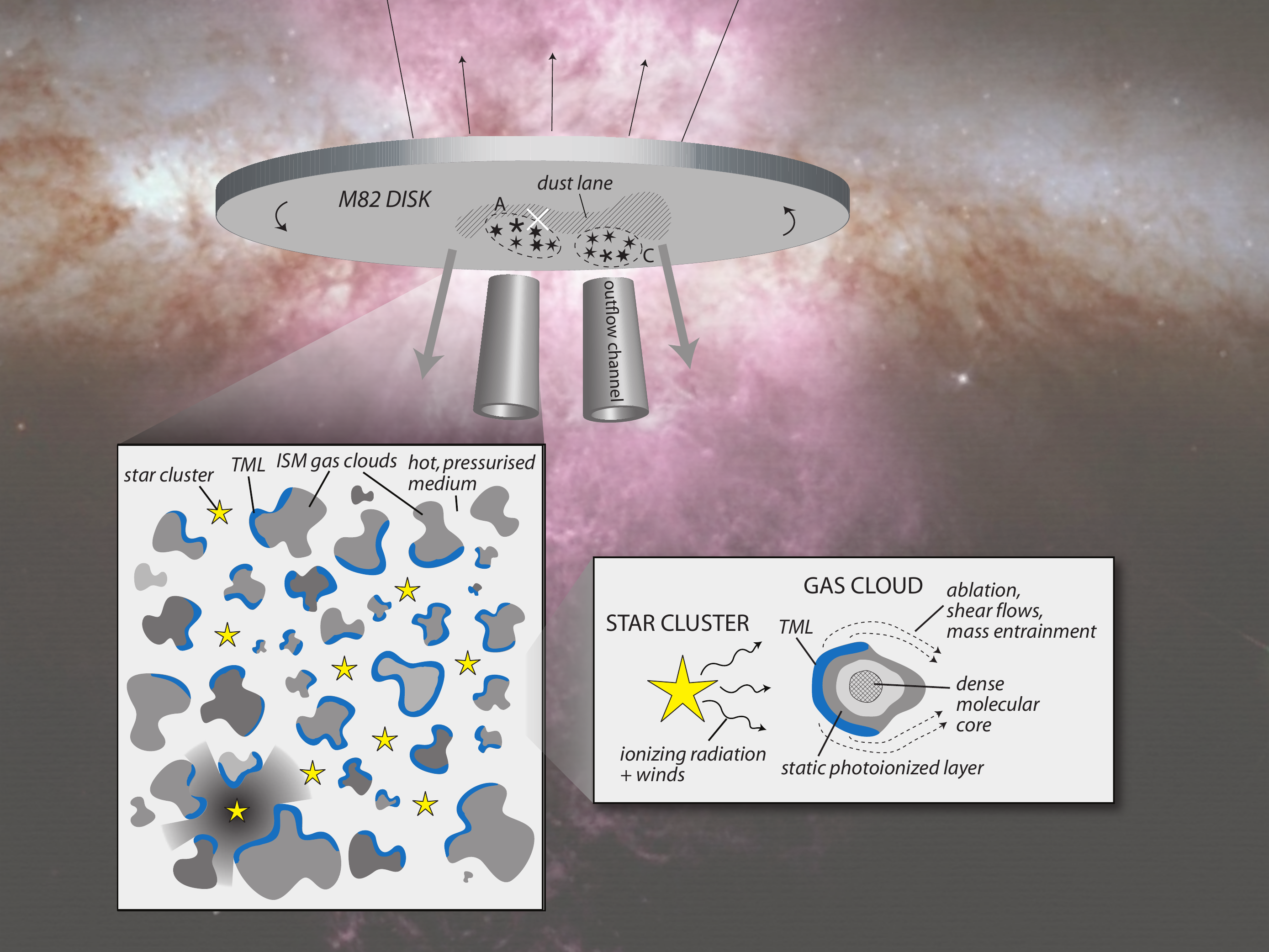}
     \caption[Dibujo esquem'atico que simula el estado del ISM en la galaxia M82]{Dibujo esquem'atico que representa las ideas desarrolladas en el trabajo de Westmoquette et al.,~(\citeyear{2009ApJ...706.1571W}) sobre el estado del ISM en la galaxia M82. El dibujo esquem'atico pertenece a dicho trabajo y la composici'on es de autor'ia de Linda Smith.}
   \end{center}
\label{fig:M82ism}
 \end{figure*}

Varios estudios han tratado el hecho de que un perfil simple Gaussiano puede no ser suficientemente realista a la hora de ajustar las l'ineas de emisi'on observadas. Y en varios estudios se ha favorecido la presencia de una componente ancha omnipresente que podr'ia estar explicando las alas del perfil integral (ver Figura 1.6) \citep{1996AJ....112.1636M,M99,1987MNRAS.226...19D,1996MNRAS.279.1219T,1997ApJ...488..652M,1999ApJ...522..199H,H07,2009MNRAS.396.2295H,Hagele+10}. Westmoquette et al.,~(\citeyear{2007MNRAS.381..894W,2007MNRAS.381..913W}) concluyen que la componente angosta representa el medio interestelar (ISM, sigla en ingl'es de Interestellar Medium) ionizado y perturbado, que surge a trav'es de la convoluci'on de los efectos de agitaci'on del gas debido al brote de formaci'on estelar y los movimientos gravitacionales del material virializado. Y por otro lado, la componente ancha resulta del campo de velocidad altamente turbulento asociado con la interacci'on de la fase caliente del ISM con nudos de gas m'as fr'io (ver Figura 1.3), lo que genera capas de mezcla turbulenta (ver Figura 1.5).
Sin embargo, otros estudios han apoyado la existencia de perfiles tipo ``shell'' con dos alas, azul y roja, a ambos lados de la componente principal \citep{1994ApJ...425..720C,Relano05,2006A&A...455..539R}.

\begin{figure*}
   \begin{center}
\includegraphics[width=.60\textwidth]{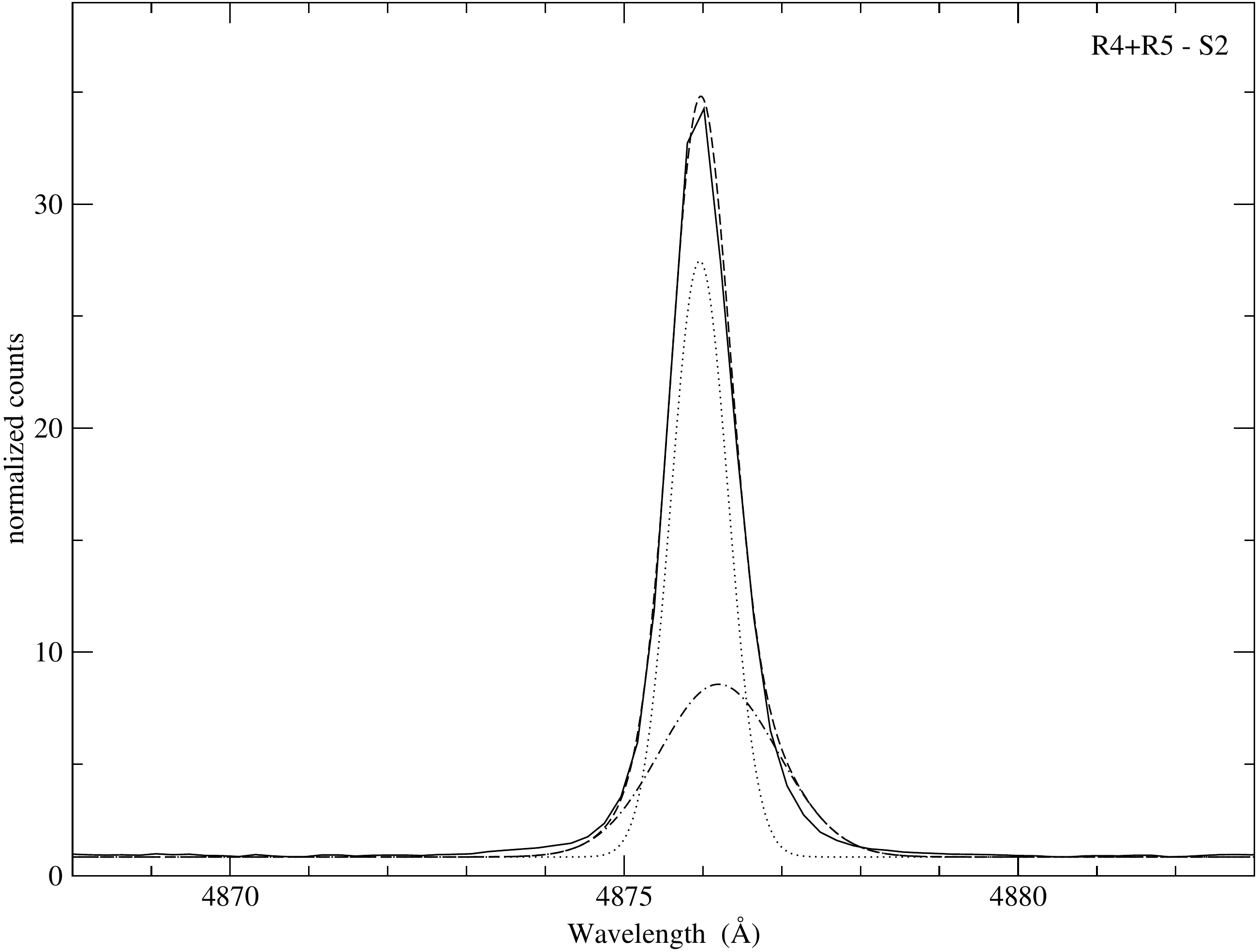}
     \caption[Ajuste de dos componentes en el perfil de emisi'on de $H\beta$ \citep{Hagele+10}]{Ajuste de dos componentes en el perfil de emisi'on de $H\beta$ en una regi'on H{\sc ii} circunnuclear de la galaxia NGC\,3310 \citep{Hagele+10}. La figura pertenece al mismo trabajo.}
   \end{center}
\label{fig:NGC3310}
 \end{figure*}
 
Para verificar la naturaleza gigante de estas regiones H{\sc ii} es necesario realizar espectroscop\'{\i}a de alta resoluci\'on. La espectroscop\'{\i}a 'echelle proporciona un medio para buscar diferentes componentes cinem\'aticas en las l\'{\i}neas de emisi\'on del gas y analizar estos efectos, ya que alcanza la resoluci\'on espectral necesaria tanto para resolver los anchos de los perfiles de las l\'{\i}neas de emisi\'on como para determinar las dispersiones de velocidades intr\'{\i}nsecas del gas ionizado y verificar si estas son supers\'onicas ($>$13 \kms). 

La principal fuente de informaci\'on sobre la metalicidad en todo tipo de objetos, desde los m\'as cercanos hasta los m\'as distantes, lo constituyen las regiones H\,{\sc ii}, desde las menos brillantes en el disco de la Galaxia, pasando por las GEH\,{\sc ii}Rs hasta las galaxias H\,{\sc ii}. Sus espectros presentan l\'{\i}neas de emisi\'on brillantes las cuales son visibles en todas las regiones donde se hayan producido episodios de formaci\'on estelar recientes. El an\'alisis de dichos espectros nebulares constituye el m\'etodo m'as efectivo, a grandes distancias, para determinar las abundancias qu\'{\i}micas de elementos tales como el helio, nitr\'ogeno, ox\'{\i}geno, ne\'on, arg\'on o azufre que dispongan de l\'{\i}neas de emisi\'on en el rango \'optico correspondientes a diferentes estados de ionizaci\'on. El conocimiento detallado de estas abundancias resulta esencial para comprender la evoluci\'on del gas y de las estrellas, y esta informaci'on permite ampliar el conocimiento que se tiene acerca de la evoluci\'on qu\'{\i}mica de las galaxias en el Universo Local. Con el advenimiento de los grandes telescopios, es posible disponer de datos para objetos menos brillantes y/o m\'as lejanos, que acrecienta el conocimiento que se tiene del Universo m\'as lejano y por tanto m\'as joven.

\section{Motivaci'on y Objetivos}
Realizando una inspecci'on en la lista de GH\,{\sc ii}Rs conocidas a comienzos de los a'nos 2000, hab'iamos encontrado que todas pertenec'ian al hemisferio norte, salvo 30 Doradus, en la Nube Mayor de Magallanes. Feinstein~(\citeyear{F97}) hab'ia estudiado la funci\'on de luminosidad de unas 10 galaxias del cielo austral, entre las cuales nosotros pudimos identificar al menos 40 regiones con luminosidades comparables a las de una GH{\sc ii}R. Ocho de esas galaxias presentan regiones H\,{\sc ii} con brillos superiores a $10^{40}$ erg s$^{-1}$, involucrando un total de 31 candidatos a RH\,{\sc ii}G. Si se extiende la muestra a las regiones con luminosidades mayores a $10^{39}$ erg s$^{-1}$, el n'umero crece hasta unos 70 objetos. Para comenzar a hacer un estudio de GH\,{\sc ii}Rs con un sentido estad'istico, esta muestra de regiones H\,{\sc ii} era ideal como puntapi'e inicial. Es as'i que de la muestra de Feinstein, seleccionamos las regiones m\'as luminosas y estas pasaron a ser nuestros primeros candidatos a estudiar. En \cite{Firpo05} obtuvimos datos de 4 candidatos, todas regiones H\,{\sc ii} muy luminosas, en las galaxias NGC\,7552 y NGC\,2997 utilizando el espectr\'ografo Magellan Inamori Kyocera Echelle (MIKE) del Telescopio Magallanes, LCO. La alta relaci\'on se\~nal-ruido sumada a la resoluci\'on del 'echelle nos permiti\'o resolver el perfil de las l\'{\i}neas de emisi\'on y calcular la dispersi\'on de velocidades del gas ionizado. Esto se hizo midiendo el ancho observado del perfil, que luego fue corregido por ensanchamiento t\'ermico, y por el ancho intr\'{\i}nseco del perfil instrumental, para obtener, as\'{\i} el ensanchamiento verdadero debido al comportamiento cinem\'atico de la Regi\'on H\,{\sc ii}. De ese an\'alisis descubrimos la naturaleza gigante de tres de las cuatro regiones candidatas y pudimos confirmar que el mecanismo de excitaci\'on de las mismas se deb'ia a fotoionizaci\'on por estrellas. Adem'as, encontramos que las regiones H\,{\sc ii} gigantes descubiertas se ubican, dentro de las incertezas, en la regresi\'on esperada para este tipo de objetos, en el plano L-$\sigma$. Uno de los principales resultados, presente en los datos publicados por \cite{Firpo05} y del an'alisis de 5 regiones H{\sc ii} observadas posteriormente, es que todas las Regiones H{\sc ii} Gigantes de las galaxias de la muestra de Feinstein presentan signos de evoluci'on qu'imica y es necesario observar regiones m'as j'ovenes.

Entonces, con el objetivo de detectar regiones m\'as j\'ovenes que permitan realizar un estudio comparativo de GH{\sc ii}Rs a distintas metalicidades y estados evolutivos, seleccionamos brotes de formaci\'on estelar intensa a partir de fotometr\'{\i}a de l\'{\i}neas de emisi\'on de galaxias m\'as j\'ovenes, tales como las enanas compactas azules estudiadas por Cair\'os et al.~(\citeyear{2001ApJS..133..321C}) y Gil de Paz et al.~(\citeyear{2003ApJS..147...29G}). Detectamos que varios de los brotes observados en dichas galaxias presentan caracter\'{\i}sticas de GH{\sc ii}Rs. De esta manera seleccionamos los brotes m\'as brillantes en algunas galaxias de baja metalicidad, siendo estos los nuevos candidatos a estudiar. 

Mediante espectroscop\'ia 'echelle de alta resoluci'on y espectroscop'ia de dispersi'on simple obtenidas en distintos telescopios del Observatorio de Las Campanas (Chile), estudiamos los brotes m\'as brillantes en diversos objetos. La muestra incluye galaxias espirales de alta metalicidad como NGC\,6070 (Figura 1.7), NGC\,7479 (Figura 1.8) \citep{2010MNRAS.406.1094F} y la galaxia de baja metalicidad Haro\,15 (Figura 1.9) del cat\'alogo de Cair\'os et al.~(\citeyear{2001ApJS..133..321C}) \citep{Firpo+11}.

\begin{figure*}
   \begin{center}
 \includegraphics[width=.65\textwidth]{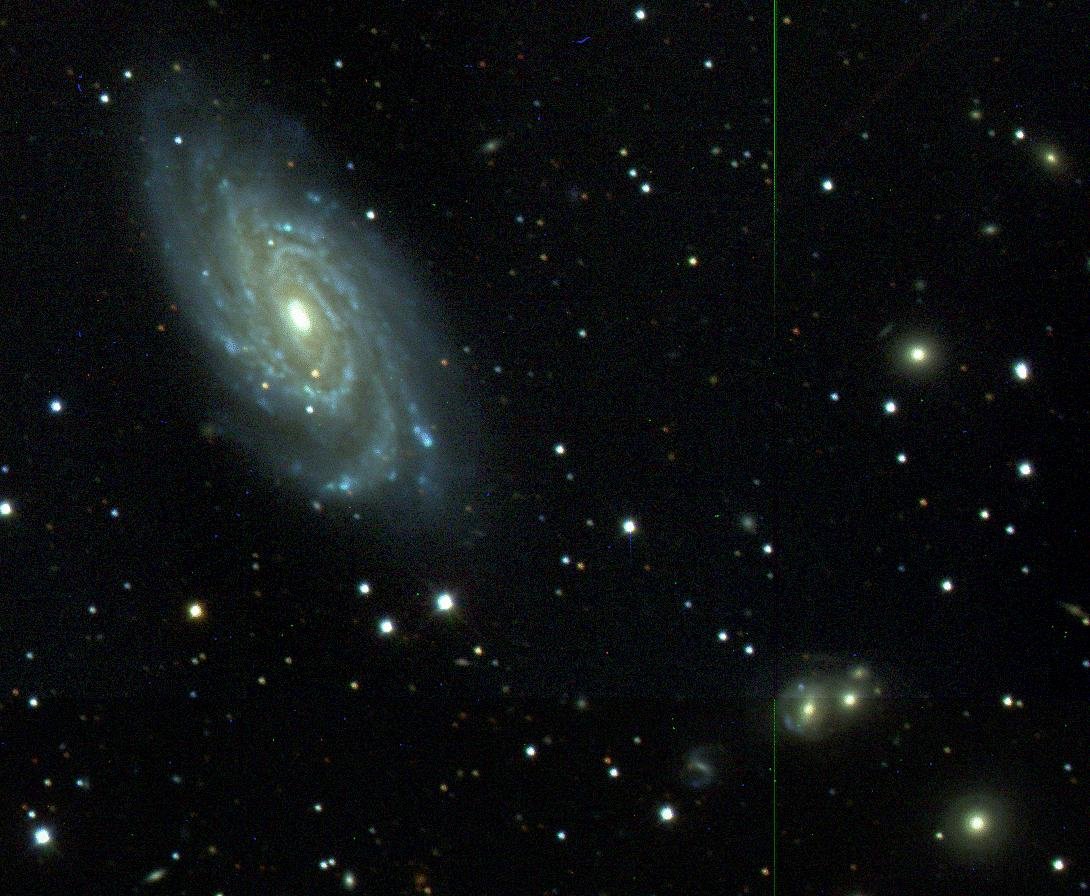}
      \caption[Imagen de la galaxia NGC\,6070]{Galaxia NGC\,6070. Esta fue la primer imagen (imagen de prueba) tomada por el Sloan Digital Sky Survey el 10 de junio de 1998.}
   \end{center}
\label{fig:ngc6070}
 \end{figure*}

\begin{figure*}
   \begin{center}
 \includegraphics[width=.55\textwidth]{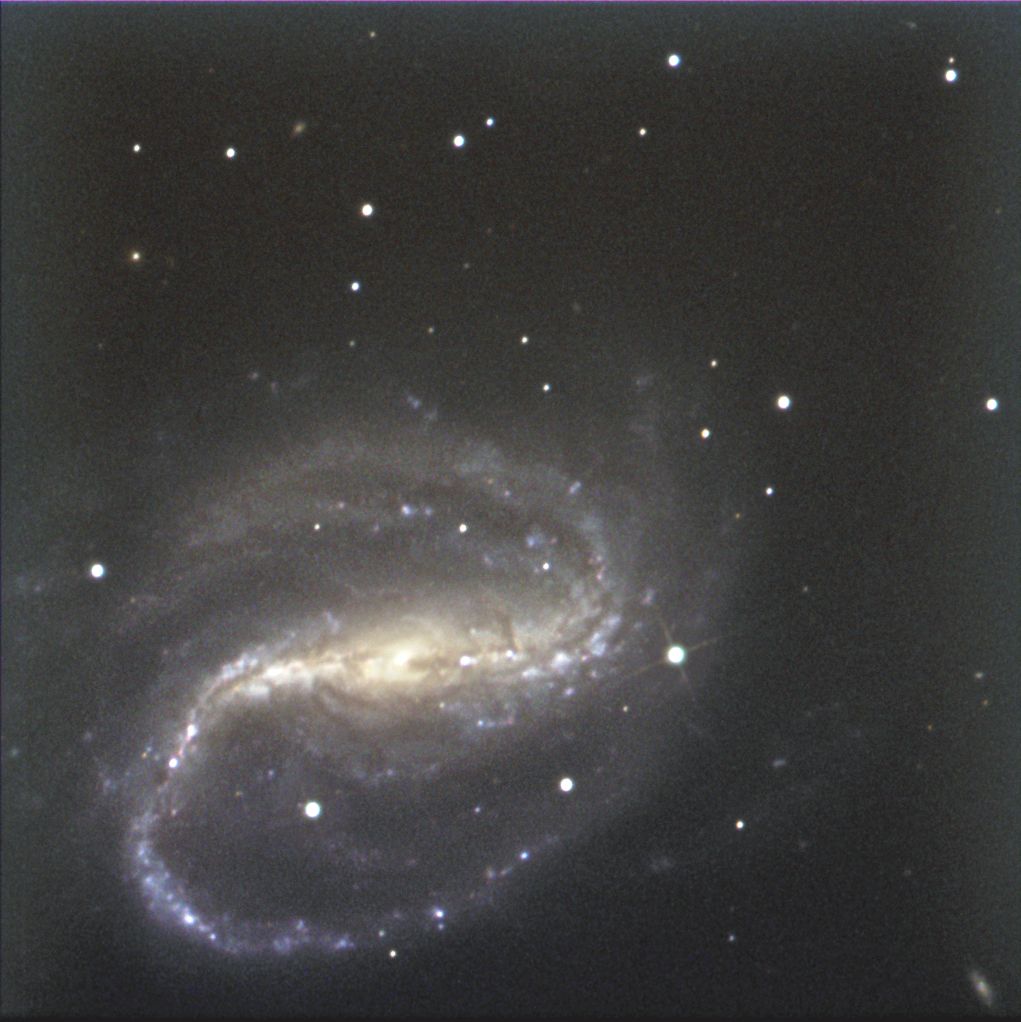}
      \caption[Imagen de la galaxia NGC\,7479]{Galaxia NGC\,7479. Tomada por Nik Szymanek y obtenida con el Telescopio Faulkes Norte, Maui, copyright FTLLC}
   \end{center}
\label{fig:ngc7479}
 \end{figure*}
 
 \begin{figure*}
   \begin{center}
 \includegraphics[width=.45\textwidth]{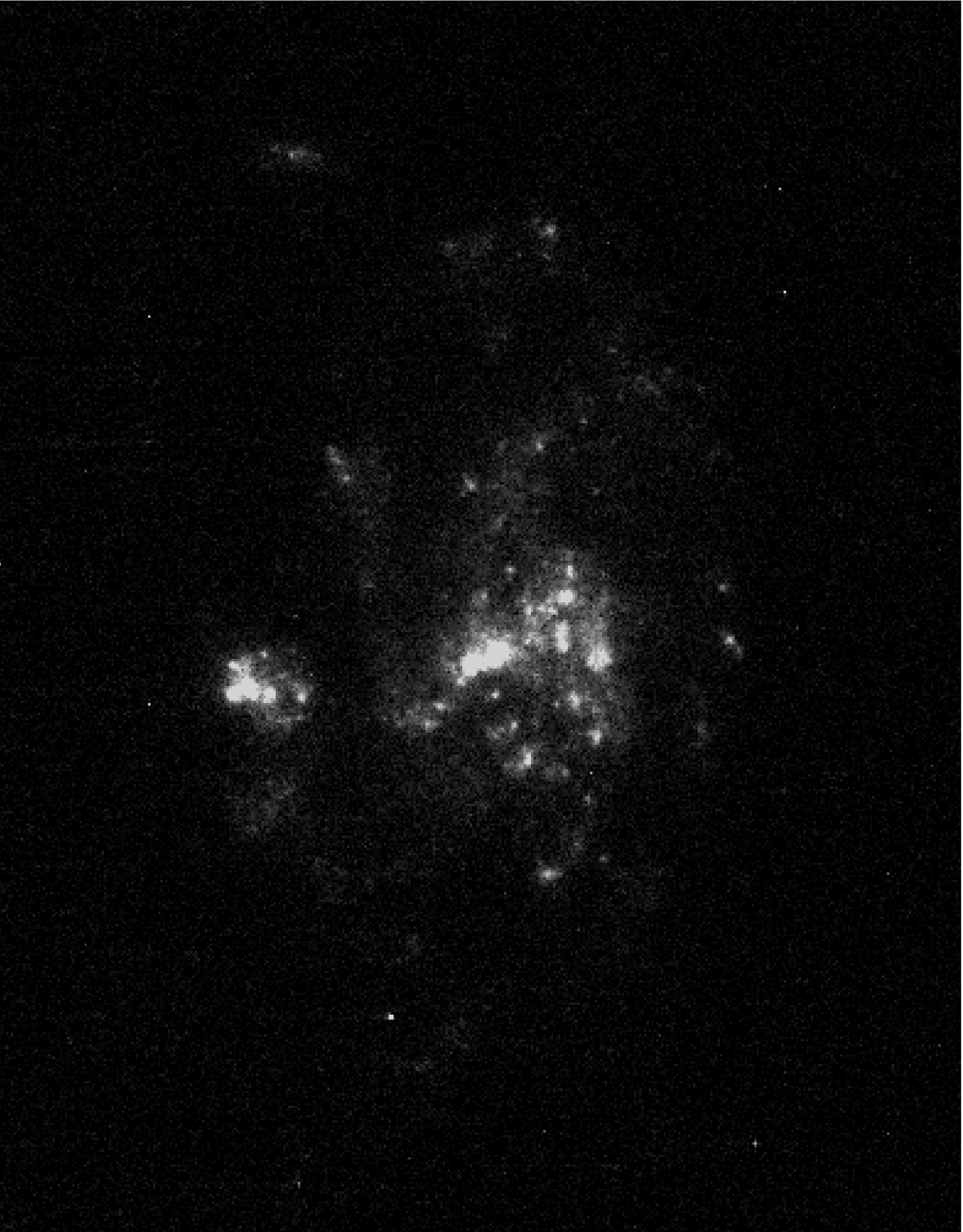}
      \caption[Imagen de la galaxia Haro\,15]{Galaxia Haro\,15. Imagen H$\alpha$ tomada con la c'amara Wide Field Planetary Camera 2, obtenida del Multimission Archive del  Space Telescope Science Institute (MAST).}
   \end{center}
\label{fig:haro15}
 \end{figure*}

A partir del ajuste de un perfil simple Gaussiano a las l\'{\i}neas de emisi\'on, observamos un residuo presente en las alas de los perfiles en varias l\'{\i}neas de emisi\'on, evidente en la l\'{\i}nea H$\alpha$ y que se confirma en otras l\'{\i}neas de emisi\'on. Para determinar el n\'umero \'optimo de componentes Gaussianas y ajustar cada perfil de l\'{\i}nea, y con el fin de minimizar las incertezas en la dispersi\'on de velocidades del gas ionizado, consideramos la informaci\'on proporcionada en la literatura a partir de la gran variedad de estudios propuestos para interpretar la existencia del ensanchamiento en las l\'{\i}neas de emisi\'on. Adem'as de la componente relativamente angosta y siempre que fue posible, evaluamos la eventual presencia de una componente ancha supers\'onica medida en el perfil de la l\'{\i}nea de emisi\'on de la GH{\sc ii}R, o la presencia de dos componentes sim\'etricas de baja intensidad en el ajuste de los anchos de los perfiles de l\'{\i}nea de emisi\'on observados. 

En esta Tesis, confirmando lo ya mencionado en la literatura, presento evidencia de la existencia de diferentes componentes cinem\'aticas en las l\'{\i}neas de emisi\'on del gas. Adem'as, en algunas regiones pudimos detectar la presencia de m\'as de una sub-componente angosta, cuyas velocidades radiales relativas no pueden explicarse mediante fluctuaciones en la rotaci\'on del disco gal\'actico \citep{2010MNRAS.406.1094F, Firpo+11}.

Para avanzar en el an\'alisis de las propiedades f\'{\i}sicas de este tipo de objetos y en las diferentes componentes encontradas, debemos poder determinar temperaturas, densidades y abundancias qu\'{\i}micas del gas. El an'alisis de dichos par'ametros f'isicos puede llevarse adelante mediante el estudio de las intensidades relativas de las l'ineas de emisi'on del gas. Para ello necesitamos tambi\'en una amplia cobertura espectral. La calidad \'unica de nuestros datos permiti\'o realizar un an\'alisis detallado de las caracter\'{\i}sticas del gas ionizado discriminado para cada componente y para el flujo global de las l'ineas. Para llevar esto a cabo, se ha aplicado el m\'etodo propuesto por el Grupo de de Astrof\'{\i}sica del Departamento de F\'{\i}sica Te\'orica de la Universidad Aut\'onoma de Madrid, Espa'na, para la determinaci\'on de los par\'ametros f\'{\i}sicos y abundancias. 

En esta Tesis presento resultados sobre la estimaci\'on de temperaturas, densidades y abundancias qu\'{\i}micas discriminadas por las diferentes componentes cinem'aticas del gas. 

\clearemptydoublepage

\clearemptydoublepage
\newpage

\clearemptydoublepage
\newpage

{\color{night}\chapter[Regiones H{\sc ii} Gigantes en NGC\,7479 \& NGC\,6070]{Regiones H{\sc ii} Gigantes en NGC\,7479 \& NGC\,6070}}
\label{capitulo2}
\begin{center}
\vskip 0.1cm
{\large \em Giant H{\sc ii} Regions in NGC\,7479 \& NGC\,6070} \\
\vskip 0.2cm
{\large \cal Firpo, V., Bosch, G., H\"agele, G.~F., and Morrell, N., 2010, MNRAS, Vol 406, Issue 2, pp. 1094}\\
\end{center}

\section{Resumen de la publicaci'on}

En este Cap'itulo se presentan nuevos resultados de la b'usqueda de Regiones H\,{\sc ii} Gigantes en dos galaxias visibles desde el hemisferio sur:\, NGC\,7479 y NGC\,6070. Usando espectros de alta resoluci'on, obtenidos con diferentes instrumentos del Observatorio Las Campanas, Chile, se han podido resolver los anchos de los perfiles de las l'ineas de emisi'on y determinar la dispersi'on de velocidades intr'inseca del gas ionizado. Se ha detectado que los anchos de los perfiles de emisi'on corresponden a dispersiones de velocidades supers'onicas en las seis regiones H\,{\sc ii} observadas.
Se ha encontrado que todos ellos muestran, al menos, dos componentes cinem'aticas diferentes: una componente relativamente angosta (entre ~11 y ~22 \kms) y otra componente m'as ancha (entre ~31 y ~77 \kms). Dos de las regiones estudiadas muestran un perfil complejo en las l'ineas de emisi'on  angostas de los diferentes iones, el cual pudo ser dividida en dos componentes con distintas velocidades radiales. Mientras tanto, el ensanchamiento de las alas del perfil global pudo ser ajustado con una componente ancha en casi todos los perfiles de las l'ineas de emisi'on observadas. En una regi'on este residuo pudo ser reproducido como perfiles tipo ``shell'' (c'ascaras) en expansi'on debido a la presencia de dos componentes separadas a ambos lados de las componentes principales. Adem'as, se ha analizado el impacto que tiene la presencia de varias componentes sobre la ubicaci'on de la regi'on H\,{\sc ii} en el plano $\log(L) - \log(\sigma)$. Aunque la distribuci'on global confirma la presencia de una regresi'on, la ubicaci'on precisa de las regiones en el plano depende fuertemente de las propiedades de las componentes derivadas en el ajuste del perfil de emisi'on.

\section{Fundamentos b'asicos}
\label{sec:Fundamentos b'asicos}
Con el fin de verificar la naturaleza de nuestras regiones candidatas a Regiones H\,{\sc ii} Gigantes se obtuvieron espectros de alta resoluci'on para poder medir los anchos en los perfiles de las l'ineas de emisi'on y as'i estimar si la dispersi'on de velocidades es, en efecto, supers'onica. Para determinar el n'umero 'optimo de componentes Gaussianas que mejor se ajusta a cada perfil de l'inea, y con el fin de minimizar las incertezas en la dispersi'on de velocidades del gas ionizado, se ha considerado la informaci'on disponible en la literatura. Algunos trabajos previos han tratado el hecho de que un perfil simple Gaussiano puede no ser suficientemente realista a la hora de ajustar las l'ineas de emisi'on observadas. Varios estudios han favorecido la presencia de una componente ancha omnipresente que podr'ia estar explicando las alas del perfil integral \citep[entre otros]{1996AJ....112.1636M,M99,1987MNRAS.226...19D,1996MNRAS.279.1219T,1997ApJ...488..652M,1999ApJ...522..199H,H07,2009MNRAS.396.2295H,Hagele+10}.
En cambio, otros estudios han apoyado la existencia de perfiles tipo shell con dos alas, azul y roja, a ambos lados de la componente principal \citep{1994ApJ...425..720C,Relano05,2006A&A...455..539R}.

En este trabajo se presentan los datos 'echelle obtenidos en el Observatorio Las Campanas (LCO) de seis regiones candidatas a GH{\sc ii}Rs en las dos galaxias mencionadas. Se ha determinado la dispersi'on de velocidades y evaluado la posible presencia de una componente ancha o dos componentes sim'etricas a ambos lados de las componentes angostas en el ajuste de los anchos de los perfiles de l'ineas de emisi'on observados. En la publicaci'on a continuaci'on se presentan las observaciones y la reducci'on de los datos en la Secci'on {\sc Observations and Reductions}; en la Secci'on {\sc Analysis of line profiles} los detalles del an'alisis realizado sobre los perfiles de las l'ineas de emisi'on y se analiza cada caso individualmente; en la Secci'on {\sc Relation between H$\alpha$ Luminosities and Velocity Dispersion} se analiza el impacto que tiene la presencia de varias componentes sobre la ubicaci'on de la regi'on H\,{\sc ii} en el plano $\log(L) - \log(\sigma)$; por 'ultimo se incluye un resumen y las conclusiones del trabajo en la Secci'on nombrada como {\sc Summary and Conclusions}.

\section{Herramientas utilizadas}

De acuerdo a la denominaci'on can'onica de GH{\sc ii}Rs, los candidatos han sido nombrados siguiendo el rango de brillo publicado en el cat'alogo de Feinstein~(\citeyear{F97}). La alta resoluci'on espectral se obtuvo con el espectr'ografo 'echelle del telescopio de 100"\ du Pont, del Observatorio Las Campanas (LCO), observados en 2006. El rango espectral abarc'o desde 3800 hasta 9500 \AA. Las condiciones de observaci'on fueron buenas con un seeing de un segundo de arco. Con el fin de minimizar la contribuci'on del ruido en la lectura del espectro final, se aplic'o un binning 2$\times$2 al CCD. La resoluci'on espectral alcanzada en los espectros 'echelle (obtenido con una ranura de 1 segundo de arco de ancho) es de R$\simeq$25000:  $\Delta\lambda$=0.25\AA\ a los $\lambda$ 6000\AA, medida a partir del ancho a potencia mitad ({\sc FWHM}) de las l'ineas de la l'ampara de comparaci'on de Thorio-Arg'on. Esto se traduce en una resoluci'on en velocidad radial de $\sim$12 \kms. 

Durantes las noches de observaci'on se tomaron espectros de estrellas est'andares, HR\,7950, HR\,4468, y una est'andar espectrofotom'etrica de CALSPEC, Feige\,110, con el prop'osito de calibrar en flujo los espectros. Adem'as, se tomaron todas las noches espectros de comparaci'on de Th-Ar, milky flats (flats de cielo obtenidos con un difusor durante la tarde), y bias. El tiempo de exposici'on para las estrellas est'andares fue de cinco segundos para las estrellas brillantes HR\,7950 y HR\,4468, y 1200 segundos para Feige\,110.

Tambi'en se han obtenido espectros de alta resoluci'on para las tres regiones H{\sc ii} de la galaxia NGC\,6070 utilizando el espectr'ografo 'echelle doble,
Magellan Inamori Kyocera Echelle (MIKE) del Telescopio de 6.5-m, Magellan II (Clay), de LCO, en julio de 2004. No se aplic'o ning'un binning al CCD de 2K$\times$4K  y la ranura utilizada fue de 1 segundo de arco. La resoluci'on espectral medida en los espectros de MIKE es de 11 km$s^{-1}$, muy similar a la obtenida en los espectros de du Pont. El tiempo de exposici'on para la estrella est'andar espectrofotom'etrica, BD+28D4211, fue de 900 segundos.
Tambi'en se tomaron l'amparas de comparaci'on de Th-Ar, milky flats y bias para calibrar los datos.

El an'alisis de los datos se realiz'o con el software IRAF\footnote{Image Reduction and Analysis Facility, distribuido por NOAO, operado por AURA, Inc., bajo contrato con NSF.}. Luego de la resta de bias y correcciones por flat fields con los Milky Flats, las im'agenes bidimensionales fueron corregidas por rayos c'osmicos con la tarea {\tt cosmicrays} de IRAF, la cual detecta y elimina los rayos c'osmicos con un algoritmo de cociente de flujos. Los datos corregidos se redujeron con las rutinas de IRAF siguiendo el mismo procedimiento descripto en Firpo et al.,~(\citeyear{Firpo05}).

\section{Metodolog\'{\i}a en la cinem\'atica}
Para determinar las velocidades radiales y las dispersiones de velocidades del gas ionizado, se ha medido la longitud de onda central y el ancho en varias l'ineas de emisi'on intensas.
La velocidad radial de cada l'inea de emisi'on se ha obtenido de su longitud de onda central determinada ajustando el perfil Gaussiano, y sus errores han sido proporcionados por el error obtenido en el ajuste realizado con la tarea \texttt{ngaussfit} de IRAF y teniendo en cuenta el error medio cuadr'atico de la calibraci'on en longitud de onda. La dispersi'on de velocidades intr'inseca ($\sigma_{int}$) de cada l'inea de emisi'on ha sido calculada como se describe en Firpo et al.,~(\citeyear{Firpo05}) y detallada aqu'i en la Secci'on \ref{Dispersi'on de velocidades}, donde se explica que la dispersi'on de velocidades intr'inseca ($\sigma_{int}$) es corregida por las contribuciones instrumental y t'ermica de cada l'inea de emisi'on.

Para ajustar las m'ultiples componentes en los perfiles observados se utiliz'o la tarea \texttt{ngaussfit} de IRAF. Esta tarea realiza un ajuste Gaussiano iterativo para varios perfiles de l'ineas espectrales. La tarea permite seleccionar distintos par'ametros a ser ajustados y ofrece una estimaci'on de la incerteza de los par'ametros derivados. La tarea necesita valores iniciales de los coeficientes, los cuales pueden ser especificados a trav'es de tablas generadas previamente.

En los casos en que m'as de una componente se hace evidente, hemos seguido un procedimiento iterativo en el que s'olo se le permite a la tarea ajustar un subconjunto limitado de par'ametros a la vez. Esto se hace para limitar el universo de posibles soluciones en el ajuste, y haciendo uso de la informaci'on disponible en cada l'inea de emisi'on. Luego de obtener una estimaci'on inicial de los par'ametros de la segunda componente, son fijados la amplitud, el centro y el ancho de cada componente, dejando los par'ametros de la otra componente sin tocar. Una vez que la longitud de onda central de cada componente se conoce relativamente bien, se establece 'esta como par'ametro fijo y se permite que la tarea realice un ajuste final sobre la amplitud y el ancho en todas las componentes y al mismo momento.

En todos los casos los ajustes de los perfiles han demostrado la presencia de un residuo en las alas de las emisiones, bien evidente en la l'inea intensa H$\alpha$. Por lo tanto, se ha evaluado la existencia de una componente ancha o dos componentes tipo shell en el perfil de la l'inea. En la primera hip'otesis, se ha introducido una componente con una anchura inicial que es tres veces el ancho de la componente angosta \citep{H07}. Se ajustaron las posiciones de los picos para las componentes angostas y, a continuaci'on, los anchos de los perfiles. Los par'ametros de las componentes anchas fueron dados luego, antes de iniciar el procedimiento iterativo nuevamente. En el segundo escenario, el ajuste del residuo presente en las alas del perfil, fue realizado siguiendo un procedimiento iterativo similar, pero en este caso se a'nade una componente angosta corrida al azul y una componente angosta corrida hacia el rojo. La validaci'on de la multiplicidad en el perfil y el ensanchamiento es controlado sobre las diferentes l'ineas de emisi'on para cada regi'on, aunque el an'alisis m'as confiable proviene del perfil de las l'ineas m'as intensas, como la l'inea de recombinaci'on H$\alpha$.

De la muestra de seis GH{\sc ii}Rs se ha encontrado que todas ellas muestran evidencia de un residuo presente en los perfiles de las l'ineas de emisi'on. S'olo en una de las regiones, donde no pudo ser ajustada una componente ancha, se han ajustado dos componentes angostas sim'etricas separadas en velocidad respecto a las componentes principales; en el resto de las regiones el ajuste de una componente ancha dio resultados satisfactorios. 
En la publicaci'on se discuten los resultados agrup'andolos seg'un los tipos de resultados de los perfiles y por regi'on.

\section{Dispersi'on de velocidades}
\label{Dispersi'on de velocidades}

El par'ametro f'isico que se desea medir es la dispersi'on de velocidades intr'inseca del gas ionizado ($\sigma_{int}$). Sin embargo, el ancho del perfil observado ($\sigma_{o}$) est'a afectado por la contribuci'on del movimiento t'ermico aleatorio ($\sigma_{t}$) y del perfil instrumental ($\sigma_{i}$) dado por el espectr'ografo. Por lo tanto, dispersi'on de velocidades verdadera estar'a dada por:
\[
\sigma_{int}{^2}\,=\,\sigma_{o}^{2}\,-\,\sigma_{i}^{2}\,-\,\sigma_{t}^{2}
\]
suponiendo que las regiones observadas tienen una temperatura cin'etica t'ipica $T\,=\,10^{4}K$ y que el perfil instrumental ($\sigma_{i}$) se puede determinar a partir del ajuste Gaussiano de las l'ineas de emisi'on angostas de la l'ampara de Th-Ar. Los errores en la determinaci'on de la dispersi'on de velocidades son calculados utilizando los errores de observaci'on en el perfil observado $\sigma_{o}$ y considerando errores insignificantes en $\sigma_{i}$ y $\sigma_t$.
Debido a la alta metalicidad de las regiones H{\sc ii} en galaxias espirales, las temperaturas son probablemente inferiores a $10^{4}$K \citep{2007MNRAS.382..251D}. En estas regiones tambi'en se ha hecho la suposici'on del caso m'as extremo donde la temperatura electr'onica sea de unos 5000 K, para analizar como afecta este cambio de temperatura a la contribuci'on de la dispersi'on de velocidades, pero los resultados no arrojaron grandes diferencias respecto de suponer T\,=\,$10^{4}$K.
Para una temperatura cin'etica de T\,=\,$10^{4}$K, la velocidad del sonido es de unos 13 \kms. Se dice que la dispersi'on de velocidades es supers'onica cuando es mayor o igual a ese valor. En este caso se est'a frente a una GH{\sc ii}R. Valores menores, corresponden a regiones H{\sc ii} cl'asicas.

\clearemptydoublepage


\includepdf[pages={1-14}]{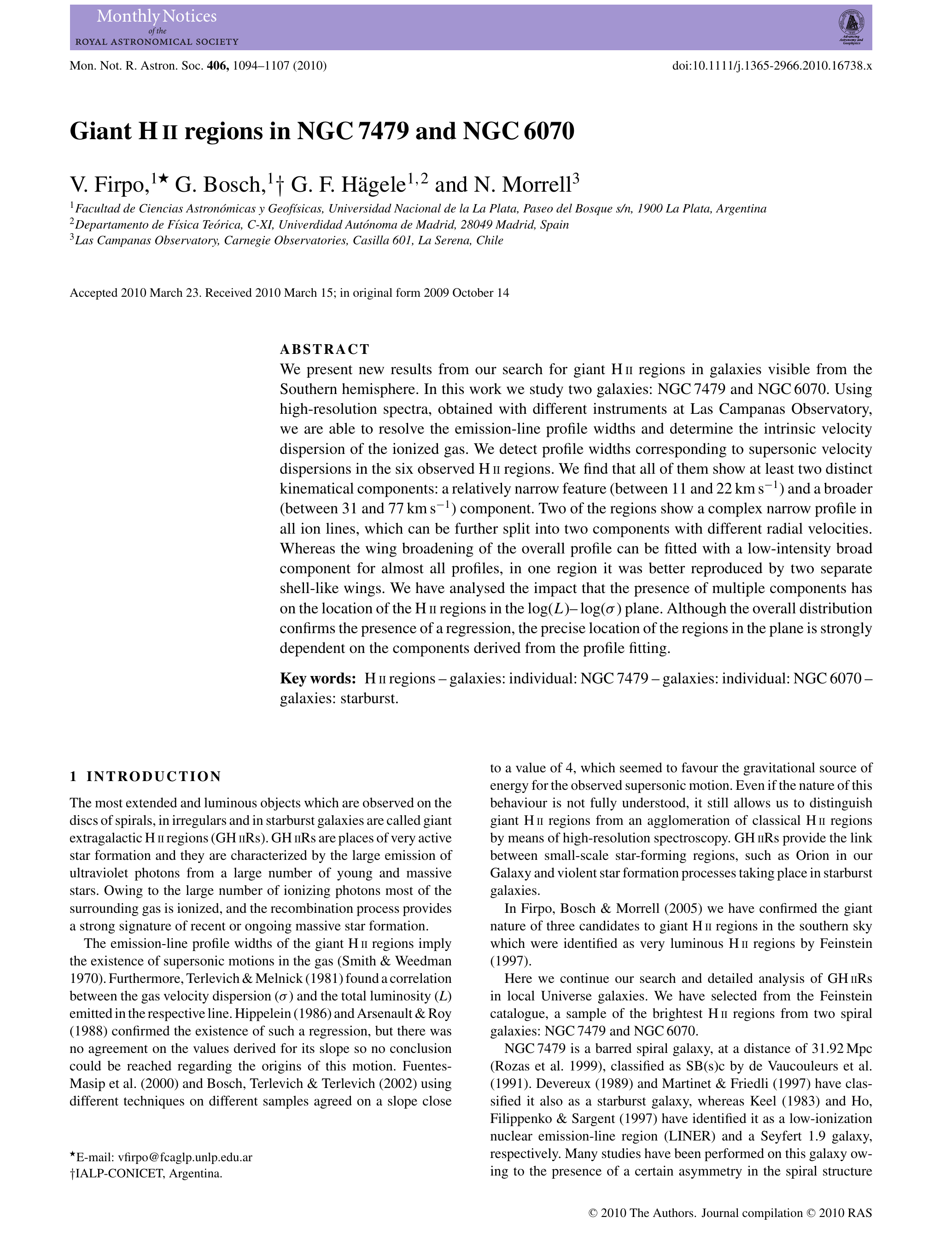}
\clearpage

\clearemptydoublepage
\newpage

\clearemptydoublepage
\newpage

{\color{night}\chapter[Espectroscop'ia de alta resoluci'on de la galaxia BCD Haro\,15: Cinem'atica interna]{Espectroscop'ia de alta resoluci'on de la galaxia BCD Haro\,15: Cinem'atica interna}}
\label{capitulo3}
\begin{center}
\vskip 0.1cm
{\large \em High resolution spectroscopy of the BCD galaxy Haro\,15: I. Internal kinematics} \\
\vskip 0.2cm
{\large \cal Firpo, V., Bosch, G., H\"agele, G.~F., \'Angeles I. D\'{\i}az, and Morrell, N., 2011, MNRAS, Vol 414, Issue 4, pp. 3288}\\
\end{center}

\section{Resumen de la publicaci'on}
En este Cap'itulo, haciendo uso de espectroscop'ia 'echelle obtenida en el Observatorio Las Campanas, se presenta un estudio detallado de la cinem'atica interna del material nebular en m'ultiples brotes de formaci'on estelar de la galaxia enana campacta azul Haro\,15.
Un an'alisis detallado de los complejos perfiles de l'ineas de emisi'on muestran la presencia de una componente ancha subyacente en casi todos los brotes, y la regi'on de formaci'on estelar m'as brillante muestra indudables signos de la presencia de dos componentes cinem'aticas angostas diferentes.
Estudiando la informaci'on que este an'alisis proporciona sobre el mov'iento de los brotes individuales en el potencial de la galaxia Haro\,15, se confirma que siguen la rotaci'on gal'actica, aunque las componentes angostas del brote A tienen velocidades relativas demasiado grandes para ser explicada por rotaci'on gal'actica. Por 'ultimo, se examina la relaci'on entre la dispersi'on de velocidades y la luminosidad, encontrando que casi todos los brotes siguen la relaci'on de sistemas virializados. Esto es v'alido para las componentes angostas identificadas en los perfiles complejos, en cambio los perfiles ajustados con una componente Gaussiana simple muestran una pendiente m'as plana.

\section{Fundamentos b'asicos}
Durante a'nos, varios estudios en diferentes frecuencias han sido llevados a cabo sobre la galaxia enana compacta azul Haro\,15.

Desde el punto de vista morfol'ogico, la imagen H$\alpha$ de Cair'os et al.~(\citeyear{2001ApJS..133..321C}) muestra que la galaxia presenta una compleja morfolog'ia resuelta en un gran n'umero de brotes dispersos por toda la galaxia (ver Figura 1 de la publicaci'on).
Haro\,15 ha sido clasificada como una galaxia peculiar (R)SB0 \citep{RC3.9} y, de hecho, en las im'agenes profundas tomadas con el telescopio de 2.2m CAHA se puede apreciar una morfolog'ia en espiral \citep{2008A&A...491..131L}. L'opez-Sanchez et al.,~(\citeyear{2010A&A...521A..63L}) encontraron una cinem'atica y abundancias qu'imicas diferentes entre las regiones A y B lo cual podr'ia estar indicando que la galaxia probablemente est'e experimentando una fusi'on menor.

En la publicaci'on se hace menci'on a que Haro\,15 cumple con los criterios de una galaxia luminosa compacta azul (LCBGs, sigla en ingl'es de Luminous Compact Blue galaxies), y es que el nombre de estas galaxias fue asignado en Jangren et al., (2004) a las galaxias luminosas (M$_{B}$\,$<$\,-17.5), compactas ($\mu$$_{B}$\,$\leq$\,21.0 mag arcsec${^2}$) y azules (B-V\,$\leq$\,0.6) las cuales est'an sometidas a un importante estallido estelar. Esta definici'on intenta incluir las galaxias BCGs locales m'as luminosas, como as'i tamb'ien a una variada familia de brotes de formaci'on estelar a redshifts intermedios. Generalmente, estos objetos son poblaciones aisladas de alto brillo superficial pudiendo ser observados a altos redshifts y tienen propiedades muy similares a las regiones de formaci'on estelar de alta luminosidad en las galaxias H\,{\sc ii} \citep{2004AJ....128.1541H}.

Como se vio en el Cap'itulo anterior, la espectroscop'ia 'echelle proporciona un medio para buscar componentes cinem'aticamente diferentes en las l'ineas de emisi'on del gas ionizado, ya que llega a la resoluci'on espectral necesaria para resolver la presencia de estructuras en el perfil de l'ineas de emisi'on, generalmente enmascarados por el gran ancho supers'onico y la baja resoluci'on espectral instrumental.

En el Cap'itulo 2 \citep{2010MNRAS.406.1094F} se observ'o que al hacer un ajuste Gaussiano simple en el perfil de la l'inea, existe un residuo presente en las alas del perfil. Basado en la literatura, y siempre que fuera posible, se evalu'o la posible presencia de una componente ancha o dos componentes sim'etricas a ambos lados de la componente angosta en el ajuste de los anchos de las emisiones observadas. En el presente trabajo, tambi'en se ha encontrado que todas las regiones de la galaxia Haro\,15 muestran este ensanchamiento siendo m'as evidente en el perfil de la l'inea H$\alpha$ y luego se confirma en el resto de las l'ineas de emisi'on. Por otro lado, se analiza la relaci'on entre la dispersi'on de velocidades en la l'inea H$\alpha$ y su luminosidad estudiando c'omo repercute la existencia de diferentes componentes cinem'aticas en el plano L(H$\alpha$) vs. $\sigma$.

En este Cap'itulo se presenta espectroscop'ia 'echelle obtenida con el telescopio de 100"\ du Pont, del Observatorio Las Campanas (LCO), en cinco brotes de formaci'on estelar de Haro\,15 con una resoluci'on en velocidad de 12 \kms. Esta resoluci'on permite la identificaci'on de las diferentes componentes cinem'aticas del gas y la medici'on de sus correspondientes dispersiones de velocidades. La identificaci'on de los brotes sigue la nomenclatura utilizada por Cair'os et al.,~(\citeyear{2001ApJS..133..321C}). Dos brotes m'as fueron observados en este trabajo, a los cuales se los ha denominado E y F, continuando la notaci'on de Cair'os.

Este estudio forma parte de un proyecto en el que, con la obtenci'on de datos de alta resoluci'on espectral 'echelle, se determina la naturaleza de las Regiones H\,{\sc ii} Gigantes visibles desde el hemisferio sur y se analizan, adem'as, las condiciones f'isicas del gas ionizado de las regiones en esta galaxia enana compacta azul. 
En la publicaci'on a continuaci'on se presentan las observaciones y la reducci'on de los datos en la Secci'on {\sc Observations and Reductions}; en la Secci'on {\sc Results and Discussion} se presenta el an'alisis de los perfiles de la l'inea de emisi'on y se analizan los resultados de cada brote individual. Adem'as se estudian las velocidades radiales y se analiza el impacto que tiene la presencia de varias componentes sobre la ubicaci'on de cada regi'on H\,{\sc ii} en el plano $\log(L) - \log(\sigma)$; por 'ultimo se incluye resumen y conclusiones del trabajo en la Secci'on nombrada como {\sc Summary}.

\section{Herramientas utilizadas}
Los espectros alta resoluci'on de Haro\,15 se obtuvieron utilizando el espectr'ografo 'echelle del telescopio de 100"\ du Pont, del Observatorio Las Campanas (LCO), Chile, en julio del 2006. El rango espectral cubierto por las observaciones fue de $\lambda$3400 a $\lambda$10000 \AA. Las condiciones de observaci'on fueron buenas, con un seeing de 1 segundo de arco. Se aplic'o un binning 2$\times$2 al CCD el cual tiene una resoluci'on espacial de 0.259"px$^{-1}$. La ranura utilizada ten'ia un ancho de 1"\, y una longitud de 4". Considerando que las exposiciones son largas, la resoluci'on espacial esta limitada no solo por seeing sino tambi'en por el guiado, y teniendo en cuenta la distancia a Haro\,15, la escala espacial de los espectros nos dice que la menor estructura que se puede ver es de al menos 0.43 kpc. La resoluci'on espectral alcanzada en los espectros 'echelle es de R$\simeq$25000:  $\Delta\lambda$=0.25\AA\ a los $\lambda$ 6000\AA, medida a partir del ancho a potencia mitad ({\sc FWHM}) de las l'ineas de la l'ampara de comparaci'on de Thorio-Arg'on. Esto se traduce en una resoluci'on de $\sim$12 \kms. 

El tiempo de exposici'on para cada regi'on fue 1800 segundos. Tambi'en se observ'o para la calibraci'on en flujo una estrella est'andar espectrofotom'etrica, Feige\,110, con un tiempo de exposici'on de 1200 segundos. Adem'as, durante todas las noches se tomaron espectros de l'ampara de comparaci'on Th-Ar, Milky Flats (flats de cielo obtenidos con un difusor durante la tarde), y bias. 

El an'alisis de los datos se realiz'o con el software IRAF\footnote{Image Reduction and Analysis Facility, distribuido por NOAO, operado por AURA, Inc., bajo contrato con NSF.}. Luego de la resta de bias, correcciones por flat fields con los Milky Flats, las im'agenes bidimensionales fueron corrigidas por rayos c'osmicos con la tarea {\tt cosmicrays} de IRAF, la cual detecta y elimina los rayos c'osmicos con un algoritmo de cociente de flujos. Los datos corregidos se redujeron con las rutinas de IRAF siguiendo el mismo procedimiento descrito en Firpo et al.,~(\citeyear{Firpo05}).

\section{Metodolog\'{\i}a en la cinem\'atica}

En el espectro 'echelle calibrado, se cort'o el rango de longitud de onda en una dada l'inea de emisi'on y este ha sido transformado al plano de velocidad aplicando la correcci'on Doppler. Midiendo la velocidad central (longitud de onda) y el ancho en varias l'ineas de emisi'on se han determinado las velocidades radiales y la dispersi'on de velocidades del gas ionizado en las diferentes regiones de formaci'on estelar. La dispersi'on de velocidades intr'inseca ($\sigma_{int}$), fue corregida por las contribuciones instrumental y t'ermica de cada l'inea de emisi'on. Los peque'nos cambios en las temperaturas electr'onicas, de unos pocos cientos de grados, no modifican sustancialmente esta correcci'on, como ya se ha discutido en el Cap'itulo anterior \citep{2010MNRAS.406.1094F}. En el Cap'itulo 4 se presentar'a un tratamiento detallado en la determinaci'on de temperaturas y densidades electr'onicas de cada uno de estos objetos.

Como fue comentado en el Cap'itulo anterior, en el presente Cap'itulo los brotes de formaci'on estelar de la galaxia BCD, Haro\,15, muestran evidencias de residuo presente en las alas de las emisiones, el cual es observado en todas las l'ineas de emisi'on. Haciendo uso del proceso iterativo, explicado en Firpo et al.,~(\citeyear{2010MNRAS.406.1094F}), para el ajuste de m'ultiples componentes Gaussianas en el perfil de l'inea, se evalu'o la presencia de una componente ancha y la presencia de m'as de una componente angosta. En este caso, se ha podido ajustar una variada gama de componentes, explicando as'i las alas del perfil integrado en cada una de las regiones.

Todas las regiones, a excepci'on del brote F, muestran m'ultiples componentes supers'onicas. En el brote F el ensanchamiento del perfil integrado solamente pudo ser detectado en la l'inea H$\alpha$. En esta l'inea de emisi'on se pudo ajustar una componente ancha supers'onica y una angosta subs'onica de 8 \kms, dispersi'on de velocidades t'ipica de regiones H\,{\sc ii} cl'asicas, aunque la presencia de una componente ancha sea casi exclusiva de las Regiones H\,{\sc ii} Gigantes.

En la publicaci'on se discuten los resultados agrup'andolos seg'un los resultados de los perfiles y por brote.
Este trabajo ha sido aceptado el d'ia 1$^{ro}$ de marzo de 2011 para su publicaci'on en la revista internacional con referato Monthly Notices of the Royal Astronomical Society.

\clearemptydoublepage


\includepdf[pages={1-10}]{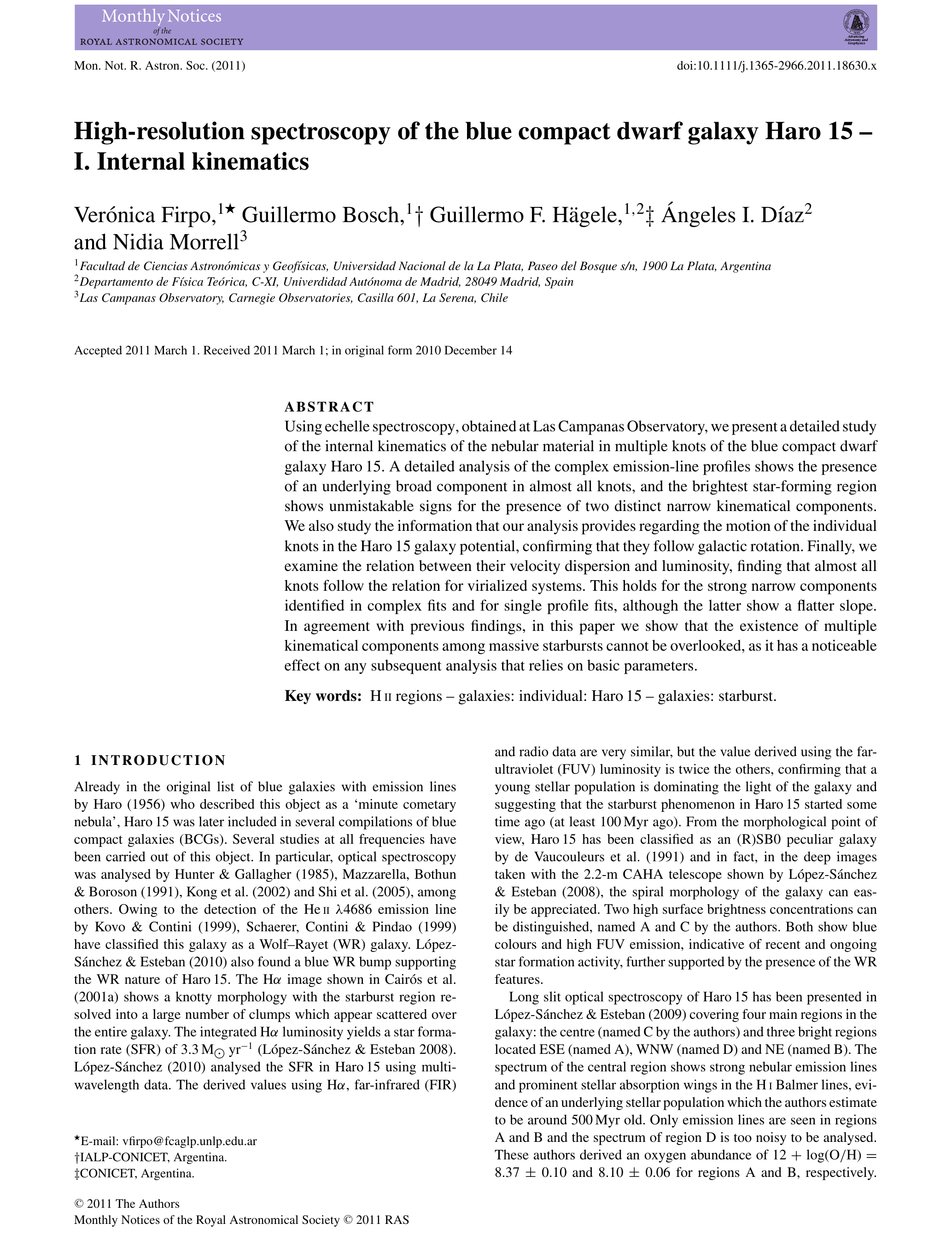}
\clearpage
\clearemptydoublepage
\newpage

\clearemptydoublepage
\newpage

{\color{night}\chapter[Espectroscop'ia de alta resoluci'on de la galaxia BCD Haro\,15: Determinaci\'on de abundancias]{Determinaci\'on de abundancias del gas ionizado para las diferentes componentes cinem\'aticas}}
\label{capitulo4}

El an\'alisis de la distribuci\'on de densidad y temperaturas electr\'onicas en el interior de una nebulosa de gas ionizado permite analizar las propiedades f\'{\i}sicas del medio interestelar y determinar las abundancias qu'imicas i\'onicas y totales de dicho gas. 
La determinaci\'on de la abundancia de cada i\'on est'a asociada a su temperatura electr\'onica. En aquellos casos en que estas temperaturas no son medibles directamente a trav'es de las l'ineas de emisi'on aurorales sensibles a la temperatura se suelen utilizar relaciones con otras temperaturas i'onicas determinadas emp'iricamente o derivadas utilizando modelos de fotoionizaci'on. El an'alisis de los par'ametros f'isicos tales como densidad, temperatura, composici'on qu'imica, etc., puede llevarse adelante mediante el estudio de las intensidades relativas de las l'ineas de emisi'on del gas. 
En este Cap\'{\i}tulo se procede a determinar las propiedades f'isicas del gas en los distintos brotes de formaci'on estelar de Haro\,15 en cada componente cinem'atica ajustada en los perfiles de las l'ineas de emisi'on. 
Si bien este tipo de an\'alisis desglosado para las diferentes componentes cinem\'aticas no es la primera vez que se realiza, este trabajo presenta una mayor completitud frente a otros trabajos publicados dada la calidad de nuestros datos y la gran cantidad de l'ineas analizadas, entre las que se encuentran las d'ebiles l'ineas aurorales, sumado a la posibilidad de descomponer sus perfiles en m'as de una componente angosta y una componente ancha. Es all'i donde radica el car'acter relevante de este Cap\'{\i}tulo para esta Tesis.

%

\section{Introducci\'on}
\label{sec: introducci'on}
Este Cap\'{\i}tulo se centra en el estudio de regiones H\,{\sc ii} de baja metalicidad como lo son las regiones ubicadas en las galaxias Enanas Compactas Azules (BCDs, sigla en ingl\'es de Blue Compact Dwarf galaxies), que pertenecen a la clase de galaxias m\'as numerosa del Universo: las galaxias enanas. Los espectros de las BCDs est\'an dominados por la emisi\'on del gas asociado a c\'umulos estelares formados por estrellas j\'ovenes y calientes que ionizan el medio interestelar, y estas galaxias se caracterizan por su color azul, aspecto compacto, alto contenido de gas, fuertes l\'{\i}neas de emisi\'on nebular y abundancias qu\'{\i}micas bajas \citep{2000A&ARv..10....1K,2001A&A...370....1S}. Gran parte de los objetos con espectros de emisi\'on en el r\'egimen de alta excitaci\'on, siendo este r'egimen de gran importancia para el estudio de la metalicidad y de las propiedades de la poblaci\'on estelar, son galaxias H\,{\sc ii}. Las galaxias H\,{\sc ii} son galaxias irregulares de baja masa con al menos un episodio de formaci\'on estelar violento concentrado en unos pocos parsecs cerca a su n\'ucleo (Melnick et al., \citeyear{1985MNRAS.216..255M}a, \citeyear{1985RMxAA..11...91M}b). Presentan un alto brillo superficial, intenso color azul y baja metalicidad con espectros de emisi\'on debidos al gas ionizado. Actualmente, se ha multiplicado el inter\'es en estudiar las propiedades de este tipo de objetos menos met\'alicos, dado que existen indicios de que pudieran tratarse de objetos j\'ovenes y, por lo tanto, proporcionar'ian importantes pistas acerca de la formaci\'on estelar en estas condiciones, al margen de la trascendencia que tienen para la b'usqueda de helio primordial.

El t\'ermino galaxia H\,{\sc ii} se utiliza cuando se hace referencia a objetos seleccionados a partir de placas obtenidas mediante la t\'ecnica del prisma-objetivo cuyo criterio de b\'usqueda son las l\'{\i}neas de emisi\'on (por ejemplo, Terlevich et al.,~\citeyear{1991A&AS...91..285T}), mientras que las BCDs son objetos seleccionados en base a su color muy azul y su aspecto compacto en las placas fotogr\'aficas. De todos modos ambos t\'erminos suelen designar a los mismos objetos, ya que sus propiedades observables en el \'optico est\'an dominadas por el espectro de la regi\'on H\,{\sc ii} en la zona de formaci\'on estelar.

 Los primeros trabajos sobre estos objetos compactos son los de Haro~(\citeyear{1956BOTT....2n...8H}) y Zwicky et al.~(\citeyear{1965ApJ...142.1293Z}) que distinguen estas galaxias de gran brillo superficial en sus placas fotogr\'aficas. Sargent \& Searle~(\citeyear{1970ApJ...162L.155S}) encuentran que casi todos estos objetos presentan un espectro dominante de l\'{\i}neas de emisi\'on como el de las regiones H\,{\sc ii} en los discos de las galaxias espirales y las denominan Regiones H\,{\sc ii} Gigantes Extragal\'acticas aisladas. Generalmente se componen de una regi\'on central de unos pocos cientos de parsecs de di\'ametro con un elevado brillo superficial en la cual se est\'an formando miles de nuevas estrellas en un proceso de formaci\'on estelar violento (starburst, de su significado en ingl\'es) \citep{1972ApJ...173...25S} y de una galaxia de cientos de parsecs de di\'ametro de baja luminosidad (M$_{v}$\,$\approx$\,-17), pudiendo alcanzar unos pocos kpc cuando se la observa en los filtros V y R, excediendo el tama'no de la regi'on de formaci'on estelar \citep{2001ApJS..133..321C}.

En general, las l\'{\i}neas de emisi\'on observadas en los espectros tanto de las galaxias H\,{\sc ii} como de las BCDs son muy similares a las de Regiones H\,{\sc ii} Gigantes Extragal\'acticas. Por lo tanto, mediante la aplicaci\'on de las mismas t\'ecnicas de medici\'on que en las regiones H\,{\sc ii}, pueden ser derivadas las temperaturas, densidades y composici\'on qu\'{\i}mica del gas interestelar en los brotes de formaci\'on estelar de este tipo de galaxias deficientes en metales \citep{1970ApJ...162L.155S,1980ApJ...240...41F,1991A&AS...91..285T}. Sometidas a intensos brotes de formaci\'on estelar, espectros con intensas y angostas l\'{\i}neas de emisi\'on, extinci\'on de polvo relativamente peque\~na, entornos de baja metalicidad y diferentes historias de formaci\'on estelar, hacen que las galaxias BCDs sean los m\'as apropiados laboratorios para estudiar la metalicidad de las galaxias \citep{2000A&ARv..10....1K}.\\

En el Cap\'{\i}tulo anterior se present\'o un estudio detallado de la cinem\'atica interna del material nebular en m\'ultiples regiones de formaci\'on estelar de la galaxia BCD Haro\,15. De la espectroscop\'{\i}a de alta resoluci\'on, obtenida en el Observatorio Las Campanas, Chile, se ha realizado un an\'alisis exhaustivo de sus l\'{\i}neas de emisi\'on, incluyendo los ajustes de m\'ultiples componentes a sus perfiles de emisi\'on. Los resultados obtenidos han demostrado que las regiones H\,{\sc ii} de Haro\,15 presentan una estructura compleja en los perfiles de sus l\'{\i}neas de emisi\'on, detectada tanto en las l\'{\i}neas de recombinaci\'on del hidr'ogeno y helio como tambi'en en las l\'{\i}neas prohibidas. La emisi\'on de la regi\'on m\'as brillante puede ser dividida en, al menos, dos componentes angostas m\'as una componente ancha. Aunque las regiones tienden a seguir la cinem\'atica de la galaxia, las componentes del brote A tienen velocidades relativas demasiado grandes para ser explicadas por la rotaci\'on gal\'actica. La mayor\'{\i}a de las regiones siguen la relaci\'on encontrada entre la luminosidad y la dispersi\'on de velocidades para sistemas virializados, ya sea cuando se considera un ajuste de perfil simple o un ajuste m\'as complejo con varias componentes, excepto cuando se considera la componente ancha del ajuste. Entre todas estas componentes, la componente simple muestra una pendiente relativamente m\'as plana que el resto.

En general, la calibraci\'on en flujo de datos 'echelle es dif\'{\i}cil de comprobar y las l\'{\i}neas d\'ebiles, sensibles a la temperatura, tienen una relaci\'on se\~nal-ruido relativamente baja para permitir una medici\'on confiable.
Los datos espectrosc\'opicos de ranura larga con resoluci\'on intermedia son m\'as sensibles para realizar un an\'alisis de las abundancias. Dichas abundancias  dependen de la precisi\'on de las temperaturas estimadas. La combinaci\'on del 'echelle y de ranura larga brinda la oportunidad de interpretar el resultado de la abundancia a la luz de la cinem\'atica interna de las regiones.

Shi et al.~(\citeyear{2005A&A...437..849S}) encuentran un promedio de 8.56 en el logar'itmo de las abundancias totales de ox'igeno de Haro\,15, para una temperatura electr'onica de 8330 K. Sin embargo, se sabe por un trabajo anterior \citep{2001ApJS..133..321C} que la imagen H$\alpha$ muestra una morfolog\'{\i}a nudosa con la regi\'on estelar resuelta en un gran n\'umero de grupos que aparecen dispersos en toda la galaxia. En L'opez-S'anchez \& Esteban~(\citeyear{2009A&A...508..615L}) los autores muestran espectroscop\'{\i}a de los diferentes nudos espacialmente resueltos y determinan densidades, temperaturas y abundancias qu\'{\i}micas de los brotes A, B y C de Haro\,15 (siguiendo la nomenclatura de Cair'os et al.,~\citeyear{2001ApJS..133..321C}). Las abundancia derivadas de ox'igeno fueron de 12+log(O/H)\,=\,8.37$\pm$0.10 y 8.10$\pm$0.06 para las regiones A y B, respectivamente.

\cite{1992ApJ...390..536E} introducen el t'ermino ``chemodynamics'' al conbinar por primera vez, informaci'on qu'imica y cinem'atica en orden de realizar un estudio qu'imico-din'amico de la nebulosa anillo W-R, NGC\,6888. Los autores combinan observaciones de alta resoluci'on espectral con observaciones de alta resoluci'on espacial para las diferentes zonas de la nebulosa, lo cual les permite analizar las diferentes componentes en velocidad de la regi'on estudiada.

M'as tarde, James et al.,~(\citeyear{2009MNRAS.398....2J}), realizaron un estudio en la galaxia BCD Mrk 996 basado en alta resoluci'on espectral con VLT VIMOS en modo IFU. Encontraron que la galaxia Mrk 996 presenta multicomponentes en los perfiles de sus l'ineas de emisi'on, descomponiendo los mismos en una componente angosta y una componente ancha, llevando a cabo un an'alisis de las propiedades f'isicas separado por componentes. Seg'un los autores esta peculiar galaxia BCD presenta en el n'ucleo zonas extremadamente densas de gas, a trav'es del cual los outflows estelares y posibles frentes de choque contribuyen a la excitaci'on de la emisi'on ancha. Ellos detectaron que las l'ineas de Balmer muestran una estructura com'un en velocidad de multicomponentes anchas que no es observable en las l'ineas prohibidas. Las l'ineas prohibidas con excepci'on de [\OIII]~$\lambda$4363 y [\NII]~$\lambda$5755, pudieron ser ajustadas con una componente angosta y una componente ancha. En cambio el perfil de emisi'on de ninguna de las l'ineas aurorales sensibles a la temperatura, como por ejemplo las l'ineas de [\OIII] y [\NII] que acab'inamos de mencionar, pudo ser descompuesto en m'as de una componente, s'olo en estas dos l'ineas aurorales los perfiles de emisi'on pudieron ser ajustados con una componente ancha. Si bien el an'alisis de abundancias por componentes presentes en los perfiles de las l'ineas de emisi'on es llevado a cabo por James y colaboradores, la diferencia fundamental con nuestro trabajo radica en que nosotros pudimos descomponer algunas de las l'ineas aurorales sensibles a la temperatura electr'onica en sus diferentes componentes cinem'aticas suponiendo que dicha soluci'on de multicomponentes cinem'aticas es la misma que se encuentra en las l'ineas m'as intensas del espectro con igual grado de ionizaci'on.

En este Cap\'{\i}tulo se realiza un an\'alisis de las caracter\'{\i}sticas del gas ionizado discriminado para cada componente y para el flujo global de las l'ineas. Para esto se estimaron temperaturas y densidades electr'onicas, y se obtuvieron abundancias i'onica y totales de las diferentes especies como ser: O, S, N, Ne, Ar. 

En lo que sigue, se explicar\'an las observaciones y la reducci\'on de los datos, y a continuaci\'on se mostrar\'an los resultados obtenidos. En la Secci\'on \ref{sec:analisis_abund} se explicar\'an los diferentes m\'etodos utilizados para determinar las condiciones f\'{\i}sicas del gas ionizado (densidades y temperaturas electr\'onicas), como las abundancias qu\'{\i}micas derivadas y sus incertezas. En la Secci'on \ref{sec:discusion} se discuten los datos dando las respectivas conclusiones. Y finalmente, en la Secci'on \ref{sec:resumen y conclusiones} se presenta un resumen y las conclusiones del Cap'itulo.
\section{Observaciones y Reducci\'on de los datos}
\label{sec:rvobs}
Se obtuvieron espectros de resoluci\'on intermedia de ranura larga usando la c\'amara Wide-Field CCD (WFCCD) montada en el telescopio du Pont de 100" del Observatorio Las Campanas (LCO) en Chile (28 de septiembre de 2005). Las observaciones obtenidas corresponden a las regiones B y C de Haro\,15 (ver Figura 1 del Cap\'{\i}tulo 3). El detector TEK5 que se utiliz\'o cubre todo el rango de longitud de onda 'optica, desde 3800 a 9300\,\AA\ (centrado en $\lambda_ {c}$\,=\,6550\,\AA). La resoluci'on espectral alcanzada en los espectros es de R$_{FWHM}$\,$\simeq$\,900:  $\Delta\lambda$$_{FWHM}$\,=\,7.5\,\AA\ a los $\lambda$\,6700\,\AA, medida a partir del ancho a potencia mitad ({\sc FWHM}) de las l'ineas de la l'ampara de comparaci'on de Thorio-Arg'on tomadas con el fin de calibraci\'on de longitud de onda. Las observaciones se realizaron en \'angulo paral\'actico, para evitar los efectos de refracci\'on diferencial en el ultravioleta, en excelentes condiciones fotom\'etricas, con un seeing de 1 segundo de arco.
Al comienzo de la noche se tomaron bias (nivel de pedestal o cuentas que tiene de base el detector a\'un sin exponer) y flat-field (del ingl\'es, imagen de campo plano) de c\'upula. Adem\'as, en esta misma noche se tomaron espectros de la l\'ampara de comparaci\'on Th-Ar. Las im\'agenes fueron procesadas y analizadas con las rutinas de IRAF\footnote{Image Reduction and Analysis Facility, distribuida por NOAO, operada por AURA, Inc., bajo acuerdo con NSF.} mediante el procedimiento usual. El procedimiento incluye la eliminaci\'on de rayos c\'osmicos, resta de bias, divisi\'on por un flat-field normalizado y calibraci\'on en longitud de onda.
Se observ\'o una estrella est\'andar, EG\,131, para la calibraci\'on en flujo y el tiempo de exposici\'on para esta estrella est\'andar de flujo fue de 180 segundos. Los espectros fueron corregidos por extinci\'on atmosf\'erica.

Se ha encontrado un cierto grado de contaminaci\'on de segundo orden en estos espectros de ranura larga. La misma comienza a estar presente en $\lambda$$\,>\,$6000\,\AA\ y es casi imposible de eliminar por completo debido a la amplia cobertura espectral usada. Por este motivo, los flujos medidos para las l'ineas nebulares que se encuentran m'as all'a de los 6000\,\AA\ han sido sistem\'aticamente subestimados, aunque nunca esta contribuci'on es importante para los cocientes de l'ineas (ver Terlevich et al.,~\citeyear{1991A&AS...91..285T}).

Por otro lado, se obtuvieron espectros de alta resoluci\'on usando el espectr\'ografo 'echelle en el telescopio du Pont de 100" de LCO, entre el 19 y el 20 de julio de 2006. Las observaciones obtenidas corresponden a varios brotes de formaci\'on estelar de la misma galaxia Haro\,15 (A, B, C, E y F) adquiridos en el modo 'echelle (ver Cap\'{\i}tulo 3). El desplazamiento al rojo calculado con estos datos es de z\,=\,0.021306. El rango espectral abarcado en estas observaciones es desde 3400 hasta 10000\,\AA. Este rango espectral garantiza la medici\'on simult\'anea de las l\'{\i}neas nebulares de [\OII]\,$\lambda\lambda$\,3727, 3729\,\AA\ a las del [\SIII]\,$\lambda\lambda$\,9069, 9532\,\AA\ en ambos extremos del espectro y en la misma regi\'on de la galaxia. Las condiciones de observaci\'on fueron buenas con un seeing de 1 segundo de arco. La estrella espectrofotom\'etrica est\'andar observada para la calibraci\'on en flujo fue Feige\,110 \citep{Bohlin01}, la cual tiene sus flujos tabulados cada 2\,\AA, y la cantidad de intervalos definidos dentro de un orden del 'echelle var'ia de cuatro a doce, dependiendo de la calidad del espectro. El tiempo de exposici\'on de la estrella est\'andar fue de 1200 segundos. La calibraci\'on en flujo obtenida en estos datos fue muy buena. Para los detalles de reducci\'on y el proceso de an'alisis ver Cap\'{\i}tulo 3. 

La Tabla \ref{Regions} lista las observaciones obtenidas, junto con los tiempos de exposici\'on y las masas de aire para cada regi\'on y su respectivo modo de observaci\'on.

 \begin{table}
\caption[Observaciones]{Observaciones de los diferentes brotes en Haro\,15. La primera columna indica el modo de observaci\'on, la segunda columna muestra la nomenclatura utilizada en el Cap'itulo 3 siguiendo Cair'os et al.,~(\citeyear{2001ApJS..133..321C}), la tercera columna corresponde a la fecha de observaci'on, la cuarta columna exhibe el tiempo de exposici\'on de los brotes, la quinta columna se'nala las masas de aire.} 
\label{Regions}
\begin{center}
\begin{tabular}{@{}ccccc@{}}
\hline 
Modo & brote & fecha & t.exp &sec z  \\
\hline
ranura larga&B,C& 2005 Sep 28 & 2x1200 + 1x900& 1.5  \\
\\
'echelle  &A& 2006 Jul 19   & 1800&1.2 \\
         & B& 2006 Jul 19 & 1800& 1.1  \\  
         & C& 2006 Jul 19 &  1800&1.1   \\
         & E& 2006 Jul 20 &1800&  1.2   \\
         & F& 2006 Jul 20 &  1800&1.1  \\
\hline 
\end{tabular}
\end{center}
\end{table}

\section{Resultados}
\label{sec:rvresults} 

En la Figura 4.1 se han marcado algunas de las l\'{\i}neas de emisi\'on m\'as relevantes observadas en los espectros WFCCD de ranura larga de los  {\bf brotes B y C} de Haro\,15. 

\begin{figure*}
\begin{center}
\label{fig:fig_BCsd}
\includegraphics[angle=0,width=.9\textwidth,height=.4\textwidth]{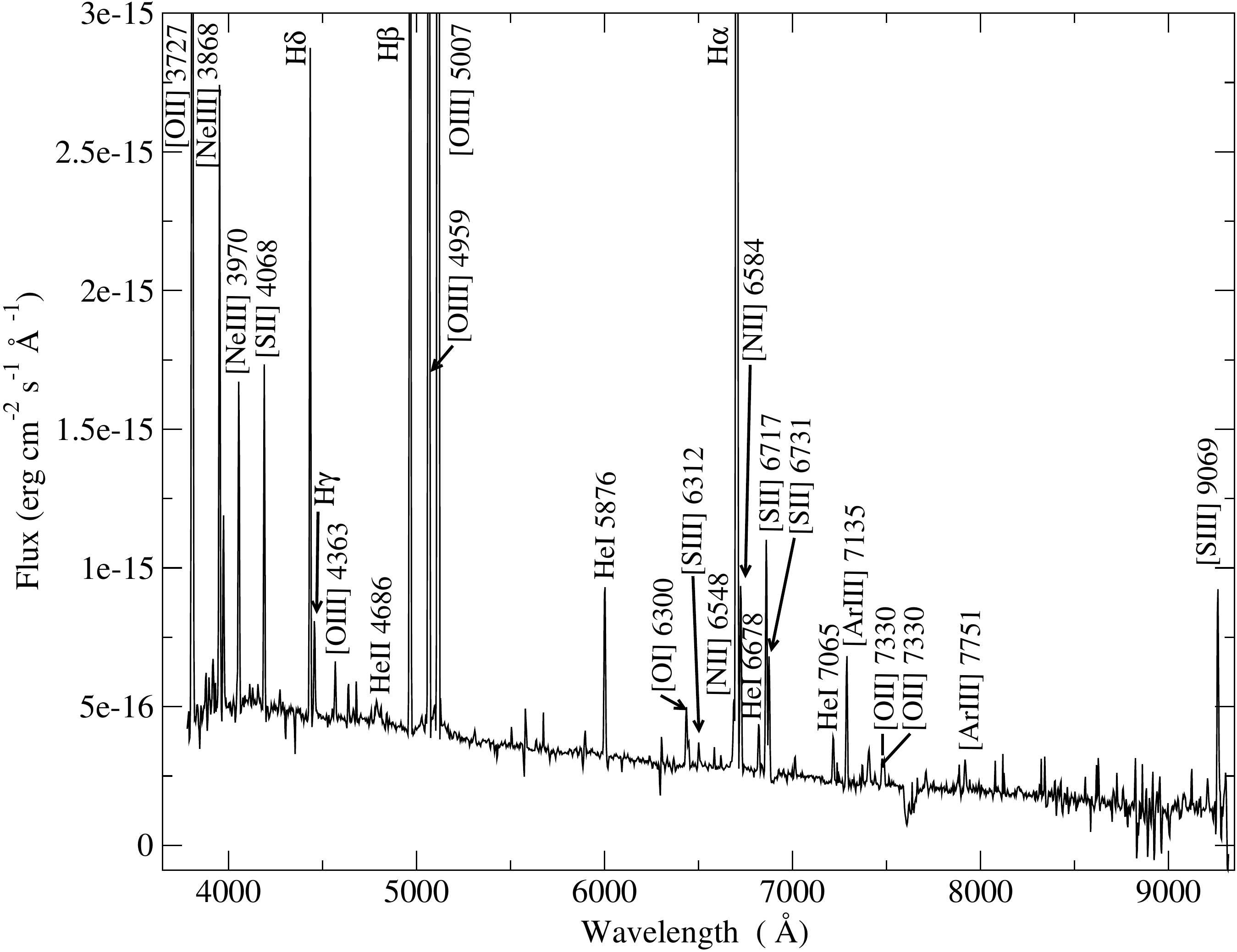}\\
\vspace{0.5cm}
\includegraphics[angle=0,width=.9\textwidth,height=.4\textwidth]{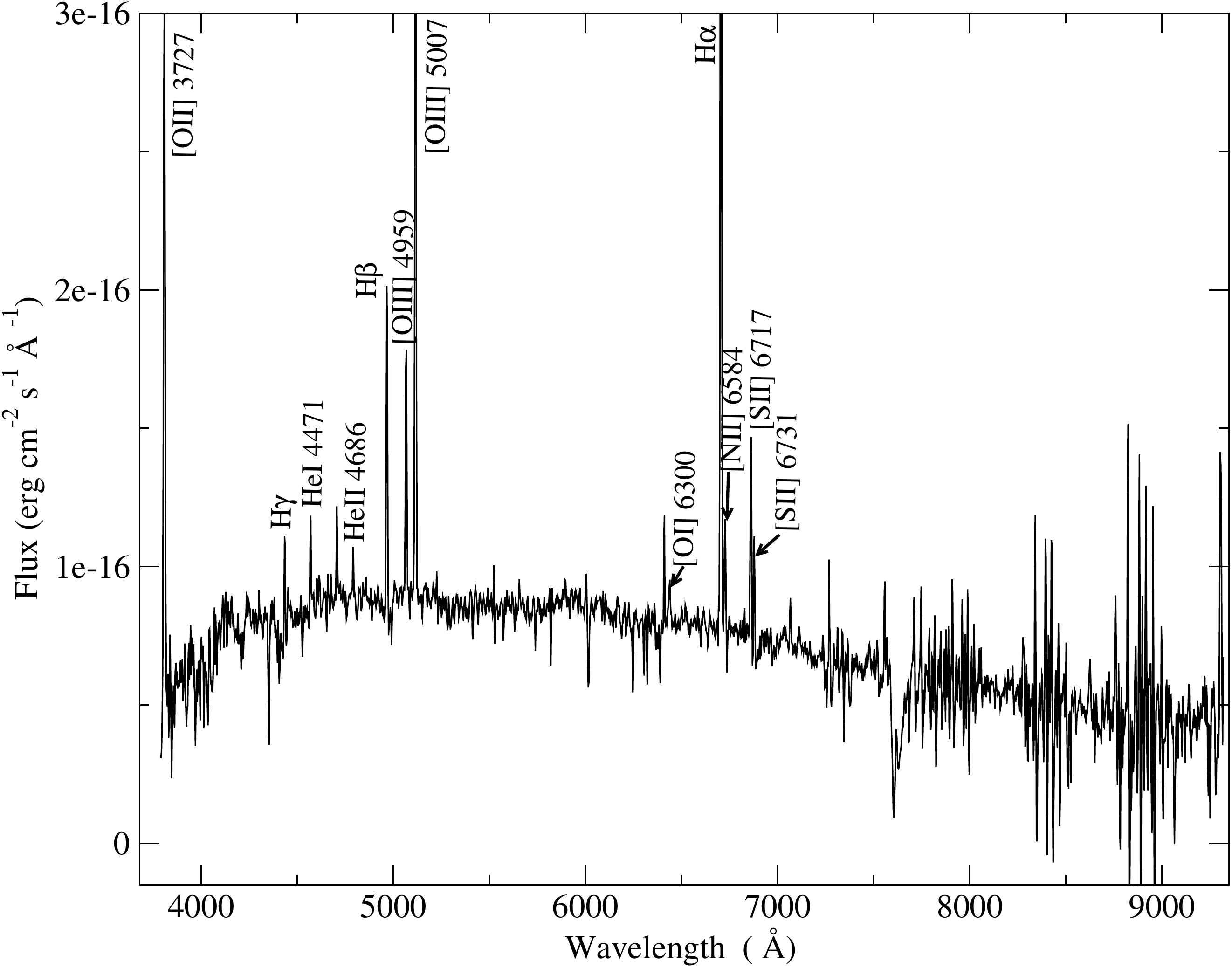}
\caption[Espectros de l\'{\i}neas de emisi\'on de ranura larga]{Espectros de l\'{\i}neas de emisi\'on de ranura larga obtenidos para los {\bf brotes B y C} (paneles superior e inferior, respectivamente) sin correcci\'on Doppler.}
\end{center}
\end{figure*}

Los flujos de las l\'{\i}neas de emisi\'on para los datos de ranura larga se midieron utilizando la tarea \texttt{splot} de {\sc IRAF} (que integra la intensidad de cada
l\'{\i}nea sobre un continuo ajustado localmente) siguiendo el mismo procedimiento descripto en \cite{2006MNRAS.372..293H}. Se utilizaron dos formas de integrar el flujo de la l\'{\i}nea de emisi\'on: (i) en el caso de una l\'{\i}nea aislada o dos l\'{\i}neas fusionadas y sin resolver, la intensidad es calculada integrando entre dos puntos dados por la posici\'on del continuo local colocados a ojo. A este flujo, que incluye todas las diferentes componentes cinem\'aticas, se lo denominar'a, de ahora en m\'as, como ``medida global"; (ii) si dos l\'{\i}neas est\'an fusionadas, pero se pueden resolver, se utiliza el procedimiento de m\'ultiples ajustes Gaussianos para estimar los flujos individuales (esto se aplic'o solamente para los datos de dispersi'on simple). Ambos procedimientos est'an explicados detalladamente en H\"agele et al.,~(\citeyear{2006MNRAS.372..293H,2008MNRAS.383..209H}).

Siguiendo P'erez-Montero \& D'iaz~(\citeyear{2003MNRAS.346..105P}), los errores estad\'{\i}sticos asociados a los flujos de las emisiones observadas se han calculado utilizando la expresi\'on 
\begin{eqnarray}
\sigma_{l}\,=\,\sigma_{c}N^{1/2}[1 + EW/(N\Delta)]^{1/2} 
\label{ecu1}
\end{eqnarray}
\noindent donde $\sigma_{l}$ es el error en el flujo de la l\'{\i}nea observada, $\sigma_{c}$ representa la
desviaci\'on est'andard en una caja cerca de la l\'{\i}nea de emisi\'on
medida y representa el error en la posici\'on del continuo, N es el n\'umero de p\'{\i}xeles en la medida del flujo de la l\'{\i}nea, EW es el ancho equivalente de la l\'{\i}nea el cual est\'a dado por el flujo del continuo cerca de la l\'{\i}nea y el flujo de la l\'{\i}nea, y $\Delta$ es la dispersi\'on de la longitud de onda en angstroms por pixel \citep{1994ApJ...437..239G}.

Cada intensidad medida debe ser corregida de enrojecimiento. La constante de enrojecimiento $c(H\beta)$ se puede calcular a partir del decremento de Balmer de las l'ineas de recombinaci\'on de hidr\'ogeno m\'as intensas. Una poblaci\'on estelar subyacente es f'acilmente apreciable debido a la presencia de alas en las absorciones que deprimen las l\'{\i}neas de emisi\'on de Balmer en los espectros de ranura larga y 'echelle. Para medir las intensidades de las l\'{\i}neas y minimizar los errores introducidos por la poblaci\'on subyacente, se define un pseudo-continuo en la base de las l\'{\i}neas de emisi\'on del hidr\'ogeno (ver discusi'on en H{\"a}gele et al.,~\citeyear{2006MNRAS.372..293H}). La presencia de alas en las l\'{\i}neas de absorci\'on implica que, a pesar de que se ha utilizado un pseudo-continuo, a\'un existe una fracci\'on absorbida del flujo emitido que no es posible medir con precisi\'on (ver la discusi\'on al respecto en D'iaz,~\citeyear{1998Ap&SS.263..143D}). Esta fracci\'on no es la misma para todas las l\'{\i}neas, ni lo son los cocientes entre las fracciones absorbidas y la emisi\'on. H{\"a}gele et al.~(\citeyear{2006MNRAS.372..293H}) estimaron que la diferencia entre los valores obtenidos usando el pseudo-continuo definido o un ajuste m'ultiple Gaussiano de las componentes de absorci\'on y emisi\'on es la misma, dentro de los errores, para todas las l\'{\i}neas de Balmer. En cualquier caso, siguiendo este trabajo, para las l\'{\i}neas de emisi\'on de Balmer se ha duplicado el error derivado, $\sigma_l$, como un enfoque conservador para incluir la incertidumbre introducida por la presencia de la poblaci\'on estelar subyacente.

La ley que permite obtener la constante de enrojecimiento a partir del decremento de Balmer de las l'ineas de recombinaci\'on m\'as intensas del hidr\'ogeno y que una vez obtenidas, permite corregir el resto de las l'ineas de emisi\'on, tiene la siguiente forma:
\begin{eqnarray}
\frac{I_0(\lambda)}{I_0(H\beta)}\,=\,\frac{I(\lambda)}{I(H\beta)} \cdot 10^{-c(H\beta) \cdot [f(\lambda)-f(H\beta)]}
\label{ecu2}
\end{eqnarray}
\noindent donde $I(\lambda)$ es la intensidad de la l'inea que se quiere corregir e $I_0(\lambda)$ es el valor corregido en relaci\'on a $H\beta$, $c(H\beta)$ es la constante de enrojecimiento y $f(\lambda)$ es la funci\'on de extinci\'on. Los valores te\'oricos de las l'ineas de recombinaci\'on del hidr\'ogeno que permiten calcular $c(H\beta)$ son funci\'on de la temperatura electr\'onica y de la densidad, lo cual implicar'a que se debe realizar un procedimiento iterativo para calcularlo.

El coeficiente de enrojecimiento ha sido calculado asumiendo la ley de extinci\'on gal\'actica de Miller \& Mathews~(\citeyear{1972ApJ...172..593M}) con $R_{\rm v}$=3.2, que es el valor para la Galaxia, y la cual toma la forma:
\[
f(\lambda)-f(H\beta)\,=\,\Big\{ \begin{array}{l} 0.477\lambda^{-1}-1.209, \:\lambda^{-1} \leq 2.29 \\
0.342\lambda^{-1}-0.645, \:\lambda^{-1} > 2.29 \end{array}
\]
\noindent donde $\lambda$ es la longitud de onda de la l'inea a calcular en unidades de micrones.

Luego, la constante de enrojecimiento, C(H$\beta$), fue obtenida mediante la realizaci\'on de un ajuste de m\'{\i}nimos cuadrados a la relaci\'on entre F($\lambda$) y F(H$\beta$) para los valores te\'oricos calculados por Storey \& Hummer~(\citeyear{1995MNRAS.272...41S}), usando el m\'etodo iterativo que aplic\'o H{\"a}gele~(\citeyear{2008PhDT........35H}) para estimar $T_e$ y $n_e$ seg'un sea el caso. Los valores iniciales tomados, cuando \'estos estan disponibles, son aquellos derivados de la medida de los flujos de las l\'{\i}neas [\SII]$\,\lambda$$\lambda$\,6717,6731\,\AA\ y [\OIII]\,$\lambda\lambda$\,4363, 4959, 5007\,\AA. Debido a los grandes errores introducidos por la presencia de la poblaci\'on estelar subyacente, solo se han tenido en cuenta las cuatro l\'{\i}neas de emisi\'on de Balmer m'as intensas (H$\alpha$, H$\beta$, H$\gamma$ y H$\delta$). 

Por otro lado, los flujos asociados a cada componente cinem\'atica identificadas en los datos de alta resoluci\'on espectral ('echelle) para las l\'{\i}neas de emisi\'on de mayor cociente se\~nal/ruido (S/N, sigla en ingl\'es de signal-to-noise ratio) han sido medidos como se detalla en el Cap\'{\i}tulo 3. Para aquellas l\'{\i}neas donde el cociente S/N no es suficiente, el proceso iterativo utilizado para deconvolucionar las diferentes componentes de los perfiles de las l\'{\i}neas no ha dado un resultado significativo. Sobre \'estas l\'{\i}neas m\'as d'ebiles y utilizando la tarea \texttt{ngaussfit} de {\sc IRAF}, se ha copiado la soluci\'on encontrada del ajuste realizado en las l\'{\i}neas m\'as intensas con un grado de ionizaci\'on comparable. Luego, fijando los centros (a la longitud de onda correspondiente de la l'inea que se quiere ajustar) y anchos de las componentes, s\'olo se le permite a la tarea variar simult'aneamente la amplitud de cada componente. 

Se han supuesto dos zonas de ionizaci\'on: una zona de baja ionizaci\'on, cuya temperatura caracter\'{\i}stica ser\'{\i}a la de [\OII] y de donde se emiten las l\'{\i}neas de recombinaci\'on de hidr\'ogeno y las l\'{\i}neas prohibidas de [\OII], [\NII], [\SII], y otra de alta ionizaci\'on caracterizada por la temperatura de [\OIII] de donde proceder\'{\i}an las emisiones de las l\'{\i}neas prohibidas de [\OIII], [\NeIII] y las l\'{\i}neas de recombinaci\'on del He{\sc i}. Se supone que tanto las l\'{\i}neas del [\SIII] como las del [\ArIII] son emitidas en una zona intermedia a las dos primeras por lo que se han realizado los ajustes a estas l'ineas utilizando ambas aproximaciones iniciales y as'i poder ver cu'al es la que produce errores m'as peque'nos. 
En los paneles superiores de las Figuras 4.2 y 4.3 se muestran el plano del Flujo-Velocidad de la l\'{\i}nea de emisi\'on m\'as intensa [\OIII]\,$\lambda$\,5007\,\AA\ en los brotes A y B, respectivamente, con las componentes ajustadas al perfil de l\'{\i}nea superpuestas. Para mostrar la buena correlaci\'on de los ajustes entre los perfiles de l\'{\i}nea de emisi\'on intensas contra los perfiles de las l\'{\i}neas con menor se\~nal-ruido, se graficaron en el panel inferior de cada figura un ejemplo del ajuste obtenido para una l\'{\i}nea de emisi\'on d\'ebil con similar grado de ionizaci\'on que [\OIII]\,$\lambda$\,5007\,\AA. En el brote A, se compara [\OIII]\,$\lambda$\,5007\,\AA\ con He{\sc i} 5876\,\AA, mostrando el buen ajuste realizado en esta l\'{\i}nea de emisi\'on d\'ebil. En el brote B, es notable la buena correlaci\'on entre las dos l\'{\i}neas de emisi\'on del ox\'{\i}geno. Gracias a este proceso es posible ajustar y deconvolucionar la  importante y d\'ebil l\'{\i}nea auroral [\OIII]\,$\lambda$\,4363\,\AA, la cual es una l'inea sensible a la temperatura electr'onica del medio donde se forma.

\begin{figure}
\begin{center}
\label{fig:figbroteAsw}
\includegraphics[angle=0,width=.67\textwidth]{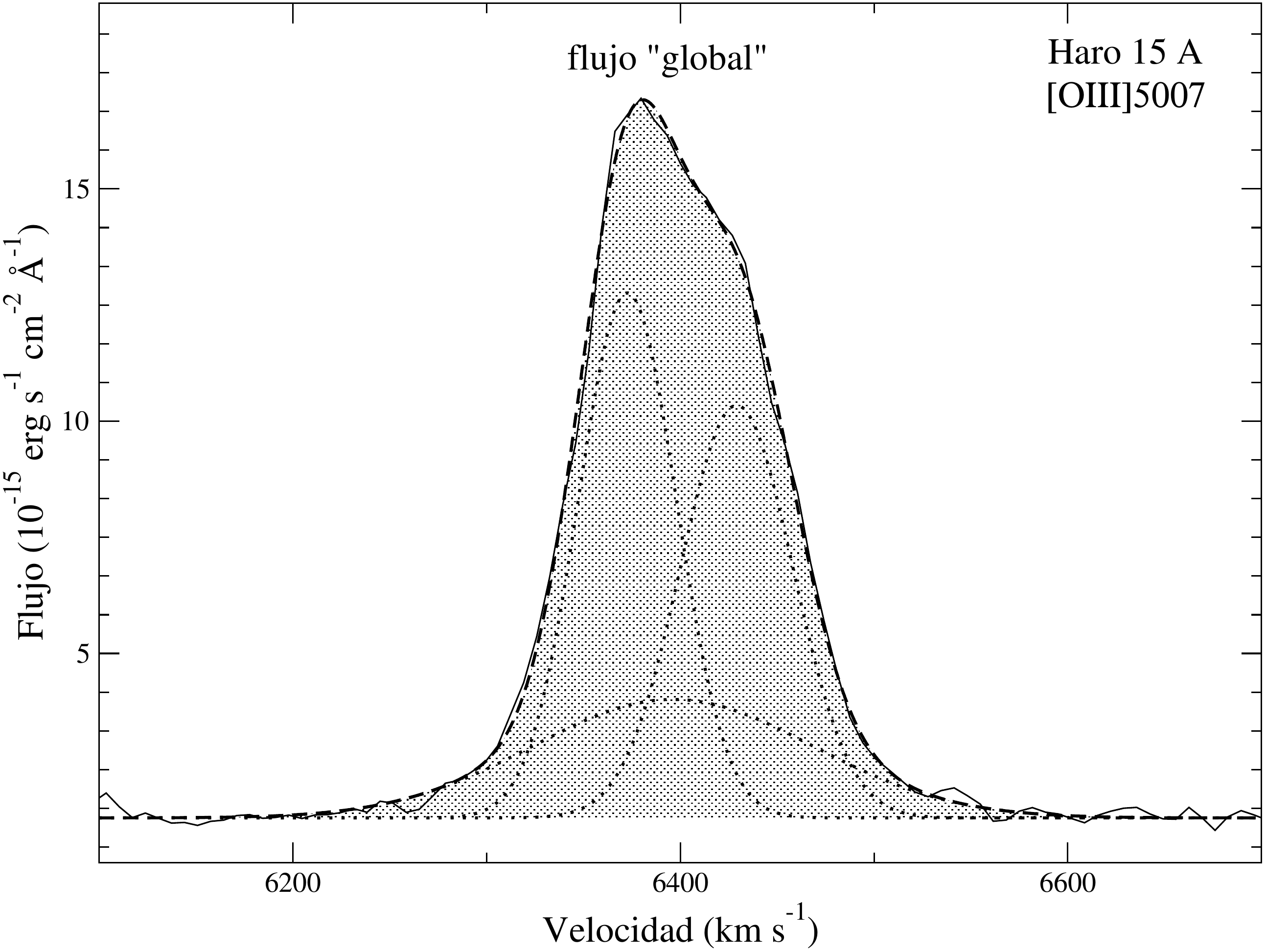}\\
\vspace{1.0cm}
\includegraphics[angle=0,width=.67\textwidth]{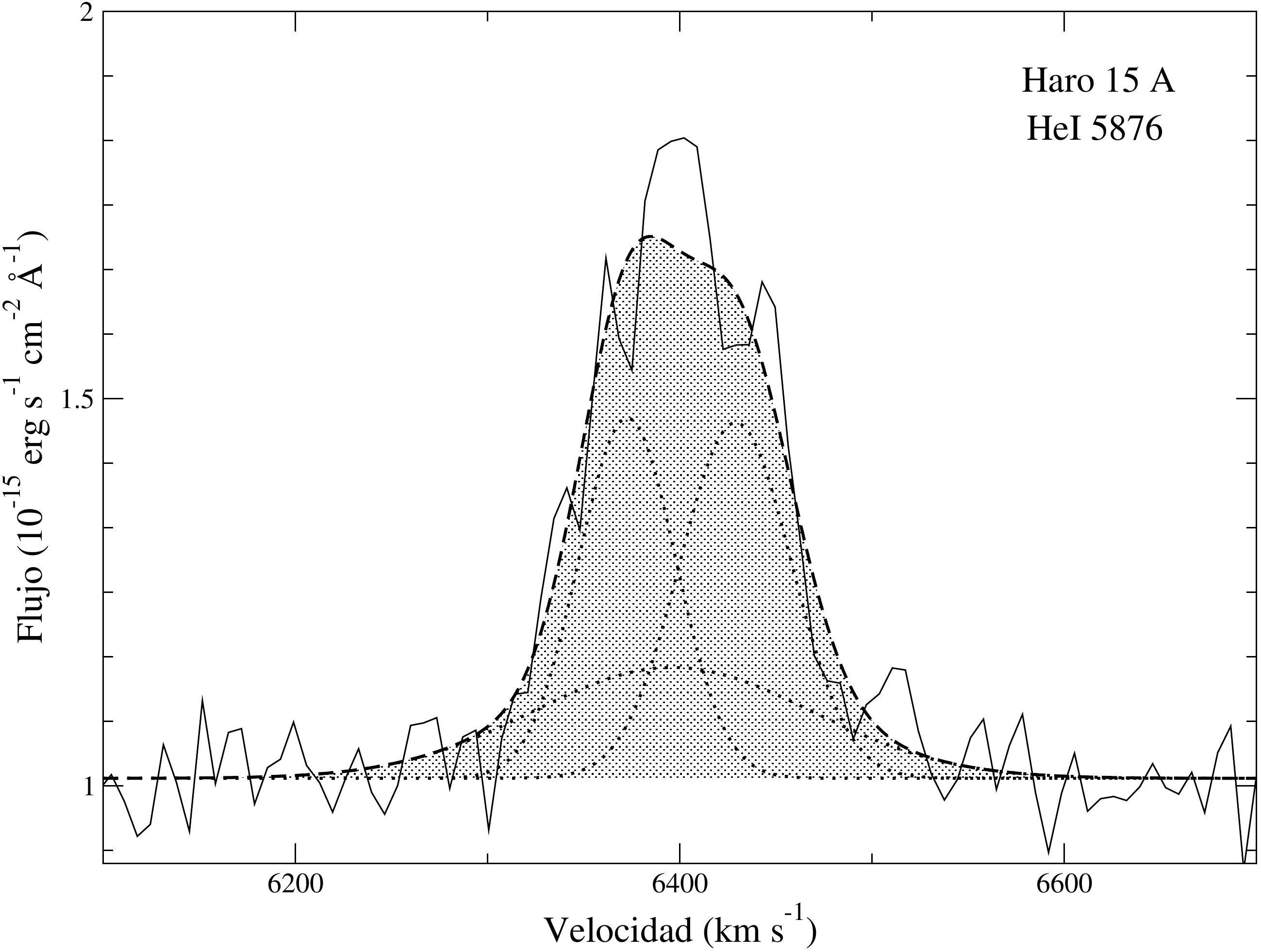}\\
\caption[Comparaci'on entre las soluciones de distintos ajustes, brote A]{Comparaci'on entre las soluciones de distintos ajustes encontrados para una l'inea intensa y otra d'ebil del {\bf brote A}}
\end{center}
\end{figure}

\begin{figure}
\begin{center}
\label{fig:figbroteBsw}
\includegraphics[angle=0,width=.67\textwidth]{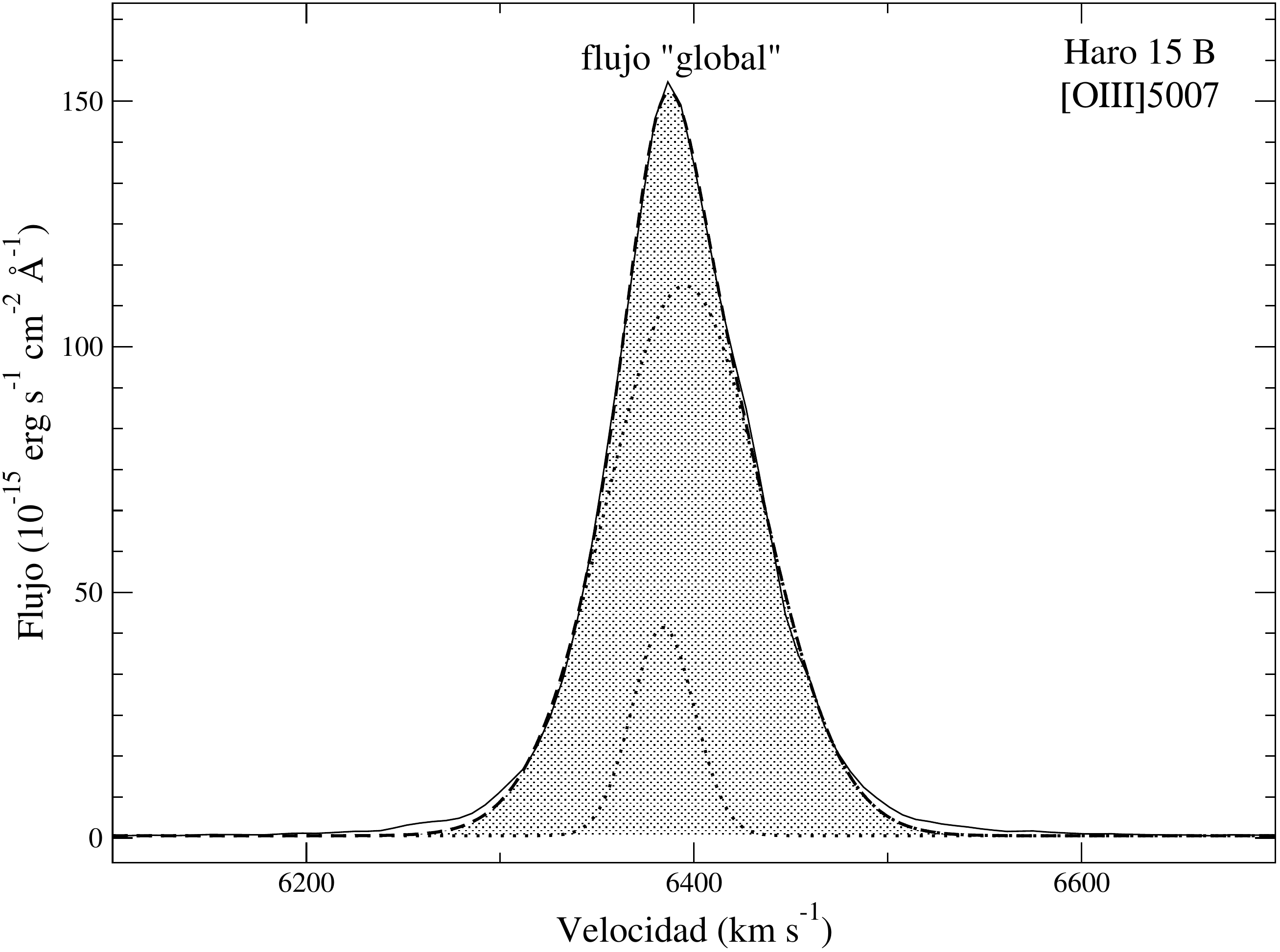}\\
\vspace{1.0cm}
\includegraphics[angle=0,width=.67\textwidth]{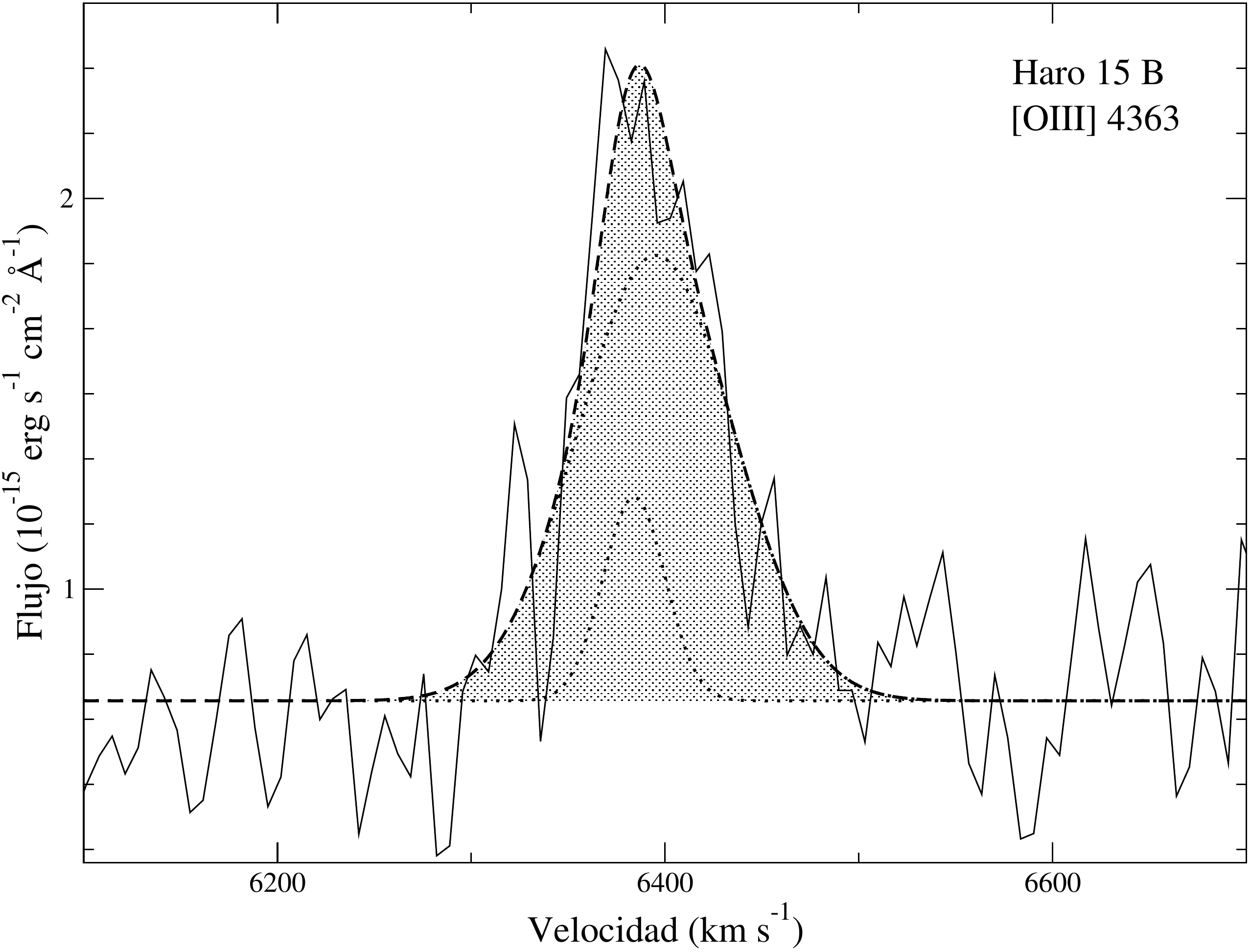}\\
\caption[Comparaci'on entre las soluciones de distintos ajustes, brote B]{Comparaci'on entre las soluciones de distintos ajustes encontrados para una l'inea intensa y otra d'ebil del {\bf brote B}}
\end{center}
\end{figure}

Los flujos de las l\'{\i}neas de emisi\'on de los cinco brotes observados son mostrados en la Tabla \ref{ratiostot 1} para los datos de ranura larga, y en las Tablas \ref{ratiostot 2}, \ref{ratiostot 3}, \ref{ratiostot 4}, \ref{ratiostot 5} y \ref{ratiostot 6} para los datos 'echelle. Cada Tabla lista la intensidad de las l\'{\i}neas de emisi\'on para cada componente corregidas por enrojecimiento, junto a la constante de enrojecimiento (C(H$\beta$)) y su correspondiente error considerado como la incerteza del ajuste de m\'{\i}nimos cuadrados y la intensidad de H$\beta$ corregida por enrojecimiento. La columna 1 muestra la longitud de onda y el nombre de la l\'{\i}nea medida. En la columna 2, la curva de enrojecimiento adoptada, f($\lambda$), normalizada a H$\beta$. La columna 3 el ancho equivalente (EW) en \AA. En la columna 4 la intensidad de la l\'{\i}nea ($I(\lambda)$) corregida por enrojecimiento y relativa a 10000 H$\beta$. Los errores relativos en las l\'{\i}neas de emisi\'on listados en la columna 5 se obtuvieron mediante la propagaci\'on cuadr\'atica de los errores observacionales en los flujos de las l\'{\i}neas y la incertidumbre en la constante de enrojecimiento. No se han tenido en cuenta los errores de las intensidades te\'oricas ya que son mucho menores que los errores de observaci\'on.
 
\begin{table*} 
\caption[L\'{\i}neas de emisi\'on medidas en el espectro de ranura larga en los brotes B y C]{L\'{\i}neas de emisi\'on medidas en el espectro de ranura larga en los {\bf brotes B y C} obtenidas para la medida global. Intensidad de la l'inea corregida por enrojecimiento [$F(H\beta)$=$I(H\beta)$=10000]}
\label{ratiostot 1}
\begin{center}
\begin{tabular}{l@{\hspace{0.1cm}}ccccccc}
\hline
 &  & \multicolumn{3}{c}{B} & \multicolumn{3}{c}{C} \\
 \multicolumn{1}{c}{$\lambda$ ({\AA})} & f($\lambda$) & -EW  & $I(\lambda)$ & Error & -EW  & $I(\lambda)$ & Error  \\
  & & (\AA) & & (\%)  & (\AA) & & (\%)\\
  \hline
 3727 [\OII]$^a$       &   0.271   &  109.1 & 11943 $\pm$  117 &  1.0 &   45.3 & 25406 $\pm$ 2104 &  8.3 \\
 3835 H9              &   0.246   &    3.8 &   465 $\pm$   23 &  5.0 &  \ldots &  \ldots & \ldots \\
 3868 [\NeIII]         &   0.238   &   34.0 &  4494 $\pm$   11 &  0.2 &  \ldots &  \ldots & \ldots \\
 3889 He{\sc i}+H8          &   0.233   &    9.4 &  1263 $\pm$   65 &  5.1 &  \ldots &  \ldots & \ldots \\
 3970 [\NeIII]+H$\epsilon$      &   0.215   &   21.1 &  2535 $\pm$    4 &  0.2 &  \ldots &  \ldots & \ldots \\
 4068 [\SII]           &   0.195   &    0.7 &    87 $\pm$   17 & 19.6 &  \ldots &  \ldots & \ldots \\
 4102 H$\delta$          &   0.188   &   21.1 &  2395 $\pm$   31 &  1.3 &  \ldots &  \ldots & \ldots \\
 4340 H$\gamma$          &   0.142   &   41.5 &  4483 $\pm$   17 &  0.4 &    6.0 &  3853 $\pm$  657 & 17.1 \\
 4363 [\OIII]          &   0.138   &    6.3 &   694 $\pm$   42 &  6.0 &  \ldots &  \ldots & \ldots \\
 4471 He{\sc i}             &   0.106   &    2.8 &   311 $\pm$   16 &  5.2 &    3.8 &  2842 $\pm$  566 & 19.9 \\
 4658 [\FeIII]         &   0.053   &    0.3 &    36 $\pm$    6 & 17.8 &  \ldots &  \ldots & \ldots \\
 4686 He{\sc ii}            &   0.045   &    1.5 &   154 $\pm$   19 & 12.1 &    1.6 &  1261 $\pm$  330 & 26.1 \\
 4713 [\ArIV]+He{\sc i}      &   0.038   &    0.6 &    64 $\pm$    7 & 11.4 &  \ldots &  \ldots & \ldots \\
 4740 [\ArIV]          &   0.031   &    0.9 &    87 $\pm$   12 & 14.1 &  \ldots &  \ldots & \ldots \\
 4861 H$\beta$           &   0.000   &  114.1 & 10000 $\pm$   25 &  0.2 &   14.6 & 10000 $\pm$  541 &  5.4 \\
 4959 [\OIII]          &  -0.024   &  221.3 & 21220 $\pm$   77 &  0.4 &    8.3 &  6682 $\pm$  209 &  3.1 \\
 5007 [\OIII]          &  -0.035   &  539.1 & 51342 $\pm$   88 &  0.2 &   27.9 & 22042 $\pm$  467 &  2.1 \\
 5876 He{\sc i}             &  -0.209   &   22.5 &  1235 $\pm$   28 &  2.3 &  \ldots &  \ldots & \ldots \\
 6300 [OI]            &  -0.276   &    6.8 &   366 $\pm$   16 &  4.3 &    2.5 &  1542 $\pm$  295 & 19.1 \\
 6312 [\SIII]          &  -0.278   &    2.3 &   118 $\pm$   18 & 15.5 &  \ldots &  \ldots & \ldots \\
 6364 [OI]            &  -0.285   &    2.2 &   117 $\pm$   11 &  9.6 &  \ldots &  \ldots & \ldots \\
 6548 [\NII]           &  -0.311   &    8.3 &   416 $\pm$   20 &  4.8 &  \ldots &  \ldots & \ldots \\
 6563 H$\alpha$           &  -0.313   &  565.8 & 27618 $\pm$   36 &  0.1 &   52.6 & 28500 $\pm$  824 &  2.9 \\
 6584 [\NII]           &  -0.316   &   23.4 &  1175 $\pm$   22 &  1.8 &    5.8 &  3230 $\pm$  372 & 11.5 \\
 6678 He{\sc i}             &  -0.329   &    5.5 &   264 $\pm$   13 &  4.7 &  \ldots &  \ldots & \ldots \\
 6717 [\SII]           &  -0.334   &   32.1 &  1453 $\pm$   24 &  1.7 &   10.2 &  5209 $\pm$  613 & 11.8 \\
 6731 [\SII]           &  -0.336   &   13.7 &   617 $\pm$   49 &  8.0 &    5.1 &  2609 $\pm$  493 & 18.9 \\
 7065 He{\sc i}             &  -0.377   &    6.3 &   251 $\pm$   18 &  7.2 &  \ldots &  \ldots & \ldots \\
 7136 [\ArIII]         &  -0.385   &   19.4 &   707 $\pm$   21 &  3.0 &  \ldots &  \ldots & \ldots \\
 7281 He{\sc i}$^b$         &  -0.402   &    1.1 &    42 $\pm$    5 & 11.9 &  \ldots &  \ldots & \ldots \\
 7319 [\OII]$^c$       &  -0.406   &    4.0 &   148 $\pm$    6 &  3.9 &  \ldots &  \ldots & \ldots \\
 7330 [\OII]$^d$       &  -0.407   &    3.1 &   116 $\pm$    4 &  3.2 &  \ldots &  \ldots & \ldots \\
 7751 [\ArIII]         &  -0.451   &    5.8 &   186 $\pm$   13 &  7.0 &  \ldots &  \ldots & \ldots \\
 9069 [\SIII]          &  -0.561   &   69.9 &  1202 $\pm$  118 &  9.8 &  \ldots &  \ldots & \ldots \\
  \hline
     I(H$\beta$)$^\ast$ &   & \multicolumn{3}{c}{4.49\,$\times$\,10$^{-14}$}  & \multicolumn{3}{c}{0.11\,$\times$\,10$^{-14}$}  \\
     c(H$\beta$) &     & \multicolumn{3}{c}{ 0.29 $\pm$ 0.01}& \multicolumn{3}{c}{ 0.24 $\pm$ 0.12 }\\
     \hline
     \multicolumn{8}{l} {$^a$\,[O{\sc ii}]\,$\lambda\lambda$\,3726\,+\,3729; $^b$\,posible doblete con una l'inea desconocida}\\
    \multicolumn{8}{l} {$^c$\,[O{\sc ii}]\,$\lambda\lambda$\,7318\,+\,7320;  $^d$\,[O{\sc ii}]\,$\lambda\lambda$\,7330\,+\,7331.}\\
\multicolumn{8}{l} {$^\ast$ erg\,seg$^{-1}$\,cm$^{-2}$}
\end{tabular}
\end{center}
\end{table*}


\begin{sidewaystable}
 {\footnotesize
\caption[L\'{\i}neas de emisi\'on medidas en el espectro 'echelle en el brote A]{L\'{\i}neas de emisi\'on medidas en el espectro 'echelle en el {\bf brote A} obtenidas en la medida global, y en las componentes angostas y en la ancha. Intensidad de la l'inea corregida por enrojecimiento [$F(H\beta)$=$I(H\beta)$=10000]}
\label{ratiostot 2}
\begin{center}
\begin{tabular}{l@{\hspace{0.1cm}}ccccccccccccc}
\hline
& & \multicolumn{3}{c}{global} & \multicolumn{3}{c}{angosta\,1} & \multicolumn{3}{c}{angosta\,2} & \multicolumn{3}{c}{ancha} \\
 \multicolumn{1}{c}{$\lambda$ ({\AA})} & f($\lambda$) & -EW  & $I(\lambda)$ & Error & -EW  & $I(\lambda)$ & Error  & -EW  & $I(\lambda)$ & Error & -EW  & $I(\lambda)$ & Error  \\
  & & (\AA) & & (\%) & (\AA) & & (\%) & (\AA) & & (\%) & (\AA) & & (\%)  \\
 \hline
   3727 [\OII]$^a$       &   0.271   &   12.8 & 30071 $\pm$ 1702 &  5.7 &    6.7 & 34386 $\pm$ 1573 &  4.6 &    2.5 & 37854 $\pm$ 4183 & 11.0 &    1.9 & 33494 $\pm$ 5979 & 17.9 \\
  3868 [\NeIII]         &   0.238   &    1.3 &  2899 $\pm$  293 & 10.1 &  \ldots &  \ldots & \ldots &  \ldots &  \ldots & \ldots &  \ldots &  \ldots & \ldots \\
 3888 H8+He{\sc i}          &   0.233   &    0.8 &  1277 $\pm$  202 & 15.8 &  \ldots &  \ldots & \ldots &  \ldots &  \ldots & \ldots &  \ldots &  \ldots & \ldots \\
  3970 [\NeIII]+H$\epsilon$     &   0.215   &    1.7 &  1937 $\pm$  363 & 18.7 &  \ldots &  \ldots & \ldots &  \ldots &  \ldots & \ldots &  \ldots &  \ldots & \ldots \\
  4102 H$\delta$          &   0.188   &    5.4 &  5342 $\pm$ 1304 & 24.4 &  \ldots &  \ldots & \ldots &  \ldots &  \ldots & \ldots &  \ldots &  \ldots & \ldots \\
 4340 H$\gamma$          &   0.142   &    6.2 &  5138 $\pm$  279 &  5.4 &    2.2 &  4590 $\pm$  190 &  4.1 &    0.8 &  4205 $\pm$  546 & 13.0 &    0.4 &  2797 $\pm$  939 & 33.6 \\
  4471 He{\sc i}             &   0.106   &    0.4 &   470 $\pm$   92 & 19.6 &  \ldots &  \ldots & \ldots &  \ldots &  \ldots & \ldots &  \ldots &  \ldots & \ldots \\
  4861 H$\beta$           &   0.000   &   12.9 & 10000 $\pm$  340 &  3.4 &    6.2 & 10000 $\pm$  156 &  1.6 &    2.9 & 10000 $\pm$  285 &  2.9 &    2.1 & 10000 $\pm$  585 &  5.8 \\
 4959 [\OIII]          &  -0.024   &    6.3 &  6302 $\pm$  123 &  2.0 &    2.3 &  4505 $\pm$  170 &  3.8 &    2.3 &  9241 $\pm$  444 &  4.8 &    1.4 &  7797 $\pm$  936 & 12.0 \\
 5007 [\OIII]          &  -0.035   &   22.4 & 19398 $\pm$  221 &  1.1 &    8.3 & 14191 $\pm$  155 &  1.1 &    7.5 & 26527 $\pm$  323 &  1.2 &    5.4 & 26718 $\pm$  988 &  3.7 \\
  5876 He{\sc i}             &  -0.209   &    1.7 &   992 $\pm$   61 &  6.1 &    0.6 &   686 $\pm$   75 & 10.9 &    0.7 &  1347 $\pm$  141 & 10.5 &    0.5 &  1539 $\pm$  346 & 22.5 \\
 6300 [\OI]            &  -0.276   &    1.7 &   764 $\pm$   43 &  5.7 &  \ldots &  \ldots & \ldots &  \ldots &  \ldots & \ldots &  \ldots &  \ldots & \ldots \\
  6548 [\NII]           &  -0.311   &    4.2 &  1857 $\pm$   67 &  3.6 &    2.1 &  1912 $\pm$   81 &  4.3 &    1.1 &  1439 $\pm$  106 &  7.3 &    0.7 &  1646 $\pm$  449 & 27.3 \\
 6563 H$\alpha$           &  -0.313   &   74.0 & 30095 $\pm$  224 &  0.7 &   31.8 & 28500 $\pm$  142 &  0.5 &   22.1 & 28500 $\pm$  211 &  0.7 &   12.9 & 28500 $\pm$  659 &  2.3 \\
 6584 [\NII]           &  -0.316   &   13.3 &  6010 $\pm$  174 &  2.9 &    5.9 &  5565 $\pm$  131 &  2.4 &    3.5 &  4812 $\pm$  201 &  4.2 &    3.2 &  7429 $\pm$  692 &  9.3 \\
 6678 He{\sc i}             &  -0.329   &    1.0 &   422 $\pm$   48 & 11.4 &  \ldots &  \ldots & \ldots &  \ldots &  \ldots & \ldots &  \ldots &  \ldots & \ldots \\
 6717 [\SII]           &  -0.334   &    8.2 &  3802 $\pm$  123 &  3.2 &    3.1 &  2944 $\pm$  102 &  3.5 &    1.4 &  1784 $\pm$  130 &  7.3 &    3.8 &  8532 $\pm$  843 &  9.9 \\
 6731 [\SII]           &  -0.336   &    8.3 &  3246 $\pm$  208 &  6.4 &    1.9 &  1889 $\pm$  122 &  6.5 &    1.1 &  1431 $\pm$  132 &  9.2 &    2.2 &  5248 $\pm$  666 & 12.7 \\
 7065 He{\sc i}             &  -0.377   &    0.6 &   272 $\pm$   89 & 32.9 &  \ldots &  \ldots & \ldots &  \ldots &  \ldots & \ldots &  \ldots &  \ldots & \ldots \\
 7136 [\ArIII]         &  -0.385   &    2.1 &   855 $\pm$   53 &  6.1 &    0.7 &   580 $\pm$   84 & 14.5 &    0.5 &   518 $\pm$   84 & 16.1 &    0.7 &  1435 $\pm$  378 & 26.4 \\
  9069 [\SIII]          &  -0.561   &    4.4 &  2173 $\pm$  136 &  6.3 &    1.6 &  1804 $\pm$  207 & 11.5 &    1.5 &  1850 $\pm$  251 & 13.6 &    0.5 &  1256 $\pm$  389 & 31.0 \\
 9532 [\SIII]          &  -0.592   &    5.6 &  3113 $\pm$  253 &  8.1 &    1.1 &  1454 $\pm$  247 & 17.0 &    2.2 &  2860 $\pm$  333 & 11.6 &    1.7 &  4631 $\pm$ 1178 & 25.4 \\
      I(H$\beta$)$^\ast$ &   & \multicolumn{3}{c}{1.56\,$\times$\,10$^{-14}$}  & \multicolumn{3}{c}{0.85\,$\times$\,10$^{-14}$}  & \multicolumn{3}{c}{0.39\,$\times$\,10$^{-14}$}  & \multicolumn{3}{c}{0.28\,$\times$\,10$^{-14}$}  \\
     c(H$\beta$) &     & \multicolumn{3}{c}{ 0.22 $\pm$ 0.04 }    & \multicolumn{3}{c}{ 0.11 $\pm$ 0.03 }    & \multicolumn{3}{c}{ 0.68 $\pm$ 0.05 }    & \multicolumn{3}{c}{ 0.37 $\pm$ 0.12 }  \\
   \hline
     \multicolumn{14}{l} {$^a$\,[O{\sc ii}]\,$\lambda\lambda$\,3726\,+\,3729}\\
\multicolumn{14}{l} {$^\ast$ erg\,seg$^{-1}$\,cm$^{-2}$}\\
\end{tabular}
\end{center}}
\end{sidewaystable}

\begin{table*} 
\caption[L\'{\i}neas de emisi\'on medidas en el espectro 'echelle en el brote B]{L\'{\i}neas de emisi\'on medidas en el espectro 'echelle en el {\bf brote B} obtenidas en la medida global, y en la componente angosta y en la ancha. Intensidad de la l'inea corregida por enrojecimiento [$F(H\beta)$=$I(H\beta)$=10000]}
\label{ratiostot 3}
{\footnotesize
\begin{center}
\begin{tabular}{l@{\hspace{0.1cm}}cccccccccc}
\hline
&  & \multicolumn{3}{c}{global} & \multicolumn{3}{c}{angosta} & \multicolumn{3}{c}{ancha} \\
 \multicolumn{1}{c}{$\lambda$ ({\AA})} & f($\lambda$)  & -EW  & $I(\lambda)$ & Error & -EW  & $I(\lambda)$ & Error  & -EW  & $I(\lambda)$ & Error  \\
  & & (\AA) & & (\%) & (\AA) & & (\%) & (\AA) & & (\%)  \\
  \hline
 3727 [\OII]$^a$       &   0.271   &   20.4 &  9049 $\pm$  236 &  2.6 &    8.1 & 24096 $\pm$  854 &  3.5 &    8.0 &  5877 $\pm$  310 &  5.3 \\
 3835 H9              &   0.246   &    1.0 &   358 $\pm$   16 &  4.4 &  \ldots &  \ldots & \ldots &  \ldots &  \ldots & \ldots \\
 3868 [\NeIII]         &   0.238   &   12.4 &  5585 $\pm$  184 &  3.3 &    1.6 &  4214 $\pm$  649 & 15.4 &    8.4 &  5845 $\pm$  271 &  4.6 \\
 3889 He{\sc i}+H8          &   0.233   &    5.1 &  1571 $\pm$   40 &  2.6 &  \ldots &  \ldots & \ldots &  \ldots &  \ldots & \ldots \\
 3968 [\NeIII]+H7      &   0.216   &    5.2 &  1498 $\pm$   30 &  2.0 &    0.5 &  1016 $\pm$  375 & 36.9 &    2.8 &  1501 $\pm$  163 & 10.8 \\
 3970 [\NeIII]+H$\epsilon$     &   0.215   &    3.8 &  1062 $\pm$   57 &  5.3 &    1.2 &  1626 $\pm$  242 & 14.9 &    3.4 &  1260 $\pm$  104 &  8.2 \\
 4102 H$\delta$          &   0.188   &    9.7 &  2766 $\pm$  118 &  4.3 &    1.4 &  2219 $\pm$  338 & 15.2 &    5.7 &  2534 $\pm$  113 &  4.5 \\
 4340 H$\gamma$          &   0.142   &   37.8 &  4827 $\pm$  122 &  2.5 &    5.6 &  5305 $\pm$  159 &  3.0 &   16.7 &  4434 $\pm$   56 &  1.3 \\
 4363 [\OIII]          &   0.138   &    4.4 &   783 $\pm$  127 & 16.3 &    0.4 &   465 $\pm$  213 & 45.8 &    2.4 &   720 $\pm$   90 & 12.5 \\
 4471 He{\sc i}             &   0.106   &    3.0 &   471 $\pm$   17 &  3.6 &  \ldots &  \ldots & \ldots &  \ldots &  \ldots & \ldots \\
 4713 [\ArIV]+He{\sc i}      &   0.038   &    2.6 &   336 $\pm$   28 &  8.4 &  \ldots &  \ldots & \ldots &  \ldots &  \ldots & \ldots \\
 4861 H$\beta$           &   0.000   &   81.3 & 10000 $\pm$   98 &  1.0 &   14.1 & 10000 $\pm$  157 &  1.6 &   45.6 & 10000 $\pm$   65 &  0.7 \\
 4959 [\OIII]          &  -0.024   &  130.0 & 22714 $\pm$  175 &  0.8 &   16.7 & 14958 $\pm$  286 &  1.9 &   91.0 & 25569 $\pm$  150 &  0.6 \\
 5007 [\OIII]          &  -0.035   &  470.0 & 69812 $\pm$  213 &  0.3 &   35.0 & 37507 $\pm$  795 &  2.1 &  233.8 & 79550 $\pm$  336 &  0.4 \\
 5876 He{\sc i}             &  -0.209   &   12.4 &  1053 $\pm$    9 &  0.9 &    1.7 &   571 $\pm$   37 &  6.5 &   10.2 &  1237 $\pm$   20 &  1.6 \\
 6300 [OI]            &  -0.276   &    2.7 &   181 $\pm$    8 &  4.4 &    2.0 &   501 $\pm$   38 &  7.5 &    0.8 &    73 $\pm$   22 & 30.3 \\
 6312 [\SIII]          &  -0.278   &    1.8 &   125 $\pm$   25 & 20.2 &    0.7 &   195 $\pm$   43 & 22.2 &    1.0 &   103 $\pm$   25 & 24.1 \\
 6548 [\NII]           &  -0.311   &    2.4 &   185 $\pm$    5 &  2.6 &    1.6 &   422 $\pm$   24 &  5.7 &    1.6 &   158 $\pm$   21 & 13.0 \\
 6563 H$\alpha$           &  -0.313   &  358.0 & 28618 $\pm$   40 &  0.1 &   73.5 & 29349 $\pm$  267 &  0.9 &  184.8 & 28008 $\pm$  130 &  0.5 \\
 6584 [\NII]           &  -0.316   &    7.7 &   623 $\pm$   12 &  1.9 &    6.0 &  1573 $\pm$   53 &  3.4 &    4.0 &   404 $\pm$   25 &  6.2 \\
 6678 He{\sc i}             &  -0.329   &    4.8 &   320 $\pm$    3 &  1.1 &    0.5 &   129 $\pm$   27 & 20.7 &    4.6 &   427 $\pm$   16 &  3.8 \\
 6717 [\SII]           &  -0.334   &   10.6 &   728 $\pm$   10 &  1.3 &    7.4 &  1921 $\pm$   46 &  2.4 &    3.6 &   362 $\pm$   29 &  8.1 \\
 6731 [\SII]           &  -0.336   &    8.4 &   585 $\pm$   22 &  3.8 &    5.2 &  1405 $\pm$   53 &  3.8 &    2.6 &   270 $\pm$   34 & 12.6 \\
 7065 He{\sc i}             &  -0.377   &    4.9 &   293 $\pm$    6 &  1.9 &  \ldots &  \ldots & \ldots &  \ldots &  \ldots & \ldots \\
 7136 [\ArIII]         &  -0.385   &   11.0 &   705 $\pm$    9 &  1.3 &    4.0 &   976 $\pm$   52 &  5.3 &    6.9 &   666 $\pm$   23 &  3.4 \\
 7751 [\ArIII]         &  -0.451   &    3.2 &   196 $\pm$   13 &  6.4 &  \ldots &  \ldots & \ldots &  \ldots &  \ldots & \ldots \\
 9069 [\SIII]          &  -0.561   &   13.9 &  1193 $\pm$  111 &  9.3 &    3.1 &  1007 $\pm$  136 & 13.5 &    9.2 &  1325 $\pm$   90 &  6.8 \\
 9532 [\SIII]          &  -0.592   &   21.3 &  2510 $\pm$  141 &  5.6 &    6.9 &  2885 $\pm$  185 &  6.4 &   13.1 &  2488 $\pm$  125 &  5.0 \\
  \hline
     I(H$\beta$)$^\ast$ &   & \multicolumn{3}{c}{3.16\,$\times$\,10$^{-14}$}  & \multicolumn{3}{c}{0.74\,$\times$\,10$^{-14}$}  & \multicolumn{3}{c}{2.40\,$\times$\,10$^{-14}$}  \\
     c(H$\beta$) &     & \multicolumn{3}{c}{ 0.13 $\pm$ 0.01 }    & \multicolumn{3}{c}{ 0.36 $\pm$ 0.02 }    & \multicolumn{3}{c}{ 0.07 $\pm$ 0.01 }  \\
   \hline
     \multicolumn{11}{l} {$^a$\,[O{\sc ii}]\,$\lambda\lambda$\,3726\,+\,3729}\\
      \multicolumn{11}{l} {$^\ast$ erg\,seg$^{-1}$\,cm$^{-2}$}
\end{tabular}
\end{center}}
\end{table*}

 \begin{table*} 
  {\footnotesize
\caption[L\'{\i}neas de emisi\'on medidas en el espectro 'echelle en el brote C]{L\'{\i}neas de emisi\'on medidas en el espectro 'echelle en {\bf brote C} obtenidas en la medida global, y en la componente angosta y en la ancha. Intensidad de la l'inea corregida por enrojecimiento [$F(H\beta)$=$I(H\beta)$=10000]}
\label{ratiostot 4}
\begin{center}
\begin{tabular}{l@{\hspace{0.1cm}}cccccccccc}
\hline
&  & \multicolumn{3}{c}{global} & \multicolumn{3}{c}{angosta} & \multicolumn{3}{c}{ancha} \\
 \multicolumn{1}{c}{$\lambda$ ({\AA})} & f($\lambda$) & -EW  & $I(\lambda)$ & Error & -EW  & $I(\lambda)$ & Error  & -EW  & $I(\lambda)$ & Error  \\
  & & (\AA) & & (\%) & (\AA) & & (\%) & (\AA) & & (\%)  \\
  \hline
 4861 H$\beta$           &   0.000   &    3.2 & 10000 $\pm$ 2188 & 21.9 &    1.0 & 10000 $\pm$ 1464 & 14.6 &    1.1 & 10000 $\pm$ 2349 & 23.5 \\
 4959 [\OIII]          &  -0.024   &    5.6 &  6879 $\pm$  391 &  5.7 &    3.4 &  9818 $\pm$  508 &  5.2 &    2.0 &  5245 $\pm$  612 & 11.7 \\
 5007 [\OIII]          &  -0.035   &    5.6 & 20467 $\pm$ 1265 &  6.2 &    3.4 & 29371 $\pm$ 1584 &  5.4 &    2.0 & 15582 $\pm$ 1859 & 11.9 \\
 6563 H$\alpha$           &  -0.313   &   18.1 & 28500 $\pm$ 1148 &  4.0 &    6.5 & 28500 $\pm$  709 &  2.5 &    8.0 & 28500 $\pm$  799 &  2.8 \\
 6584 [\NII]           &  -0.316   &    1.3 &  2334 $\pm$  740 & 31.7 &    0.9 &  3852 $\pm$  942 & 24.4 &    0.4 &  1412 $\pm$  588 & 41.7 \\
  6717 [\SII]           &  -0.334   &    2.6 &  5015 $\pm$ 1591 & 31.7 &    1.2 &  5400 $\pm$ 1357 & 25.1 &    1.4 &  5316 $\pm$ 1890 & 35.6 \\
 6731 [\SII]           &  -0.336   &    1.8 &  2891 $\pm$ 1071 & 37.0 &    0.7 &  3186 $\pm$ 1295 & 40.6 &    1.0 &  3891 $\pm$ 1606 & 41.3 \\
  \hline
      I(H$\beta$)$^\ast$ &   & \multicolumn{3}{c}{0.08\,$\times$\,10$^{-14}$}  & \multicolumn{3}{c}{0.04\,$\times$\,10$^{-14}$}  & \multicolumn{3}{c}{0.04\,$\times$\,10$^{-14}$}  \\
     c(H$\beta$) &     & \multicolumn{3}{c}{ 0.21 $\pm$ 0.40 }    & \multicolumn{3}{c}{ 0.14 $\pm$ 0.25 }    & \multicolumn{3}{c}{ 0.38 $\pm$ 0.41 }  \\
  \hline
  \multicolumn{11}{l} {$^\ast$ erg\,seg$^{-1}$\,cm$^{-2}$}
\end{tabular}
\end{center}}
\end{table*}

\begin{table*} 
 {\footnotesize
\caption[L\'{\i}neas de emisi\'on medidas en el espectro 'echelle en el brote E]{L\'{\i}neas de emisi\'on medidas en el espectro 'echelle en {\bf brote E} obtenidas en la medida global, y en la componentes angostas. Intensidad de la l'inea corregida por enrojecimiento [$F(H\beta)$=$I(H\beta)$=10000]} 
\label{ratiostot 5}
\begin{center}
\begin{tabular}{l@{\hspace{0.1cm}}cccccccccc}
\hline
 &  & \multicolumn{3}{c}{global} & \multicolumn{3}{c}{angosta\,1} & \multicolumn{3}{c}{angosta\,2} \\
 \multicolumn{1}{c}{$\lambda$ ({\AA})} & f($\lambda$) & -EW  & $I(\lambda)$ & Error& -EW  & $I(\lambda)$ & Error  & -EW  & $I(\lambda)$ & Error\\
  & & (\AA) & & (\%)  & (\AA) & & (\%)& (\AA) & & (\%)\\
  \hline
 3727 [\OII]$^a$       &   0.271   &    8.2 & 32110 $\pm$ 9745 & 30.3 &  \ldots & \ldots &\ldots &  \ldots & \ldots &\ldots \\
 4861 H$\beta$           &   0.000   &    7.5 & 10000 $\pm$ 1530 & 15.3 &    1.8 & 10000 $\pm$ 1267 & 12.7 &    1.9 & 10000 $\pm$ 1816 & 18.2 \\
 4959 [\OIII]          &  -0.024   &    2.9 &  5747 $\pm$  729 & 12.7 &    2.2 &  3443 $\pm$  351 & 10.2 &    4.4 &  6900 $\pm$  383 &  5.6 \\
 5007 [\OIII]          &  -0.035   &    6.9 & 12160 $\pm$  921 &  7.6 &    2.2 & 10157 $\pm$ 1047 & 10.3 &    4.4 & 20671 $\pm$ 1214 &  5.9 \\
 5876 He{\sc i}             &  -0.209   &    1.6 &  1733 $\pm$  561 & 32.4 &  \ldots & \ldots &\ldots &  \ldots & \ldots &\ldots \\
 6300 [OI]            &  -0.276   &    2.1 &  2017 $\pm$  563 & 27.9 &  \ldots & \ldots &\ldots &  \ldots & \ldots &\ldots \\
 6548 [\NII]           &  -0.311   &    1.4 &  1419 $\pm$  468 & 33.0 &  \ldots & \ldots &\ldots &  \ldots & \ldots &\ldots \\
 6563 H$\alpha$           &  -0.313   &   32.0 & 28500 $\pm$ 1087 &  3.8 &   15.9 & 28500 $\pm$  637 &  2.2 &   11.0 & 28500 $\pm$  964 &  3.4 \\
 6584 [\NII]           &  -0.316   &    5.8 &  5268 $\pm$ 1157 & 22.0 &    1.7 &  2810 $\pm$  620 & 22.0 &    3.7 &  9025 $\pm$ 2224 & 24.6 \\
 6717 [\SII]           &  -0.334   &    6.0 &  6212 $\pm$ 1443 & 23.2 &    1.4 &  2509 $\pm$  512 & 20.4 &    4.3 & 11233 $\pm$ 2878 & 25.6 \\
 6731 [\SII]           &  -0.336   &    5.2 &  4448 $\pm$ 1400 & 31.5 &    0.8 &  1875 $\pm$  789 & 42.1 &    1.8 &  6636 $\pm$ 2630 & 39.6 \\
  \hline
      I(H$\beta$)$^\ast$ &   & \multicolumn{3}{c}{0.18\,$\times$\,10$^{-14}$}  & \multicolumn{3}{c}{0.07\,$\times$\,10$^{-14}$}  & \multicolumn{3}{c}{0.07\,$\times$\,10$^{-14}$}  \\
     c(H$\beta$) &     & \multicolumn{3}{c}{ 0.00 }    & \multicolumn{3}{c}{ 0.59 $\pm$ 0.22 }    & \multicolumn{3}{c}{0.00 }  \\
 \hline
     \multicolumn{11}{l} {$^a$\,[O{\sc ii}]\,$\lambda\lambda$\,3726\,+\,3729}\\
     \multicolumn{11}{l} {$^\ast$ erg\,seg$^{-1}$\,cm$^{-2}$}\\
\end{tabular}
\end{center}}
\end{table*}
 
\begin{table*} 
\caption[L\'{\i}neas de emisi\'on medidas en el espectro 'echelle en el brote F]{L\'{\i}neas de emisi\'on medidas en el espectro 'echelle en  {\bf brote F} obtenidas en la medida global. Intensidad de la l'inea [$F(H\beta)$=$I(H\beta)$=10000]. $^\dagger$Las l'ineas no fueron corregidas por enrojecimiento, ver texto}
\label{ratiostot 6}
\begin{center}
\begin{tabular}{l@{\hspace{0.1cm}}cccc}
\hline
 &  & \multicolumn{3}{c}{global} \\
 \multicolumn{1}{c}{$\lambda$ ({\AA})} & f($\lambda$) & -EW  & $F(\lambda)$ & Error \\
  & & (\AA) & & (\%)  \\
  \hline
 4861 H$\beta$          &   0.000   &   -2.5 & 10000 $\pm$ 3213 & 32.1 \\
 4959 [\OIII]          &  -0.024   &   -1.2 &  7274 $\pm$  597 &  8.2 \\
 5007 [\OIII]          &  -0.035   &   -3.6 & 21797 $\pm$ 1981 &  9.1 \\
 6563 H$\alpha$          &  -0.313   &  -10.6 & 28500 $\pm$ 2593 &  9.1 \\
 6584 [\NII]           &  -0.316   &   -1.1 &  3415 $\pm$ 2153 & 63.0 \\
 6717 [\SII]          &  -0.334   &   -2.6 &  7317 $\pm$ 4024 & 55.0 \\
 6731 [\SII]          &  -0.336   &   -3.1 &  9011 $\pm$ 5317 & 59.0 \\
  \hline
      F(H$\beta$)$^\ast$&   & \multicolumn{3}{c}{0.04\,$\times$\,10$^{-14}$}  \\
     c(H$\beta$) &     & \multicolumn{3}{c}{ $^\dagger$ }  \\
 \hline
\multicolumn{5}{l} {$^\ast$ erg\,seg$^{-1}$\,cm$^{-2}$}
\end{tabular}
\end{center}
\end{table*}
 

En el espectro de ranura larga para el caso del brote B, todos los par\'ametros f\'{\i}sicos que dependen de la intensidad de la l\'{\i}nea [\SIII]\,$\lambda$\,9532\,\AA\ fueron calculados utilizando la relaci\'on te\'orica entre esta l\'{\i}nea y la l\'{\i}nea [\SIII]\,$\lambda$9069\,\AA, I(9069)\,$\approx$\,2.44$\times$\,I(9532) \citep{O89}, ya que esta l'inea est'a fuera del rango espectral de observaci'on. En el brote C fue imposible cuantificar una medida de estas l\'{\i}neas. En este caso, no se incluyeron estas l\'{\i}neas ni en las tablas ni en nuestros c\'alculos. Tambi\'en para este mismo brote, es imposible medir la l\'{\i}nea [\NII]\,$\lambda$\,6548\,\AA, ya que se solapa con H$\alpha$. As\'{\i} que se tuvo que corregir el flujo de H$\alpha$ de la contaminaci\'on proveniente de esta l\'{\i}nea de nitr\'ogeno estimando su contribuci'on a partir de [\NII]\,$\lambda$\,6584\,\AA\ utilizando su relaci'on te'orica, I(6584)\,$\approx$\,2.9$\,\cdot$\,I(6548).

Debido al desplazamiento al rojo que presenta esta galaxia, en todos los espectros 'echelle, la l\'{\i}nea [\SII]\,$\lambda$\,6731\,\AA\ est\'a afectada por la emisi\'on de l\'{\i}neas tel\'uricas de cielo de OH. Entonces, se procedi'o a utilizar la estrella estandar Feige\,110 (sdOB) como estandar tel\'urica, dado que la misma no presenta l\'{\i}neas estelares en ese rango, que fue observada con la misma masa de aire que la galaxia y que pudo ser muy bien calibrada en flujo. As'i es que, haciendo uso de la rutina \texttt{telluric} de {\sc IRAF}, se pudieron corregir los espectros de los brotes observados en el modo 'echelle por el efecto que producen las l\'{\i}neas tel'uricas. Luego esta l\'{\i}nea fue ajustada con la tarea \texttt{ngaussfit} copiando la soluci\'on del ajuste obtenido para la l\'{\i}nea [\SII]\,$\lambda$\,6717\,\AA, y usando el mismo procedimiento que se ha usado para las l\'{\i}neas de bajo S/N.
Luego, este ser\'a un importante punto a tener en cuenta al momento de analizar las densidades electr\'onicas derivadas para las diferentes componentes en estos brotes.
Aquellas l\'{\i}neas que presentaban errores relativos en flujo mayores al 45\% no fueron tenidas en cuenta para los c\'alculos. La 'unica excepci'on ha sido la l\'{\i}nea [\OIII]\,$\lambda$\,4363\,\AA\ en la componente angosta del brote B que tiene un error mayor al 45\% pero se la consider\'o para poder c\'alcular de temperatura del [\OIII].
En el brote F, solo se ha podido obtener la medida global dado que las l\'{\i}neas no tienen el suficiente S/N para realizar un ajuste por componentes.
En ning\'un caso fue posible medir las l\'{\i}neas del [\OIII]\,$\lambda\lambda$\,7319, 7330\,\AA.
Por \'ultimo, en los brotes C, E y F, debido al bajo S/N de estos tres espectros, las l\'{\i}neas [\SIII]\,$\lambda\lambda$\,9069, 9532\,\AA\ no pudieron ser medidas y, por otro lado, el flujo de la l'inea [\OIII]\,$\lambda$\,4959\,\AA\ fue calculado a partir de la l'inea [\OIII]\,$\lambda$\,5007\,\AA\ de acuerdo a la relaci'on te'orica entre ambas l'ineas, I(5007)\,$\approx$\,3$\times$\,I(4959) \citep{O89}. 

Para el brote A, no se tuvo en cuenta la l\'{\i}nea H$\delta$ debido a que la misma es muy d\'ebil y no puede ser descompuesta en las diferentes componentes cinem\'aticas. En el brote E, la segunda componente angosta y la medida global presentan un valor de C(H$\beta$) un poco menor a cero, pero compatible con \'este dentro de los errores observacionales, por eso fue tomado igual a cero. Adem'as en el brote F, la l'inea de recombinaci'on H$\beta$ tiene un bajo cociente S/N, por lo tanto las intensidades de las l'ineas en este brote no fueron corregidas por enrojecimiento.

\section{An\'alisis de Abundancias}
\label{sec:analisis_abund}
La espectrocop'ia es una poderosa herramienta que permite la determinaci'on de las abundancias de regiones H\,{\sc ii} en el Universo Local haciendo uso de los m'etodos basados en la medici'on de las intensidades de las l'ineas de emisi'on y en la f'isica at'omica. Esto es bien conocido como el m'etodo ``directo''. En el caso de las galaxias m'as distantes o intr'insecamente d'ebiles, el bajo cociente S/N obtenido con los telescopios actuales se contrapone a la aplicaci'on de este m'etodo y por ello es necesario la aplicaci'on de m'etodos emp'iricos, los cuales se basan en las l'ineas de emisi'on m'as intensas. La base fundamental de estos m'etodos emp'iricos se conoce bastante bien (ver por ejemplo P'erez-Montero \& D'iaz~\citeyear{2005MNRAS.361.1063P}). La exactitud de los resultados depender'a de la bondad de su calibraci'on, que a su vez depende de un conjunto de abundancias derivadas con precisi'on a trav'ez del m'etodo ``directo"\, para que el proceso de interpolaci'on sea fiable. 

Sin embargo, derivar de forma precisa las abundancias de cada elemento no es un asunto sencillo. En primer lugar, porque es necesario medir las l'ineas de emisi'on con exactitud. Y en segundo lugar, porque se requiere de un cierto conocimiento de la estructura de ionizaci'on de la regi'on con el fin de obtener las abundancias i'onicas de los diferentes elementos y, en algunos casos, se necesita de los modelos de fotoionizaci'on para corregir los estados de ionizaci'on inobservables en el rango espectral en cuesti'on. Para ello es necesario cubrir un amplio rango espectral que incluya desde las l'ineas del ultravioleta, el doblete de [\OII]\,$\lambda$$\lambda$\,3727,3729\,\AA, hasta el infrarrojo, con el doblete [\SIII]\,$\lambda$$\lambda$\,9069,9532\,\AA. Para tener un diagn'ostico preciso del medio se requiere de la medici'on de las l'ineas d'ebiles aurorales.  Y adem'as, la precisi'on en la medici'on de los cocientes entre estas l'ineas aurorales y las l'ineas de recombinaci'on de Balmer debe ser superior al 5\%, ya que en general, estas l'ineas son d'ebiles, alrededor del 1\% de la intensidad H$\beta$. 
Estos requerimientos permiten derivar las diferentes temperaturas de l'ineas: T$_e$([\OIII]), T$_e$([\OII]), T$_e$([\SIII]), T$_e$([\SII]), T$_e$([\NII]), necesarias para estudiar la estructura de temperatura y la ionizaci'on de cada brote o galaxia H\,{\sc ii} considerados como una regi'on ionizada multizona.

En esta secci\'on ser\'an explicados los diferentes m\'etodos utilizados para estimar las densidades y temperaturas electr\'onicas con las que luego ser\'an derivadas las abundancias i\'onicas y totales de las principales especies qu\'{\i}micas presentes en las regiones H\,{\sc ii} de Haro\,15. Las condiciones f'isicas del gas como las abundancias qu\'{\i}micas son derivadas y analizadas para cada brote de formaci\'on estelar (medida global) y para cada componente cinem\'atica estudiada en el Cap\'{\i}tulo 3. Se realizar'a todo este an'alisis siguiendo el procedimiento descripto en H{\"a}gele et al.~(\citeyear{2008MNRAS.383..209H}) y que se describir'a a continuaci'on.

\subsection{Propiedades f\'{\i}sicas del gas ionizado}
\label{sec:Prop_fisicas}
Las propiedades f\'{\i}sicas del gas ionizado, como las densidades y temperaturas electr\'onicas, pueden ser determinadas a partir de los cocientes de los flujos de las l\'{\i}neas observadas de algunos iones. Seg\'un Osterbrock~(\citeyear{O89}), la temperatura puede ser calculada a partir del cociente de transiciones colisionales con energ'ia similar pero originadas en diferentes niveles. El ajuste entre este cociente y la temperatura 
electr\'onica se ha obtenido utilizando la tarea \texttt{temden} de {\sc IRAF}. Dicha tarea est\'a basada en un modelo de equilibrio estad\'{\i}stico de un \'atomo de 5 niveles \citep{1987JRASC..81..195D,1995PASP..107..896S}. Las probabilidades de transici\'on y niveles de energ\'ia son los que dicho programa trae por defecto. Como fuente de error de las densidades y temperaturas se han tomado las incertezas asociadas a la medici\'on de los flujos de las l\'{\i}nea de emisi\'on y la correcci\'on de enrojecimiento, y han sido propagados a trav\'es de nuestros c\'alculos. Los coeficientes at\'omicos de colisi\'on adoptados son los mismos que en el trabajo de H{\"a}gele et al.~(\citeyear{2008PhDT........35H}) (ver Tabla 2.5 de dicho trabajo).
En la Tabla \ref{cocientes} se muestran los cocientes de l\'{\i}neas de emisi\'on necesarios para calcular las densidades y cada temperatura. 

\begin{table}
\centering
\caption[Cocientes de l\'{\i}neas de emisi\'on]{Cocientes utilizados para derivar las densidades y temperaturas electr\'onicas}
\vspace{0.3cm}
\label{cocientes}
\begin{tabular}{@{}ll@{}}
\hline
 & Cociente de l\'{\i}neas \\
\hline
$n_e$[\SII] &$R_{S2}$ = I(6717)/I(6731)\\
$T_e$[\OIII] &$R_{O3}$ = I(4959,5007)/I(4363)\\
$T_e$[\OII] &$R_{O2}$ = I(3727)/I(7319,7330)\\
$T_e$[\SIII] &$R_{S3}$ = I(9069,9532)/I(6312)\\
$T_e$[\SII] &$R'_{S2}$ = I(6717,6731)/I(4068,4076)\\
$T_e$[\NII]& $R_{N2}$ = I(6548,6584)/I(5755)\\
\hline
\end{tabular}
\end{table}

\subsubsection*{Densidades electr\'onicas}
\label{sec:Densidades}
El valor de la densidad electr'onica, N$_e$, fue determinado a partir del cociente $R_{S2}$ de las l\'{\i}neas [\SII]\,$\lambda\lambda$\,6717, 6731\,\AA, el cual es el m\'as usado para determinar la densidad electr\'onica en la zona de baja excitaci\'on del gas ionizado. 
Aqu\'{\i} abajo se muestra la expresi'on para calcular las densidad a partir del mejor ajuste encontrado entre el cociente y la temperatura electr\'onica:
\begin{eqnarray}
n([S{\textsc{ii}}])\,=\,10^3\cdot\frac{R_{S2} \cdot a_0(t)+a_1(t)}{R_{S2} \cdot a_2(t) + a_3(t)}
\label{ecu3}
\end{eqnarray}
donde
$$a_0(t) = 2.21-1.3/t-1.25t+0.23t^2$$
$$a_1(t) = -3.35+1.94/t+1.93t-0.36t^2$$
$$a_2(t) = -4.33+2.33/t+2.72t-0.57t^2$$
$$a_3(t) = 1.84-1/t-1.14t+0.24t^2$$

\noindent donde n([S{\sc ii}])\,=\,n$_e$\,=\,10$^{-4}$N$_e$ y $t$ suele ser generalmente, t$_e$([\OIII]), donde t$_e$\,=\,10$^{-4}$T$_e$, aunque se puede hacer un proceso iterativo para calcular con t$_e$([\SII]), ya que esta temperatura, al igual que ocurre a con t$_e$([\OII]), al ser un ion de tipo np$^3$ depende de la densidad. En este caso el cociente que se debe utilizar es el R$_{S2}'$.

Las densidades derivadas en los brotes, tanto para las medidas globales (datos de ranura larga y del 'echelle), como para cada componente (datos 'echelle) pueden ser enmarcadas en el l\'{\i}mite de baja densidad (n$_e$$\leq$500cm$^{-3}$), excepto en unos pocos casos donde la densidad presenta un error grande. Estos casos ser'an discutidos m'as adelante.

En la Secci\'on \ref{sec:discusion} se tratar\'a cada caso en particular.

\subsubsection*{Temperaturas electr\'onicas}
\label{sec:Temperaturas}

Dado que las temperaturas electr'onicas son cantidades f'isicas importantes para la determinaci'on de las abundancias qu'imicas del gas ionizado, es necesario derivar su valor para poder obtener una estimaci'on de la metalicidad del medio donde se forman las l'ineas de emisi'on. Sin embargo, en ocasiones, las l'ineas aurorales sensibles a la temperatura no son detectadas en el espectro de emisi'on de la nebulosa como ocurre, por ejemplo, en las regiones ricas en ox\'{\i}geno, donde la l\'{\i}nea auroral del [\OIII]\,$\lambda$\,4363\,\AA\ es muy d\'ebil y dif\'{\i}cil de detectar \citep{2007MNRAS.382..251D}. En general, este hecho puede deberse a una mala relaci'on se\~nal-ruido del espectro o a una elevada metalicidad del gas que produce un enfriamiento tan eficaz que la temperatura es muy baja. En este 'ultimo caso, las l'ineas no son detectables debido a que sus flujos son inversamente proporcionales a la exponencial de la temperatura y son intr'insecamente d'ebiles. 
Existen alternativas que consisten en la determinaci'on de la abundancia del ox'igeno haciendo uso de las calibraciones emp\'{\i}ricas con las l\'{\i}neas de emisi\'on intensas f\'acilmente observables. 
En este trabajo se ha encontrado que varias de las regiones observadas en Haro\,15 no presentan una o varias de las l'ineas aurorales necesarias para derivar las temperaturas con el m'etodo directo. Es por eso que a continuaci'on se detallan separadamente las diferentes v'ias por las que uno puede optar para derivar las temperaturas, tanto por el m'etodo directo o recurriendo a los diferentes modelos te'oricos o relaciones emp'iricas disponibles en la literatura, los cuales predicen relaciones entre temperaturas electr'onicas y distintos par'ametros emp'iricos y las temperaturas. 

La temperatura electr'onia est'a definida como la temperatura de la distribuci'on de Maxwell-Boltzmann asociada a las velocidades de los electrones que siguen dicha distribuci'on. Entonces, por convenci'on, cada vez que se escriba el t'ermino ``temperatura'' o ``temperatura de l\'inea'' estaremos refiri\'endonos al valor aproximado de la temperatura determinada por la energ'ia cin'etica media de los electrones libres en la regi'on donde se forma las l\'ineas de emisi'on del i\'on estudiado.

\subsubsection*{M'etodo directo}
\begin{itemize}
\item {Ox'igeno}
\end{itemize}

Para el c'alculo de la temperatura de [\OIII] se usa el cociente de l'ineas R$_{O3}$.
En este caso, el ajuste obtenido, al igual que para el resto de iones de tipo np$^2$, no depende de la densidad electr\'onica y es, por lo tanto, un buen indicador de temperatura.
El ajuste entre cociente y temperatura es:
\begin{eqnarray}
t_e([O{\textsc{iii}}])\,=\,0.8254 - 0.0002415 \cdot R_{O3} + \frac{47.77}{R_{O3}}
\label{ecu4}
\end{eqnarray}
\noindent
donde $t_{e}$ es la temperatura en unidades de 10$^4$ K

En el caso del [\OII], la temperatura electr\'onica se calcula a partir del cociente R$_{O2}$. En este caso hay que tener algo de cuidado dado que las l'ineas aurorales [\OII]\,$\lambda\lambda$\,7319, 7330\,\AA\  suelen tener algo de contaminaci\'on por emisi\'on producida por recombinaci\'on. Dicha emisi\'on, no obstante, se puede cuantificar y corregir. Seg\'un Liu et al. (2000) dicha contribuci\'on puede ser ajustada en el rango 0.5 $\leq$ t $\leq$ 1.0 mediante la funci\'on:
\begin{eqnarray}
\frac{I_R(7319+7330)}{I(H\beta)}\,=\,9.36 \cdot t^{0.44} \cdot \frac{O^{2+}}{H^+}
\label{ecu5}
\end{eqnarray}
expresi'on v'alida s'olo en el rango de temperaturas entre los 5000 y los 10000 K.
Adem\'as, el cociente de las l'ineas de [\OII] depende fuertemente de la densidad
electr\'onica. La situaci'on ideal es tener la densidad de [\OII] mediante el cociente 
I(3726\AA)/I(3729\AA), pero generalmente no hay bastante resoluci\'on para resolver este 
doblete, por lo que suele ser necesario recurrir a la densidad de [\SII], tambi\'en representativa de la zona de baja excitaci\'on. En este trabajo, a pesar de poder resolver las l'ineas del [\OII] en el espectro 'echelle, ambas se encuentran en el borde del orden ultravioleta del 'echelle con lo cual se prefiri'o calcular la densidad con las l'ineas prohibidas del azufre dos veces ionizado.
El ajuste obtenido es:
\begin{eqnarray}
t_e([O{\textsc{ii}}])\,=\,0.23 + 0.0017\cdot R_{O2} + \frac{38.3}{R_{O2}}+f_{1}(n_{e})
\label{ecu6}
\end{eqnarray}
 \noindent
donde f$_{1}$(n$_{e}$) es una funci'on dependiente de la densidad.

\begin{itemize}
\item {Azufre}
\end{itemize}
La temperatura de [S{\sc ii}] se calcula con el ajuste:
\begin{eqnarray}
t_e([S{\textsc{ii}}])\,=\,1.92 - 0.0375 \cdot R_{S2}' - \frac{14.5}{R_{S2}'} +  \frac{105.64}{R_{S2}'^2}+f_{2}(n_{e})
\label{ecu7}
\end{eqnarray}  
\noindent
donde f$_{2}$(n$_{e}$) es una funci'on dependiente de la densidad.

En el caso de las l\'{\i}neas aurorales de [\SII] basta con poder medir una de las dos, ya que entre ellas hay una relaci\'on fija te\'orica, de modo que I(4068\AA)\,$\approx$\,3\,$\cdot$\,I(4076\AA). De esta manera, es posible calcular la temperatura de [\SII].

En relaci\'on a la temperatura de [\SIII] su medida directa se ha comenzado a utilizar cuando han empezado a ser accesibles las l'ineas colisionales en el infrarrojo cercano. El cociente de l'ineas que se usa en este caso es el R$_{S3}$.
En ese caso, la temperatura de [\SIII] se puede calcular mediante el ajuste:
\begin{eqnarray}
t_e([S{\textsc{iii}}])\,=\,\frac{R_{S3} + 36.4}{1.8 \cdot R_{S3} - 3.01}
\label{ecu8}
\end{eqnarray} 
 \noindent

\begin{itemize}
\item {Nitr'ogeno}
\end{itemize}
La temperatura de [\NII] puede ser calculada mediante el cociente de las l'ineas de nitr'ogeno, R$_{N2}$.
Las l'ineas nebulares de [\NII] est\'an muy pr\'oximas a H$\alpha$ por lo que a veces aparecen fusionadas con \'esta y no es posible medirlas o medir la m\'as d\'ebil de ellas. En ese caso se suele recurrir a la relaci\'on te\'orica entre ellas, de tal manera que I(6584)\,$\approx$\,2.9$\,\cdot$\,I(6548).
Asimismo, la emisi'on de la l'inea auroral de [\NII]\,$\lambda$\,5755\,\AA\ est\'a afectada por procesos de recombinaci\'on, que puede ser corregida mediante la siguiente expresi\'on propuesta por Liu et al. (2000):
\begin{eqnarray}
\frac{I_R(5755)}{I(H\beta)}\,=\,3.19 \cdot t^{0.30} \cdot \frac{N^{2+}}{H^+}
\label{ecu9}
\end{eqnarray} 
en el rango de temperaturas entre 5000 y 20000 K.
H\"agele et al.~(\citeyear{2008MNRAS.383..209H}) usando el c'alculo de temperatura electr'onica [\OIII], estimaron las contribuciones por recombinaci'on tanto en las l'ineas aurorales del [\OII] como en la del [\NII] y encontraron que en todos los casos son menores al 4\%. De esta forma si se opta por no corregir esta contribuci'on, es necesario saber que se cuenta con un peque\~no error adicional.
En estas condiciones, es posible determinar la temperatura de [N{\sc ii}] mediante la expresi\'on:
\begin{eqnarray}
t_e([N{\textsc{ii}}])\,=\,0.537 + 0.000253 \cdot R_{N2} + \frac{42.126}{R_{N2}}
\label{ecu10}
\end{eqnarray}  
\noindent que no depende de la densidad electr\'onica debido a la naturaleza $np^2$ del N$^+$.

En ninguno de los espectros obtenidos en este trabajo, y para ninguna de las regiones observadas fue posible medir la l'inea auroral del [\NII] en 5755\AA, por lo cual se tuvo que recurrir a modelos o relaciones entre temperaturas para poder derivar un valor de esta temperatura.

Las expresiones anteriores son v'alidas en el rango de temperaturas electr'onicas entre 7000 y 23000 K y los errores involucrados en los ajustes son siempre menores que los errores observacionales por factores entre 5 y 10. Los errores en las temperaturas son los errores calculados a partir de las incertezas en las medidas de las intensidades de las l'ineas, aplicando la f'ormula de propagaci'on de errores, sin asignarle error alguno a la expresi'on del ajuste con el que se determina la temperatura electr'onica. 

\subsubsection*{Relaciones entre temperaturas}
\label{sec:Relaciones}

 En ocasiones es posible medir s'olo alguna de las l'ineas de emisi'on aurorales de los iones m'as representativos de la parte 'optica del espectro. En esos casos, es necesario suponer una estructura interna de ionizaci'on de la regi'on H\,{\sc ii} de estudio. Las primeras aproximaciones, consideraban a la regi'on H\,{\sc ii} como isoterma, pero algunos estudios posteriores demostraron la existencia, en su interior, de una estructura interna de ionizaci'on que apunta a una variaci'on espacial de la temperatura electr'onica, adem'as de la posible presencia de fluctuaciones de temperatura \citep{1967ApJ...150..825P}. Esto implica que, en la determinaci'on de cada abundancia i'onica, hay que tener en cuenta la temperatura media de la zona en la que el i'on se encuentra.
Una mejor aproximaci'on es considerar dos zonas: una de alta ionizaci'on, la m'as interna y cercana al c'umulo ionizante, cuya temperatura caracter'istica sea la de [\OIII], y otra de baja ionizaci'on, m'as externa, caracterizada por la temperatura de [\OII]. Con la temperatura de la primera zona, se calculan las abundancias de iones como O$^{2+}$, S$^{2+}$ o Ne$^{2+}$ y en la segunda las de O$^+$, S$^+$ o N$^+$ \citep{1992MNRAS.255..325P}. Los trabajos de Garnett~(\citeyear{1992AJ....103.1330G}) refinaron un poco esta visi'on al demostrar por un lado, que la temperatura de [\SIII] ocupa un valor intermedio entre ambas zonas y, por otro, que las temperaturas deducidas a partir de las l'ineas 'opticas de emisi'on sobreestiman el valor de la temperatura electr'onica para toda la nube, ya que hay un efecto de selecci'on hacia las zonas m'as calientes.

P\'erez-Montero \& D\'iaz~(\citeyear{2003MNRAS.346..105P,2005MNRAS.361.1063P}) examinaron las relaciones entre las temperaturas que se utilizan para la determinaci'on de la abundancia de cada i'on, adapt'andola a los resultados de los 'ultimos modelos de fotoionizaci'on y la utilizaci'on de los coeficientes at'omicos m'as recientes.
Siguiendo H\"agele et al.,~(\citeyear{2006MNRAS.372..293H,2008MNRAS.383..209H}), estas modificaciones han sido tenidas en cuenta en el presente trabajo, y fueron utilizadas en la medida que eran requeridas para derivar temperaturas. 
Cabe mencionar que en las relaciones entre temperaturas electr\'onicas no es posible obtener una estimaci\'on realista de las incertidumbres asociadas a los modelos. Los errores de los modelos no se pueden cuantificar y por ello el error que aparecer'a en las tablas ser'a una cota inferior dada por las l'ineas de emisi'on.

\begin{itemize}
\item {t$_e$[\OIII] vs. t$_e$[\OII]}
\end{itemize}
En la mayor'ia de las regiones, las l\'ineas aurorales de [\OII] no son observadas con suficiente relaci'on se'nal-ruido como para derivar la temperatura utilizando el m'etodo directo. Por el contrario, las l'ineas aurorales del [\OIII] son generalmente ``muy intensas"\, en este tipo de objetos de baja metalicidad y se las suele medir con una complicaci'on menor pudiendo as'i determinar la T$_e$([\OIII]) observacionalmente a trav'es del m'etodo directo. En estos casos, se suele utilizar alguna relaci\'on basada en modelos de fotoionizaci\'on para deducir T$_e$([\OII]) a partir de T$_e$([\OIII]).
Una relaci'on que se usa com'unmente en la bibliograf'ia y con bastante aceptaci\'on es la expresi\'on dada por Pagel et al.,~(\citeyear{1992MNRAS.255..325P}):
\begin{eqnarray}
t_e([O{\textsc{ii}}])\,=\,\frac{2}{t_e([O{\textsc{iii}}])^{-1} + 0.8}
\label{ecu11}
\end{eqnarray}  
basada en los modelos de \cite{1990A&AS...83..501S}. 
Sin embargo dicha relaci\'on obvia la dependencia de T$_e$([\OII]) con la densidad \citep{1999ApJ...527..110L}, hecho confirmado en parte por la gran dispersi\'on observada al representar los objetos cuyas dos temperaturas han podido ser medidas, aunque no se ha encontrado una clara tendencia observacional con la densidad N$_{e}$ \citep{2008PhDT........35H}. El ajuste obtenido por P\'erez-Montero \& D\'iaz~(\citeyear{2003MNRAS.346..105P}) es:
\begin{eqnarray}
t_e([O{\textsc{ii}}])\,=\,\frac{1.2 + 0.002 \cdot n_{e} + \frac{4.2}{n_{e}}}{t([O{\textsc{iii}}])^{-1} + 0.08 + 0.003 \cdot n_{e} + \frac{2.5}{n_{e}}}
\label{ecu12}
\end{eqnarray} 
y es la que se usa en este trabajo cuando la situaci'on lo requiera, y se tenga una estimaci'on de T$_e$[\OIII], y siempre teniendo en cuenta que en esta relaci'on T$_e$([\OII]) depende expl'icitamente de la densidad electr'onica.

\begin{itemize}
\item {t$_e$[\SIII] vs. t$_e$[\OIII]}
\end{itemize}
Existen trabajos que relacionan la temperatura de [\OIII] con la temperatura de [\SIII]. La motivaci\'on de dicha relaci\'on aparece con Garnett~(\citeyear{1992AJ....103.1330G}) quien sugiri'o que la T$_e$([\SIII]) ocupa un valor medio entre la de [\OIII] y la de [\OII], y calcul'o la abundancia de S$^{2+}$ s\'olo con la l\'inea auroral [\SIII]\,$\lambda$\,6312 \AA. 
La relaci\'on que da Garnett es:
\begin{eqnarray}
t_e([S{\textsc{iii}}])\,=\,0.83 \cdot t_e([O{\textsc{iii}}]) + 0.17
\label{ecu13}
\end{eqnarray} 
\noindent 
En los \'ultimos a\~nos, la importancia de esta relaci\'on ha tomado m'as relevancia, dado que en objetos de muy alta metalicidad se ha detectado con mayor facilidad la l\'inea auroral de [\SIII] que las l'ineas aurorales de otros iones y elementos. 
Sin embargo, la aparici\'on de coeficientes at\'omicos m'as nuevos para las transiciones del azufre \citep{1999ApJ...526..544T} hizo que la relaci\'on subestime la T$_e$([\SIII]). En su lugar, con los nuevos coeficientes de los modelos de fotoionizaci'on descritos en P\'erez-Montero \& D\'iaz~(\citeyear{2005MNRAS.361.1063P}) la relaci'on tom'o la forma:
\begin{eqnarray}
t_e([S{\textsc{iii}}])\,=\,1.05 \cdot t_e([O{\textsc{iii}}]) - 0.08
\label{ecu14}
\end{eqnarray}
\noindent
Dado que la relaci\'on entre temperaturas electr\'onicas no permite obtener una estimaci\'on realista de las incertidumbres asociadas a los modelos y con la aparici\'on de un cada vez mayor n\'umero de observaciones en el rango espectral donde se puede observar esta temperatura, ha sido posible obtener un nuevo ajuste emp'irico que tiene la forma:
\begin{eqnarray}
t_e([S{\textsc{iii}}])\,=\,(1.19 \pm 0.08) t_e([O{\textsc{iii}}]) - (0.32 \pm 0.10)
\label{ecu15}
\end{eqnarray} 
\noindent definida por H{\"a}gele et al.~(\citeyear{2006MNRAS.372..293H}), de ahora en m'as H06. Este 'ultima relaci'on emp'irica es la que se ha optado por aplicar en este trabajo de Tesis cuando la situaci'on lo requiera.

\begin{itemize}
\item {An'alisis cl'asico: t$_e$[\OII] vs. t$_e$[\SII] vs. t$_e$[\NII]}
\end{itemize}
En general, es com'unmente aceptado que los tres iones, O$^+$, S$^+$ y N$^+$, se forman en la zona de baja excitaci'on de una nebulosa y, por lo tanto, deber'ian tener temperaturas muy similares: T$_e$([\OII])\,$\approx$\,T$_e$([\SII])\,$\approx$\,T$_e$([\NII]). 
Dado que es en las zonas m'as externas de la nebulosa donde se producen las variaciones locales m'as significativas de la temperatura, ha sido necesario profundizar en las relaciones entre temperaturas para mejorar las determinaciones de las abundancias de estos tres iones, teniendo siempre en consideraci'on la dependencia de T$_e$([\OII]) con la densidad.

En aquellos casos en que las l\'ineas aurorales de [\SII] no est'an accesibles en el espectro, se suele suponer que T$_e$([\SII])\,$\approx$\,T$_e$([\OII]). Sin embargo, existen evidencias que apuntan a que esta T$_e$([\SII]) es levemente menor a la T$_e$([\OII]). 
De los modelos utilizados por P\'erez-Montero \& D\'iaz~(\citeyear{2003MNRAS.346..105P}), se obtiene un ajuste lineal que resulta ser:
\begin{eqnarray}
t_e([S{\textsc{ii}}])\,=\,0.71 \cdot t_e([O{\textsc{ii}}]) + 0.12
\label{ecu16}
\end{eqnarray} 
para una densidad electr'onica de 10 part\'iculas por cm$^3$. Para densidades electr'onicas de 100 part\'iculas por cm$^3$ el ajuste da:
\begin{eqnarray}
t_e([S{\textsc{ii}}])\,=\,0.86 \cdot t_e([O{\textsc{ii}}]) + 0.08
\label{ecu17}
\end{eqnarray}

Las observaciones confirman que en general la temperatura electr'onica de [\SII] es menor a la de [\OII] para la mayor parte de las galaxias H\,{\sc ii} , sin embargo, muchas de las galaxias H\,{\sc ii} presentan un valor de T$_e$([\SII]) menor que el predicho por los modelos. Esto podr'ia deberse a la contribuci'on del gas difuso ionizado por un posible escape de fotones o a una mayor contribuci'on del [\SII] en las zonas de baja excitaci'on \citep{2003MNRAS.346..105P}. A pesar de estas evidencias, no existe una certeza de dicha dependencia desde el punto de vista observacional (ver Figura 2.17 de H\"agele~(\citeyear{2008PhDT........35H}). Luego, la T$_e$([\SII]) se ha obtenido a partir de la primera relaci'on arriba mencionada.

Por otro lado, como se mencion'o m'as arriba, la l'inea [\NII]\,$\lambda$\,5755\,\AA\ lamentablemente suele tener muy poca se\~nal-ruido o es inobservable en nuestros espectros, por lo que es habitual considerar la aproximaci\'on T$_e$([\NII])\,$\approx$\,T$_e$([\OII]) como v\'alida. Esta relaci\'on, confirmada por los modelos, es sensible a la densidad y a la estructura de ionizaci\'on interna de la nebulosa, pudiendo alcanzar valores m\'as cercanos a T$_e$([\SII]), por lo que, en todo caso, debe ser tomada con su grado de incerteza. De los modelos de fotoionizaci'on propuestos por P\'erez-Montero \& D\'iaz~(\citeyear{2003MNRAS.346..105P}) resulta que la T$_e$([\SII]) es sensiblemente menor que T$_e$([\OII]) y que T$_e$([\NII]), lo cual puede deberse, seg'un los autores, a la contribuci'on de las zonas de baja excitaci'on. Este efecto puede no ser importante en objetos de alta excitaci'on, en que las fracciones de S$^+$ son menores, pero s'i puede ser relevante a la hora de determinar la abundancia total de azufre en objetos menos excitados (ver discusi'on en P'erez-Montero et al.,~\citeyear{2010MNRAS.404.2037P}).

Con las relaciones hasta aqu'i expuestas, se puede determinar las temperaturas faltantes debido a la no detecci'on de alguna de las l'ineas aurorales correspondientes. En los casos en que ninguna l'inea auroral sea detectable, y con s'olo la medida de las l'ineas m'as intensas, se puede recurrir al uso de los dos m'etodos emp'iricos abajo descriptos.

\begin{itemize}
\item {M'etodo emp'irico D07}
\end{itemize}
El m'etodo utilizado para derivar la temperatura de [\SIII] cuando no se tienen las l'ineas aurorales de este mismo elemento ni las de ning'un otro i'on o elemento, es el desarrollado por D'iaz et al.~(\citeyear{2007MNRAS.382..251D}) basado en el par'ametro emp'irico SO$_{23}$\footnote{{\tiny $SO_{23}\,=\,S_{23}/O_{23}$}} definido por D'iaz \& P'erez-Montero~(\citeyear{2000MNRAS.312..130D}) y recientemente reivindicado como un buen indicador de metalicidad para regiones H\,{\sc ii} de alta metalicidad. Dicho par'ametro es la combinaci'on de los par'ametros emp'iricos O$_{23}$\footnote{{\tiny $O_{23}\,=\,I(3727,29{\textsc \AA})+I(4959{\textsc \AA})+I(5007{\textsc \AA})/I(H\beta)$}} \citep{1979MNRAS.189...95P} y del S$_{23}$\footnote{{\tiny $S_{23}\,=\,I(6717{\textsc \AA})+I(6731{\textsc \AA})+I(9069{\textsc \AA})+I(9532{\textsc \AA})/I(H\beta)$}} \citep{1996MNRAS.280..720V} usados para la determinaci'on de las abundancias de ox'igeno, de los cuales se hablar'a en la Secci'on \ref{sec:Calibradores}. La relaci'on emp'irica entre la temperatura y el par'ametro emp'irico esta dada por el siguiente ajuste: 
\begin{eqnarray}
t_e([S{\textsc{iii}}])\,=\,0.596 - 0.283 log SO_{23} + 0.199(\log SO_{23})^{2}
\label{ecu18}
\end{eqnarray} 
\noindent
Esta relaci'on es v'alida para regiones de alta metalicidad. Pero de la Figura 9 de D'iaz et al.,~(\citeyear{2007MNRAS.382..251D}) se desprende que esta relaci'on es v'alida tambi'en para el extremo de la zona donde se encuentran las galaxias H\,{\sc ii}, donde la dispersi'on es baja, y que es el rango de valores del par'ametro emp'irico SO$_{23}$ donde se encuentra Haro\,15.

Para poder estimar as'i la T$_e$([\SIII]), es necesaria la detecci\'on y medida de las l\'{\i}neas intensas del ox'igeno y del azufre. Luego, haciendo uso de la relaci'on emp'irica de H06 se puede estimar la temperatura del [\OIII] y a partir de esta temperatura la de [\OII] a trav'es de los modelos, y posteriormente la de [\SII] y la de [\NII] igual'andolas a T$_e$([\OII]).
El m'etodo para derivar las abundancias con esta temperatura y usando las l'ineas intensas se denomina m'etodo semi-emp'irico. 

\begin{itemize}
\item {M'etodo emp'irico P07}
\end{itemize}
En ocasiones, las l'ineas intensas del [\SIII] no son observadas con buena calidad, o est\'an demasiado hacia el rojo para el rango espectral obtenido. Pilyugin~(\citeyear{2007MNRAS.375..685P}) propuso una calibraci'on para la relaci'on de intensidades entre las l'ineas intensas y la auroral del [\NII] en t'erminos de las l'ineas del ox'igeno, [\OII]\,$\lambda\lambda$\,3727, 3729\,\AA\ y [\OIII]\,$\lambda\lambda$\,4959, 5007\,\AA, y de la l'inea de recombinaci'on H$\beta$. Entonces, la temperatura electr'onica del [\NII], que aparentemente caracteriza a la zona de baja ionizaci'on de la nebulosa, puede ser derivada (ver Pilyugin,~\citeyear{2007MNRAS.375..685P}). 
\begin{eqnarray}
t_2\,=\,\frac{1.111}{\log Q_{N{\textsc{ii}}} - 0.892 - 0.144 \cdot \log t_2 + 0.023 \cdot t_2}
\label{}
\end{eqnarray} 
 \noindent
donde t$_{2}$ es la temperatura representativa de la zona de baja excitaci'on y Q$_{N{\textsc{ii}}}$ es la aproximaci'on al cociente de l'ineas de [\NII] y sale de la relaci'on emp'irica
\[
 \log Q_{N{\textsc{ii}}}\,=\,2.619 - 0.609 \cdot \log R_2 - 0.010 \cdot [\log R_2]^2 + 1.085 \cdot \log(1 - P) + 0.382 \cdot [\log(1- P)]^2
\] 
 \noindent
donde P es el par'ametro de excitaci'on que se expresa como $$P\,=\,R_3/(R_3 + R_2)$$
con R$_2$ y R$_3$ definidos como $$R_2\,=\,I[O{\textsc{ii}}]\,\lambda\lambda\,3727,3729 / I(H\beta) \hspace{0.5cm}  R_3\,=\,I[O{\textsc{iii}}]\,\lambda\lambda\,4959,5007 / (H\beta)$$ 

Siguiendo el m'etodo descripto por Pilyugin, D'iaz et al.,~(\citeyear{2007MNRAS.382..251D}) encuentran para las regiones circumnucleares de formaci'on estelar (CNSFR) una correlaci'on entre las temperaturas del [\SIII] y del [\NII], las cuales son muy similares, siendo en promedio T$_e$([\NII]) 500K mayor que T$_e$([\SIII]) (ver Figura 15 de D'iaz et al.,~\citeyear{2007MNRAS.382..251D}). 
Una vez estimada la temperatura del nitr'ogeno, es posible aproximar un valor a la temperatura del [\SIII] con la aproximaci'on T$_e$([\NII])\,$\approx$\,500K + T$_e$([\SIII]). Y as'i, una vez m'as haciendo uso de la relaci'on emp'irica de H06, se puede estimar la temperatura del [\OIII]. Luego, con esta temperatura y los modelos se determina T$_e$([\OII]) y posteriormente la de T$_e$([\SII]) suponiendo que es igual a la temperatura electr'onica de [\OII].

\begin{itemize}
\item {T=10$^4$K}
\end{itemize}

Cuando no s'olo no se pueden medir las l'ineas aurorales, sino que adem'as no se cuenta con algunas de las l'ineas intensas para el c'alculo de temperaturas electr'onicas a trav'es de los m'etodos emp'iricos que se acaban de describir, y considerando que para esta clase de objetos la temperatura suele estar alrededor de 10000 K, se tomar'a este valor para la temperatura electr'onica del [\OIII], y con esta temperatura y los modelos se determinar'a T$_e$([\OII]), y posteriormente T$_e$([\SII]) y T$_e$([\NII]), suponiendo ambas igual a la temperatura electr'onica de [\OII], y de T$_e$([\SIII]) se determinar'a a trav'es de su relaci'on con T$_e$([\OIII]) derivada por H06.

\subsection{Derivaci\'on de las abundancias qu\'{\i}micas}
\label{sec:Abundancias}
Para estudiar las abundancias qu'imicas globales y discriminadas por componentes cinem'aticas en cada brote de formaci'on estelar de Haro\,15, se han derivado las abundancias de las diferentes especies i'onicas usando las l'ineas de emisi'on m'as intensas disponibles y detectadas en los espectros analizados. Para ello se utiliz'o la tarea \texttt{ionic} del paquete STSDAS de {\sc IRAF}, tal como se describe en \cite{2008MNRAS.383..209H}.
Este paquete tambi'en se basa en la aproximaci'on de un 'atomo de cinco niveles en equilibrio estad'istico \citep{1987JRASC..81..195D,1995PASP..107..896S}.
%
 %
 Uno de los problemas para la determinaci'on de abundancias es no poder medir las l'ineas de emisi'on de todos los estados de ionizaci'on de los elementos cuya abundancia se quiere calcular. En general, una manera de solucionar este problema es recurrir a la cantidad llamada factor de correcci'on de ionizaci'on (ICF sigla en ingl\'es de ionization correction factors) que da cuenta de la contribuci'on a la abundancia de las especies  i'onicas que no han sido observadas.
En este Cap'itulo las abundancias totales han sido obtenidas teniendo en cuenta, cuando fuese necesario, los estados de ionizaci'on inobservables de cada elemento, recurriendo a los factores de correcci'on de ionizaci'on ampliamente aceptados para cada especie, X/H\,=\,ICF(X$^{+i}$)\,X$^{+i}$/H${^+}$ \citep{2007MNRAS.381..125P,2008MNRAS.383..209H}. 

A continuaci'on se describen las expresiones de los ajustes usados para cada i\'on, una vez que se dispone de las densidades y temperaturas electr\'onicas derivadas adecuadamente para cada componente y medida global en cada uno de los brotes estudiados. 

\subsubsection{Helio}
\label{sec:Helio}
Las l'ineas de helio, al igual que las de hidr\'ogeno son l'ineas de recombinaci\'on. Son bastante intensas y numerosas, aunque muchas de ellas suelen estar fusionadas con otras l'ineas. En general, est'an afectadas por la absorci\'on de la poblaci\'on estelar subyacente, por fluorescencia, y tienen una contribuci\'on por excitaci'on colisional \citep[para un completo tratamiento de estas contribuciones ver][]{2001NewA....6..119O,2004ApJ...617...29O}. Generalmente, para estimar las abundancias de helio (una y dos veces ionizado) se usan las l'ineas de He{\sc  i}\,$\lambda\lambda$\,4471, 5876, 6678 y 7065\AA, y He{\sc  ii}\,$\lambda$\,4686\,\AA, respectivamente.
En este trabajo se ha tomado la temperatura electr'onica T$_e$([\OIII]) como representativa de la zona en que estas l'ineas son emitidas. Usando las ecuaciones dadas por Olive \& Skillman se puede derivar el valor de He$^{+}$/H$^{+}$, usando la emisividad te'orica escalada a H$\beta$ de Benjamin~(\citeyear{1999ApJ...514..307B}) y las expresiones para los factores de correcci'on colisional de Kingdon \& Ferland~(\citeyear{1995ApJ...442..714K}). Para calcular la abundancia de helio dos veces ionizado, He$^{2+}$/H$^+$, se ha utilizado la ecuaci'on (9) de Kunth \& Sargent~(\citeyear{1983ApJ...273...81K}). Dado que los objetos observados tienen densidades electr'onicas bajas, la leve dependencia con la profundidad 'optica que tienen tres de las l'ineas medidas del He ser'a ignorada y por eso no se ha hecho ninguna correcci'on por fluorescencia. Tampoco se ha hecho ninguna correcci'on por la posible presencia de alguna poblaci'on estelar subyacente dado que las l'ineas de absorci'on de este elemento son m'as angostas y no se pueden ver sus alas a los lados de la emisi'on.

Con todo esto, es v\'alida la aproximaci'on para la abundancia total del helio:
\begin{eqnarray}
\label{ecu19}
\frac{He}{H}\,=\,\frac{He^++He^{2+}}{H^+}
\end{eqnarray} 

Seg'un sea el caso, los resultados obtenidos para cada l'inea del helio junto a sus correspondientes errores son presentados en las Tablas \ref{abundancesBC} y \ref{abundancesABCEF}. Se adjunta el valor ``adoptado'' para He$^+$ / H$^+$ que es el promedio ponderado por los errores de las diferentes abundancias i'onicas derivadas para cada l'inea de emisi'on de He\,I. Este valor es conocido como ``adopted"\,(del ingl'es, valor considerado o adoptado). De aqu'i en adelante se utilizar'a el t'ermino valor ``adoptado'' teniendo en cuenta esta definici'on.

\subsubsection{Abundancias qu'imicas i'onicas y totales determinadas con las l'ineas prohibidas}
\label{sec:ionicas_totales}

Se han calculado las abundancias i'onicas y totales de O, S, N, Ne, Ar y Fe, cuando las l'ineas est'an presentes, utilizando las expresiones dadas por H\"agele et al.,~(\citeyear{2008MNRAS.383..209H}) y que son incluidas en la Tabla \ref{ecuaciones}. 
Muchas veces suele ser 'util expresar el cociente de algunos elementos respecto del ox'igeno, por ejemplo, N/O, S/O, Ne/O, Ar/O, en unidades logar'itmicas. Esto tiene la ventaja que como las abundancias i'onicas se han determinado a partir de dos l'ineas de excitaci'on colisional (por ejemplo, N$^{+}$/O$^{+}$), estos cocientes no dependen tanto de la temperatura electr'onica.
  \begin{table*}
  \begin{minipage}{175mm}
  \vspace{-0.3cm}
  \normalsize
  \caption[Ecuaciones para el c'alculo de las abundancias de las l'ineas prohibidas.]{Ecuaciones para el c'alculo de las abundancias de las l'ineas prohibidas      \citep{2008MNRAS.383..209H}.}
  \label{ecuaciones}
  \begin{center}
  \begin{tabular}{ll}
  \hline
  I'on$^*$ & Ecuaci'on \\
  \hline
  O$^+$     & $\log\left(\frac{I(3727)}{I(H\beta)}\right)+5.992+\frac{1.583}{t_{e}}-0.681 \cdot \log t_{e} +\log (1+2.3 \cdot n_{e} )$ \\
  O$^{2+}$  & $\log\left(\frac{I(4959)+I(5007)}{I(H\beta)}\right)+6.144+\frac{1.251}{t_{e}}-0.55 \cdot \log t_{e} $\\
  S$^+$     & $\log\left(\frac{I(6717)+I(6731)}{I(H\beta)}\right)+5.423+\frac{0.929}{t_{e}}-0.28 \cdot \log t_{e} +1.0 \cdot n_{e})$ \\
  S$^{2+}$  & $\log\left(\frac{I(9069)+I(9532)}{I(H\beta)}\right)+5.8+\frac{0.771}{t_{e}}-0.22 \cdot \log t_{e} $\\
  N$^+$     & $\log\left(\frac{I(6548)+I(6584)}{I(H\beta)}\right)+6.273+\frac{0.894}{t_{e}}-0.592 \cdot \log t_{e}$ \\
  Ne$^{2+}$  & $\log\left(\frac{I(3868)}{I(H\beta)}\right)+6.486+\frac{1.558}{t_{e}}-0.504 \cdot \log t_{e} $\\
  Ar$^{2+}$  & $\log\left(\frac{I(7137)}{I(H\beta)}\right)+6.157+\frac{0.808}{t_{e}}-0.508 \cdot \log t_{e} $\\
  Ar$^{3+}$  & $\log\left(\frac{I(4740)}{I(H\beta)}\right)+5.705+\frac{1.246}{t_{e}}-0.156 \cdot \log t_{e} $\\
  Fe$^{2+}$  & $\log\left(\frac{I(4658)}{I(H\beta)}\right)+3.504+\frac{1.298}{t_{e}}-0.483 \cdot \log t_{e} $\\
   \hline
   \multicolumn{2}{l} {$^*$ Como 12+log(X/H$^+$)}\\
 \multicolumn{2}{l} {t$_{e}$ es la temperatura electr'onica de la l'inea, en unidades de 10$^4$ K}\\
  \multicolumn{2}{l} {n$_{e}$ es la densidad electr'onica, en unidades de 10$^4$ cm$^{-3}$}\\
  \end{tabular}
  \end{center}
  \end{minipage}
  \end{table*}
 En los casos en los que no existen mediciones, ni de la l'inea de [\SII]\,$\lambda$\,4068\,\AA\ ni de la l'inea de [\NII]\,$\lambda$\,5755\,\AA\ (esta l'inea no se ha podido medir en ninguno de los brotes), y bajo la suposici'on de una temperatura electr'onica homog'enea en la zona de baja excitaci'on, se ha considerado la aproximaci'on T$_e$([\NII])\,$\approx$\,T$_e$([\SII])\,$\approx$\,T$_e$([\OII])\, explicada anteriormente en el an'alisis cl'asico.

\begin{description}
\item[(i)] Los cocientes de abundancia i'onica del ox'igeno, O$^{+}$/H$^{+}$ y O$^{2+}$/H$^{+}$, se pueden determinar a partir de las intensidades de las l'ineas [\OII]\,$\lambda$$\lambda$\,3727,3729\,\AA\ y [\OIII]\,$\lambda$$\lambda$\,4959,5007\,\AA, respectivamente, usando para cada i'on su correspondiente temperatura electr'onica.
Sin considerar la contribuci'on de ox'igeno neutro que, en proporci'on, es igual a la fracci'on de hidr'ogeno neutro, gracias a la reacci'on de intercambio de carga \citep{O89}:
\[{O^+}+{H^0} \rightarrow {O^0}+{H^+}\] 
Dado que a la temperatura a la que se encuentran los objetos aqu'i observados, la mayor parte del ox'igeno se encuentra en forma de O$^{+}$ y O$^{2+}$, la abundancia total de ox'igeno es bien representada por la expresi\'on:
\begin{eqnarray}
\label{ecu20}
\frac{O}{H} = \frac{O^++O^{2+}}{H^+}
\end{eqnarray} 

\item[(ii)] Las abundancias i'onicas del azufre, S$^+$/H$^{+}$ y S$^{2+}$/H$^{+}$, est'an bien representadas por las l'ineas [\SII]\,$\lambda\lambda$\,6717, 6731\,\AA\ y las l'ineas [\SIII]\,$\lambda\lambda$\,9069, 9532\AA, respectivamente.
Dependiendo del grado de excitaci'on de la regi'on, se podr'ia esperar una contribuci'on relativamente importante de S$^{3+}$. Para esos estados de ionizaci'on inobservables, la abundancia total de azufre se puede calcular usando un ICF para S$^+$+S$^{2+}$. Una buena aproximaci\'on es la f\'ormula de Barker~(\citeyear{1980ApJ...240...99B}), basada en los modelos de fotoionizaci'on de Stasinska~(\citeyear{1978A&A....66..257S}) y toma la forma:
\begin{eqnarray}
\label{ecu21}
ICF(S^{+} + S^{2+}) = \left[ 1-\left( 1-\frac{O^{2+}}{O^{+}+O^{2+}}\right)^\alpha\right]^{-1/\alpha} 
\end{eqnarray} 
con $\alpha$\,$\approx$\,2.5, medida para una muestra de objetos con la l'inea de [\SIV] a 10.5 $\mu$ por P\'erez-Montero et al.,~(\citeyear{2006A&A...449..193P}).
Utilizando este ICF como una funci'on del cociente O$^{2+}$/O en lugar del cociente O$^+$/O se reduce el error propagado por esta cantidad, ya que los errores asociados a O$^{2+}$ son considerablemente menores que O$^+$ \citep{2008MNRAS.383..209H}.
\item[(iii)] La abundancia i'onica del nitr'ogeno, N$^{+}$/H$^{+}$, se deriva a partir de la intensidad de las l'ineas [\NII]\,$\lambda\lambda$\,6548, 6584\,\AA. En caso de no disponer de alguna de ellas por su proximidad a la l'inea H$\alpha$, se recurre a la relaci\'on te\'orica entre ambas l'ineas del [\NII] antes citada. 
Para el c'alculo de la abundancia total de nitr\'ogeno se parte de la relaci'on:
\begin{eqnarray}
\label{ecu22}
\frac{N}{O} = \frac{N^+}{O^+}
\end{eqnarray} 
luego la abundancia total del nitr'ogeno, N/H, es calculada como:
\begin{eqnarray}
\label{ecu23}
log(N/H)\,=\,log(N/O)+log(O/H)
\end{eqnarray}

\item[(iv)] La abundancia i'onica del ne'on, Ne$^{2+}$, se deriva usando la l'inea [\NeIII]\,$\lambda$\,3868\,\AA. Dado que su estructura de ionizaci\'on es bastante similar a la del ox'igeno, cl'asicamente la abundancia total del ne'on ha sido calculada asumiendo:
\begin{eqnarray}
\label{ecu24}
\frac{Ne}{O} = \frac{Ne^{2+}}{O^{2+}}
\end{eqnarray}

La temperatura que se considera para este i'on es la representativa de la zona de alta ionizaci'on T$_e$([\NeIII])\,$\approx$\,T$_e$([\OIII]) \citep{1969BOTT....5....3P}.
Izotov et al.~(\citeyear{2004A&A...415...87I}) indicaron que el hacer esta suposici'on puede llevar a una sobreestimaci'on de Ne/H en objetos con baja excitaci'on, donde la transferencia de carga entre O$^{2+}$ y H$^0$ se vuelve importante. Por este motivo aqu'i se calcula con el ICF dado por P\'erez-Montero et al.~(\citeyear{2007MNRAS.381..125P}). Dada la alta excitaci'on de los objetos observados, no hay diferencias significativas entre la abundancia total derivada para el ne'on usando este ICF y la estimada con el modelo cl'asico \citep{2008MNRAS.383..209H, 2011arXiv1101.4140H}.
\item[(v)]La abundancia i'onica del arg'on, Ar$^{2+}$, se calcula usando la l'inea [\ArIII]\,$\lambda$\,7136\,\AA\ asumiendo que T$_e$([\ArIII])\,$\approx$\,T$_e$([\SIII]) \citep{1992AJ....103.1330G}. 
Tambi\'en es posible medir, en algunos espectros, las l'ineas del [\ArIV]$\lambda$ 4713 y 4740 \AA. En ese caso la abundancia i'onica del Ar$^{3+}$, se calcula asumiendo que T$_e$([\ArIV])\,$\approx$\,T$_e$([\OIII]). No obstante, la primera de las l'ineas del [\ArIV] suele aparecer fusionada con otra l'inea de He\,I$\lambda$4711\,\AA, dif'icil de corregir, por lo que es mejor usar la segunda, m\'as intensa, para calcular la abundancia de Ar$^{3+}$. 
La abundancia total del arg'on es calculada usando el ICF(Ar$^{2+}$) y el ICF(Ar$^{2+}$ + Ar$^{3+}$) derivados a partir de modelos de fotoionizaci'on por P'erez-Montero et al.~(\citeyear{2007MNRAS.381..125P}). 


\item[(vi)]La abundancia i'onica del hierro se calcula usando la l'inea [\FeIII]\,$\lambda$\,4658\,\AA\ y la temperatura asumida es la del [\OIII]. 
La abundancia total es posible calcularla usando el ICF(Fe$^{2+}$) de Rodriguez \& Rubin~(\citeyear{2004IAUS..217..188R}). En este trabajo 'unicamente fue posible medir la l'inea [\FeIII]\,$\lambda$\,4658\,\AA\ en el espectro de ranura larga del brote B. 
 \end{description}
 
Las abundancias i'onicas de los elementos m'as pesados que el helio, los factores de correcci'on de ionizaci'on ICFs, las abundancias totales con sus  correspondientes errores son mostrados en la Tabla \ref{abundancesBC} para los datos de ranura larga, y \ref{abundancesABCEF} para los datos 'echelle.

\begin{table*}[p]
{\scriptsize 
\caption[Propiedades f'isicas, abundancias qu'imicas i'onicas y totales, brotes B y C de dispersi'on simple]{Propiedades f'isicas, abundancias qu'imicas i'onicas y totales derivadas de las l'ineas prohibidas y de la l'inea de recombinaci'on del helio para la medida global en los datos de ranura larga: {\bf brotes B y C}}
\label{abundancesBC}
\begin{center}
\begin{tabular}{lcc}
\hline
 &B&C\\
\hline
  n([S{\sc ii}]) &100: & 100:         \\
  T$_e$([O{\sc iii}]) &        1.26$\pm$0.03    & 1.01$\pm$0.02${^h}$    \\
  T$_e$([S{\sc iii}]) & 1.19$\pm$0.12 & 0.88$\pm$0.13${^d}$${^p}$\\
  T$_e$([O{\sc ii}]) & 1.20$\pm$0.04 & 1.21$\pm$0.01${^m}$\\
  T$_e$([S{\sc ii}]) &  0.79$\pm$0.11  &1.21$\pm$0.01${^m}$\\ 
  T$_e$([N{\sc ii}]) & 1.20$\pm$0.04${^m}$  &0.93$\pm$0.01${^p}$\\ 
  \hline
12+$\log(O^+/H^+)$        &  7.34$\pm$0.05 &  7.82$\pm$0.06${^m}$
  \\
 12+$\log(O^{2+}/H^+)$     &  7.95$\pm$0.03 &  7.84$\pm$0.03${^h}$ 
  \\
 \bf{12+log(O/H)}          &  8.04$\pm$0.03 &  8.13$\pm$0.04${^*}$
 \\[2pt]
  
\hline

 12+$\log(S^+/H^+)$        &  5.95$\pm$0.20 &  6.16$\pm$0.07${^m}$
  \\
 12+$\log(S^{2+}/H^+)$     &  6.17$\pm$0.23 &  \ldots
  \\
 ICF($S^++S^{2+}$)         &  1.41$\pm$0.03 &   \ldots
  \\
 \bf{12+log(S/H)}          &  6.52$\pm$0.22 &   \ldots
  \\
 \bf{log(S/O)}           & -1.52$\pm$0.22 &  \ldots
  \\[2pt]
  
\hline

 12+$\log(N^+/H^+)$       &  6.18$\pm$0.04${^m}$ &  6.70$\pm$0.07${^p}$
  \\
   \bf{12+log(N/H)}             & 6.88$\pm$0.20${^m}$ & 7.02$\pm$0.23${^p}$
  \\
 \bf{log(N/O)}             & -1.16$\pm$0.07${^m}$ & -1.11$\pm$0.09${^p}$
 \\[2pt]
  
\hline

 12+$\log(Ne^{2+}/H^+)$    &  7.33$\pm$0.03 &  \ldots
  \\
 ICF($Ne^{2+}$)          &  1.08$\pm$0.01 &  \ldots
  \\
 \bf{12+log(Ne/H)}         &  7.36$\pm$0.03 &  \ldots
  \\
 \bf{log(Ne/O)}           & -0.68$\pm$0.04 &  \ldots
  \\[2pt]
  
\hline

 12+$\log(Ar^{2+}/H^+)$    &  5.65$\pm$0.10 &  \ldots
  \\
 12+$\log(Ar^{3+}/H^+)$    &  4.62$\pm$0.08 &  \ldots
  \\
 ICF($Ar^{2+}$+$Ar^{3+}$)  &  1.03$\pm$0.01 &  \ldots
  \\
 \bf{12+log(Ar/H)}        &  5.70$\pm$0.10 &  \ldots
 \\
 \bf{log(Ar/O)}            & -2.34$\pm$0.11 &\ldots
 \\ [2pt]
  
\hline

 12+$\log(Fe^{2+}/H^+)$    &  5.04$\pm$0.10 &  \ldots
  \\
 ICF($Fe^{2+}$)            &  5.21$\pm$0.62 &  \ldots
  \\
 \bf{12+log(Fe/H)}         &  5.76$\pm$0.11 &  \ldots
  \\[2pt]
  
\hline

 $He^+/H^+$($\lambda$\,4471)    & 0.064$\pm$0.003 & \ldots
  \\
  $He^+/H^+$($\lambda$\,5876)    & 0.096$\pm$0.002 & \ldots
  \\
   $He^+/H^+$($\lambda$\,6678)    & 0.072$\pm$0.003 & \ldots
  \\
 $He^+/H^+$($\lambda$\,7065)    & 0.100$\pm$0.008 & \ldots
  \\
  $He^+/H^+$adopted    & 0.086$\pm$0.015 & \ldots
  \\
$He^{2+}/H^+$($\lambda$\,4686)    & 0.0013$\pm$0.0002 & \ldots
  \\
\bf{(He/H)}                & 0.087$\pm$0.015 & \ldots
  \\
  \hline
\multicolumn{3}{@{\hspace{0.1cm}}l}{densidades en $cm^{-3}$ y temperaturas en 10$^4$\,K}\\
\end{tabular}
\end{center}}
\vspace{0.1cm}
{\scriptsize \texttt{${^m}$} resultados usando las temperaturas predichas por los modelos, como se describe en el texto. \texttt{${^d}$} resultados usando las temperaturas derivadas con el m'etodo emp'irico D07. \texttt{${^h}$} resultados usando las temperaturas estimadas con el m'etodo emp'irico encontrado por H\"agele et al. (2006). \texttt{${^p}$} resultados usando las temperaturas derivadas con el m'etodo emp'irico P07. ${^*}$ resultados usando los datos necesarios para su c'alculo que fueron derivados de modelos o relaciones emp'iricas}.\\
\end{table*}

\begin{sidewaystable}
\begin{minipage}[t]{\columnwidth}
 {\tiny
\caption[Propiedades f'isicas, abundancias qu'imicas i'onicas y totales, brotes A, B, C, E y F de 'echelle]{{\scriptsize Propiedades f'isicas, abundancias qu'imicas i'onicas y totales derivadas de las l'ineas prohibidas y de la l'inea de recombinaci'on del helio para la medida global y para las diferentes componentes cinem'aticas en los datos 'echelle: {\bf brotes A, B, C, E y F}.}}
\label{abundancesABCEF}
\begin{center}
\begin{tabular}{l@{} | @{}c@{\hspace{0.1cm}}c@{\hspace{0.1cm}}c@{\hspace{0.1cm}}c@{} | @{}c@{\hspace{0.1cm}}c@{\hspace{0.1cm}}c@{} |@{}c@{\hspace{0.1cm}}c@{\hspace{0.1cm}}c@{} | @{}c@{\hspace{0.1cm}}c@{\hspace{0.1cm}}c@{} | @{}c}
\hline
&\multicolumn{4}{@{\hspace{0.1cm}}c}{A}&\multicolumn{3}{c}{B}&\multicolumn{3}{c}{C}&\multicolumn{3}{c}{E}&\multicolumn{1}{c}{F}\\
\hline
 & global&  angosta\,1 & angosta\,2  & ancha&global &angosta &ancha& global &angosta &ancha &global &angosta\,1 &angosta\,2\ &global \\
 \hline
 n([S{\sc ii}])  &300: & 100:   & 200: & 100:    &150$\pm$50 & 100:   & 150:  &400: & 400:   & 1400: &500: & 900:   & 400:    &100:         \\     
 T([O{\sc iii}]) &1.00$\pm$0.02${^h}$    & 1.08$\pm$0.03${^h}$&     1.16$\pm$0.04${^h}$  & 0.95$\pm$0.05${^h}$  &1.20$\pm$0.06 &1.22$\pm$0.18 & 1.12$\pm$0.04& 1.0$\dagger$ &1.0$\dagger$&1.0$\dagger$& 0.97$\pm$0.03${^h}$  & 1.0$\dagger$  &   1.0$\dagger$&1.0$\dagger$\\
 T([S{\sc iii}])  & 0.87$\pm$0.05${^d}$    &0.96$\pm$0.06${^d}$&    1.06$\pm$0.10${^d}$    & 0.8$\pm$0.12${^d}$ &1.31$\pm$0.17 & 1.71$\pm$0.30&1.16$\pm$0.16 & 0.87$\pm$0.13${^m}$ &0.87$\pm$0.13${^m}$ & 0.87$\pm$0.13${^m}$ & 0.83$\pm$0.13${^d}$${^p}$& 0.87$\pm$0.13${^m}$   & 0.87$\pm$0.13${^m}$&0.87$\pm$0.13${^m}$\\
 T([O{\sc ii}])${^m}$  &0.93$\pm$0.01 & 1.28$\pm$0.01& 1.06$\pm$0.01& 1.08$\pm$0.03& 1.11$\pm$0.02 &1.28$\pm$0.10&      1.07$\pm$0.01  &  0.87$\pm$0.01 &0.87$\pm$0.01 & 0.76$\pm$0.01  &   0.84$\pm$0.01       & 0.80$\pm$0.01  &  0.88$\pm$0.01&1.21$\pm$0.01\\
  T([S{\sc ii}])${^m}$ &0.93$\pm$0.01 & 1.28$\pm$0.01& 1.06$\pm$0.01&1.08$\pm$0.03&   1.11$\pm$0.02  &1.28$\pm$0.10&       1.07$\pm$0.01& 0.87$\pm$0.01 & 0.87$\pm$0.01 & 0.76$\pm$0.01   &  0.84$\pm$0.01       & 0.80$\pm$0.01  &  0.88$\pm$0.01 &1.21$\pm$0.01\\
 T([N{\sc ii}])  & 0.93$\pm$0.01${^m}$& 1.28$\pm$0.01${^m}$ &  1.06$\pm$0.01${^m}$&  1.08$\pm$0.03${^m}$& 1.11$\pm$0.02${^m}$  &  1.28$\pm$0.10${^m}$&    1.07$\pm$0.01${^m}$& 0.87$\pm$0.01${^m}$  & 0.87$\pm$0.01${^m}$  & 0.76$\pm$0.01${^m}$     & 0.88$\pm$0.01${^p}$   & 0.80$\pm$0.01${^m}$   & 0.88$\pm$0.01${^m}$&1.21$\pm$0.01${^m}$\\ [2pt]
 \hline
 
 12+$\log(O^+/H^+)$${^m}$          &  7.92$\pm$0.06 &  7.87$\pm$0.05 &  7.87$\pm$0.10 &  8.00$\pm$0.15 &  7.21$\pm$0.08 &  7.61$\pm$0.21 &  7.09$\pm$0.08 &  \ldots &  \ldots &  \ldots &  8.01$\pm$0.19 &  \ldots &  \ldots &  \ldots \\
  12+$\log(O^{2+}/H^+)$     &  7.80$\pm$0.03${^h}$ &  7.55$\pm$0.04${^h}$ &  7.75$\pm$0.05${^h}$ &  8.01$\pm$0.11${^h}$ &  8.11$\pm$0.07 &  7.84$\pm$0.19 &  8.26$\pm$0.05 &  7.83$\pm$0.03${^m}$ &  7.99$\pm$0.02${^m}$ &  7.71$\pm$0.05${^m}$ &  7.70$\pm$0.08${^h}$ &  7.53$\pm$0.04$\dagger$ &  7.84$\pm$0.02$\dagger$ &  7.86$\pm$0.04$\dagger$
  \\
 \bf{12+log(O/H)}          &  8.17$\pm$0.05${^*}$ &  8.04$\pm$0.05${^*}$ &  8.11$\pm$0.08${^*}$ &  8.31$\pm$0.13${^*}$ &  8.16$\pm$0.07${^*}$ &  8.04$\pm$0.20${^*}$ &  8.29$\pm$0.05${^*}$ &  \ldots &  \ldots &  \ldots &  8.18$\pm$0.16${^*}$ &  \ldots &  \ldots &  \ldots
  \\[2pt]
  
\hline

 12+$\log(S^+/H^+)$${^m}$        &  6.14$\pm$0.04 &  5.90$\pm$0.04 &  5.71$\pm$0.07 &  6.44$\pm$0.09 &  5.30$\pm$0.05 &  5.68$\pm$0.12 &  5.02$\pm$0.08 &  6.20$\pm$0.17 &  6.24$\pm$0.16 &  6.37$\pm$0.28 &  6.36$\pm$0.17 &  5.99$\pm$0.20 &  6.56$\pm$0.16 &  6.48$\pm$0.20  \\
 12+$\log(S^{2+}/H^+)$     &  6.42$\pm$0.18${^d}$ &  6.11$\pm$0.18${^d}$ &  6.20$\pm$0.16${^d}$ &  6.54$\pm$0.28${^d}$ &  5.93$\pm$0.12 &  5.79$\pm$0.13 &  6.03$\pm$0.13 &  \ldots &  \ldots &  \ldots &  \ldots &  \ldots &  \ldots &  \ldots
  \\
 ICF($S^++S^{2+}$)         &  1.05$\pm$0.01${^*}$ &  1.02$\pm$0.01${^*}$&  1.05$\pm$0.01${^*}$ &  1.08$\pm$0.02${^*}$ &  1.72$\pm$0.02${^*}$ &  1.16$\pm$0.01${^*}$ &  2.13$\pm$0.07${^*}$ & \ldots & \ldots & \ldots &  \ldots & \ldots & \ldots & \ldots
  \\
 \bf{12+log(S/H)}            &  6.62$\pm$0.13${^*}$ &  6.33$\pm$0.13${^*}$ &  6.34$\pm$0.14${^*}$ &  6.83$\pm$0.21${^*}$ &  6.26$\pm$0.11${^*}$ &  6.11$\pm$0.13${^*}$ &  6.40$\pm$0.12 ${^*}$ & \ldots &  \ldots &  \ldots &  \ldots &  \ldots &  \ldots &  \ldots
  \\
 \bf{log(S/O)}                  & -1.54$\pm$0.14${^*}$ & -1.71$\pm$0.14${^*}$ & -1.77$\pm$0.16${^*}$ & -1.48$\pm$0.25${^*}$ & -1.90$\pm$0.13${^*}$ & -1.93$\pm$0.23${^*}$ & -1.88$\pm$0.13${^*}$ &  \ldots &  \ldots &  \ldots & \ldots &  \ldots &  \ldots &  \ldots
  \\[2pt]
  
\hline

 12+$\log(N^+/H^+)$        &  6.97$\pm$0.03${^m}$&  6.89$\pm$0.03${^m}$ &  6.78$\pm$0.05${^m}$ &  7.06$\pm$0.10${^m}$ &  5.87$\pm$0.05${^m}$ &  6.25$\pm$0.13${^m}$ &  5.75$\pm$0.06${^m}$ &  6.57$\pm$0.15${^m}$ &  6.78$\pm$0.12${^m}$ &  6.35$\pm$0.19${^m}$ &  6.92$\pm$0.12${^p}$ &  6.65$\pm$0.11${^m}$&  7.15$\pm$0.12${^m}$ &  6.73$\pm$0.27${^m}$
  \\
   \bf{12+log(N/H)}             & 7.21$\pm$0.23${^m}$ &  7.06$\pm$0.22${^m}$ & 7.06$\pm$0.22${^m}$ & 7.02$\pm$0.39${^m}$ & 6.82$\pm$0.32${^m}$  &  6.68$\pm$0.79${^m}$ & 6.95$\pm$0.28${^m}$ &  \ldots &  \ldots &  \ldots & 7.09$\pm$0.70${^p}$ &  \ldots &  \ldots &  \ldots 
   \\
 \bf{log(N/O)}             & -0.96$\pm$0.06${^m}$ & -0.98$\pm$0.06${^m}$ & -1.09$\pm$0.11${^m}$ & -0.94$\pm$0.18${^m}$ & -1.34$\pm$0.09${^m}$ & -1.36$\pm$0.25${^m}$ & -1.34$\pm$0.10${^m}$ & \ldots &  \ldots &  \ldots & -1.09$\pm$0.23${^p}$ & \ldots &  \ldots &  \ldots
  \\[2pt]
  
\hline

 12+$\log(Ne^{2+}/H^+)$    &  7.50$\pm$0.07${^h}$ &  \ldots &  \ldots &  \ldots &  7.49$\pm$0.09 &  7.34$\pm$0.28 &  7.62$\pm$0.07 &  \ldots &  \ldots &  \ldots &  \ldots &  \ldots &  \ldots &  \ldots
  \\
 ICF($Ne^{2+}$)            &  1.21$\pm$0.01${^h}$ &  \ldots &  \ldots &  \ldots &  1.07$\pm$0.01 &  1.11$\pm$0.01 &  1.07$\pm$0.01 &  \ldots &  \ldots &  \ldots &  \ldots &  \ldots &  \ldots &  \ldots
  \\
 \bf{12+log(Ne/H)}         &  7.59$\pm$0.07${^h}$ &  \ldots &  \ldots &  \ldots &  7.52$\pm$0.09 &  7.39$\pm$0.28 &  7.65$\pm$0.07 &  \ldots &  \ldots &  \ldots &  \ldots &  \ldots &  \ldots &  \ldots
  \\
 \bf{log(Ne/O)}            & -0.58$\pm$0.09${^h}$ &  \ldots &  \ldots &  \ldots & -0.64$\pm$0.11 & -0.65$\pm$0.35 & -0.64$\pm$0.09 &  \ldots &  \ldots &  \ldots &  \ldots &  \ldots &  \ldots &  \ldots
  \\[2pt]
  
\hline

 12+$\log(Ar^{2+}/H^+)$    &  6.05$\pm$0.19${^d}$ &  5.76$\pm$0.20${^d}$ &  5.63$\pm$0.20${^d}$ &  6.36$\pm$0.31${^d}$ &  5.56$\pm$0.11 &  5.50$\pm$0.14 &  5.65$\pm$0.14 &  \ldots &  \ldots &  \ldots &  \ldots &  \ldots &  \ldots &  \ldots
  \\
 ICF($Ar^{2+}$)            &  1.15$\pm$0.01${^d}$ &  1.19$\pm$0.01${^d}$ &  1.15$\pm$0.01${^d}$ &  1.13$\pm$0.01${^d}$ &  1.38$\pm$0.02 &  1.11$\pm$0.01 &  1.79$\pm$0.08 &  \ldots &  \ldots &  \ldots &  \ldots &  \ldots &  \ldots &  \ldots
  \\
 \bf{12+log(Ar/H)}         &  6.11$\pm$0.19${^d}$ &  5.83$\pm$0.20${^d}$ &  5.69$\pm$0.20${^d}$&  6.41$\pm$0.31${^d}$ &  5.70$\pm$0.11 &  5.55$\pm$0.14 &  5.90$\pm$0.14 &  \ldots &  \ldots &  \ldots &  \ldots &  \ldots &  \ldots &  \ldots
  \\
  \bf{log(Ar/O)}            & -2.06$\pm$0.20${^d}$ & -2.21$\pm$0.21${^d}$ & -2.43$\pm$0.21${^d}$ & -1.90$\pm$0.34${^d}$ & -2.46$\pm$0.13 & -2.49$\pm$0.24 & -2.39$\pm$0.15 &  \ldots &  \ldots &  \ldots &  \ldots &  \ldots &  \ldots &  \ldots  \\  [2pt]
\hline

$He^+/H^+$($\lambda$\,4471)${^m}$    & 0.094$\pm$0.018 & \ldots & \ldots & \ldots & 0.096$\pm$0.003 & \ldots & \ldots & \ldots & \ldots & \ldots & \ldots & \ldots & \ldots & \ldots
  \\
 $He^+/H^+$($\lambda$\,5876)${^m}$    & 0.072$\pm$0.004 & 0.051$\pm$0.005 & 0.101$\pm$0.010 & 0.111$\pm$0.024 & 0.079$\pm$0.001 & 0.044$\pm$0.001 & 0.092$\pm$0.001 & \ldots & \ldots & \ldots & 0.124$\pm$0.040 & \ldots & \ldots & \ldots
  \\
 $He^+/H^+$($\lambda$\,6678)${^m}$    & 0.109$\pm$0.012 & \ldots & \ldots & \ldots & 0.086$\pm$0.001 & 0.035$\pm$0.006 & 0.113$\pm$0.003 & \ldots & \ldots & \ldots & \ldots & \ldots & \ldots & \ldots
  \\
 $He^+/H^+$($\lambda$\,7065)${^m}$    & 0.107$\pm$0.036 & \ldots & \ldots & \ldots & 0.109$\pm$0.005 & \ldots & \ldots & \ldots & \ldots & \ldots & \ldots & \ldots & \ldots & \ldots
  \\
 $He^+/H^+$adopted${^m}$    & 0.077$\pm$0.023 & 0.051$\pm$0.005 & 0.101$\pm$0.010 & 0.111$\pm$0.024 & 0.080$\pm$0.017 & 0.043$\pm$0.006 & 0.094$\pm$0.014 & \ldots & \ldots & \ldots & 0.124$\pm$0.040 & \ldots & \ldots & \ldots
\\[2pt]
\hline
\multicolumn{15}{@{\hspace{0.1cm}}l}{densidades en $cm^{-3}$ y temperaturas en 10$^4$\,K}\\
\end{tabular}
\end{center}}
\end{minipage}
\vspace{0.1cm}
{\scriptsize  \texttt{${^m}$} resultados usando las temperaturas predichas por los modelos, como se describe en el texto. \texttt{${^d}$} resultados usando las temperaturas derivadas con el m'etodo emp'irico D07. \texttt{${^h}$} resultados usando las temperaturas estimadas con el m'etodo emp'irico encontrado por H\"agele et al. (2006). \texttt{${^p}$} resultados usando las temperaturas derivadas con el m'etodo emp'irico P07. ${^*}$ resultados usando los datos necesarios para su c'alculo que fueron derivados de modelos o relaciones emp'iricas. $\dagger$ temperatura\,=\,10$^4$\,K}.\\
\end{sidewaystable}

\section{Discusi\'on}
\label{sec:discusion}
\subsection{Datos de ranura larga}
\label{sec:ranura larga}

Cuatro temperaturas electr'onicas, T$_e$([\OIII]), T$_e$([\OII]), T$_e$([\SIII]) y T$_e$([\SII]), han sido derivadas de mediciones directas en el brote B. La buena calidad de los datos permiti'o alcanzar precisiones del orden de 2\%, 3\%, 10\% y 14\%, respectivamente para cada una de las temperaturas. La obtenci'on de las cuatro temperaturas fue gracias a la buena relaci'on S/N del espectro que permiti'o la detecci'on de las l'ineas d'ebiles aurorales, como [\OIII]\,$\lambda$\,4363\,\AA, [\SII]\,$\lambda$\,4068\,\AA, [\SIII]\,$\lambda$\,6312\,\AA\ y  [\OII]\,$\lambda\lambda$\,7319, 7330\,\AA. Dado que la l'inea auroral [\NII]\,$\lambda$\,5755\,\AA\ no ha sido posible medirla en el espectro, se ha considerado la aproximaci'on  T$_e$([\NII])\,$\approx$\,T$_e$([\OII]) para la determinaci'on de la temperatura T$_e$([\NII]).

Se ha calculado la densidad electr'onica a partir del cociente de l'ineas del [\SII], obteniendo un valor de 100 part'iculas por cm$^3$, aunque es necesario destacar que el error de la misma es muy grande, por lo cual este valor sirve para determinar el orden de magnitud de la densidad. 

Las abundancias i'onicas del O$^{+}$, O$^{2+}$, S$^{+}$, S$^{2+}$, N$^{+}$, Ne$^{2+}$, Ar$^{2+}$ y Ar$^{3+}$ se han deducido utilizando las expresiones dadas en la Secci'on \ref{sec:Abundancias}, las cuales dependen de la intensidad de las l'ineas intensas del i'on cuya abundancia se quiere calcular, as'i como de la densidad (en algunos casos) y temperaturas electr'onicas. Como en este brote fue posible medir la l'inea del hierro 4658\AA\, se obtuvo tambi'en Fe$^{2+}$ considerando la temperatura del [\OIII] para su c'alculo. Adem'as, fueron calculadas las abundancias totales del O, S, N, Ne, Ar y Fe, y los cocientes logar'itmicos N/O, S/O, Ne/O y Ar/O.
Tanto las abundancias i'onicas como las abundancias totales y los cocientes logar'itmicos respecto al ox'igeno, junto con sus respectivos errores, est'an tabulados en la Tabla \ref{abundancesBC}.\\
%

El brote B tiene una abundancia de ox'igeno 0.22 veces el valor solar \citep[][ 12+log(O/H)\,$\odot$\,=\,8.69]{2001ApJ...556L..63A}. Este valor de abundancia (8.04$\pm$0.03) es similar, dentro de los errores, al encontrado por L'opez-S'anchez \& Esteban~(\citeyear{2009A&A...508..615L}) que es igual a 8.10$\pm$0.06. La abundancia de azufre es 0.15 veces el solar \citep[][12+log(S/H)\,$\odot$\,=\,7.33]{1998SSRv...85..161G}, y la abundancia de nitr'ogeno es 0.09 veces el solar \citep[][12+log(N/H)\,$\odot$\,=\,7.93]{2001AIPC..598...23H}.
Respecto a las abundancias relativas (ver Figura 4.4), se estim'o un valor de log(N/O)\,= -1.16 el cual es 0.28 dex menor al valor solar \citep[][log(N/O)\,$\odot$\,= -0.88]{2005ASPC..336...25A} y 0.44 dex mayor al valor t'ipico para este tipo de objetos \cite[el valor del plateau se encuentra en log(N/O)\,=\,-1.6, ver][]{2010ApJ...715L.128A,Perez-Montero+11} y un valor de log(S/O)\,= -1.52$\pm$0.22 que est'a de acuerdo, dentro de los errores, con el valor solar \citep[][log(S/O)\,$\odot$\,= -1.36]{1998SSRv...85..161G}.\\
Ha sido posible estimar la abundancia total de He a partir de la obtenci'on de las abundancias i'onicas He$^{+}$ y He$^{2+}$ dado que fue posible medir las l'ineas, He{\sc  i}\,$\lambda\lambda$\,4471, 5876, 6678 y 7065\AA, y He{\sc  ii}\,$\lambda$\,4686\,\AA\, respectivamente. El valor ``adoptado" para He$^{+}$/H$^{+}$ es 0.086$\pm$0.015 y 0.0013$\pm$0.0002 para el He$^{2+}$/H$^{+}$ dando un valor de la abundancia total del He estimada en esta regi'on de 0.087$\pm$0.015, valores t'ipicos para las galaxias H {\sc ii} \citep{2008MNRAS.383..209H}. Los resultados pueden verse en la misma Tabla antes mencionada.

%

En el brote C no fue posible obtener de forma directa las temperaturas electr'onicas, dado que no se pudo medir las l'ineas aurorales correspondientes para sus c'alculos. En este caso, se recurri'o tanto a los modelos como a las relaciones emp'iricas para determinar las diferentes temperaturas. Con el m'etodo de P07 descripto en la Secci'on \ref{sec:Relaciones}, se estim'o la temperatura del [\NII], encontrando un valor para la misma de 0.93$\pm$0.01. Y haciendo uso de la aproximaci'on T$_e$([\NII])\,$\approx$\,500K + T$_e$([\SIII]) dada por D07 se estim'o un valor para la T$_e$([\SIII]), con el que se deriv'o T$_e$([\OIII]) a partir de la relaci'on de H06. La temperatura electr'onica T$_e$([\OII]) fue calculada a partir de T$_e$([\OIII]) utilizando los modelos de P'erez-Montero \& D'iaz~(\citeyear{2003MNRAS.346..105P}) y de la aproximaci'on T$_e$([\SII])\,$\approx$\,T$_e$([\OII]) se obtuvo la temperatura de [\SII]. Tambi'en en este brote se estim'o una densidad de 100 part'iculas por cm$^3$ con errores muy grandes.
Los resultados de las abundancias i'onicas y totales usando las temperaturas estimadas con los modelos y los m'etodos emp'iricos, y los ICFs son expuestos en la Tabla \ref{abundancesBC}. En este brote no se ha medido la l'inea  [\FeIII]\,$\lambda$4658\,\AA, por lo que no se pudo derivar una abundancia de este elemento. Tampoco, como se ha comentado en la Secci'on \ref{sec:rvresults}, se ha podido medir la l'inea  [\NII]\,$\lambda$\,6548\,\AA, con lo cual para el c'alculo de las abundancias del nitr'ogeno, se ha recurrido a su relaci\'on te\'orica con la l'inea [\NII]\,$\lambda$\,6584\,\AA. 
Para este brote se encontr'o un valor de la abundancia de ox'igeno de 8.13$\pm$0.04, que es 0.28 veces el solar. 

Comparando los resultados de los datos de ambos brotes en los espectros de ranura larga, se puede ver que la abundancia de ox'igeno es mayor en el brote C que en el B pero ambas por debajo del valor solar, con valores t'ipicos en esta clase de objetos \citep[ver por ejemplo,][]{1991A&AS...91..285T,2006MNRAS.365..454H}. Las temperaturas representativas de la zona de alta ionizaci'on son mayores en la regi'on B. Las temperaturas correspondientes a la zona de baja ionizaci'on, en el caso de T$_e$([\OII]) aparentan ser similares en ambos brotes, mientras que para T$_e$([\SII]) y T$_e$([\NII]) las temperaturas fueron derivadas utilizando distintas aproximaciones. En el brote B, la temperatura T$_e$([\SII]), obtenida de las medidas directas y es 4200 K mayor que en la del brote C, donde se deriv'o de suponer la igualdad con T$_e$([\OII]). Es de notar que en el brote B, T$_e$([\OII]) es 4100 K m'as alta que T$_e$([\SII]). En cuanto a la temperatura de [\NII] en el brote B, el valor obtenido es mayor que la del brote C en 2700 K, pero nuevamente fueron estimados utilizando dos aproximaciones distintas. A modo de comparaci'on se ha derivado la temperatura del [\NII] para el brote B usando el m'etodo P07. El valor encontrado para B es T$_e$([\NII])\,=\,10500$\pm$100 K, que comparado con el valor de T$_e$([\NII]) encontrado para el brote C usando el mismo m'etodo P07 es $\sim$1200 K mayor que 'este.
El valor estimado para N/O es un poco mayor en el brote C que en el B, aunque igual dentro de los errores (ver Figura 4.4).

El grado de excitaci'on, log(O$^{2+}$/O$^{+}$), calculado para el brote B es 0.61 y para C es 0.02. Sin embargo, debe tenerse en cuenta que, para el brote C las abundancias i'onicas se han deducido utilizando las temperaturas derivadas de los modelos y relaciones emp'iricas. Sin embargo, igualmente se puede considerar que es mayor el grado de excitaci'on en el brote B que en el C, concordando este resultado con lo estimado por L'opez-S'anchez \& Esteban~(\citeyear{2009Ap&SS.324..355L}).
La extinci'on observada en ambos brotes B y C es similar dentro de los errores y consistente con baja extinci'on.

\subsection{Datos 'echelle}
\label{sec:'echelle}
En esta subsecci'on se analizar'an los resultados alcanzados para las regiones estudiadas con los datos de los espectros 'echelle. Cada brote ser'a analizado por separado, estudiando las temperaturas, densidades, abundancias i'onicas y totales estimadas para la medida global y para cada componente cinem'atica separadamente. Luego se dar'a un an'alisis comparativo entre brotes.

\subsubsection*{Haro\,15 A}
El an'alisis que a continuaci'on se detalla, se refiere tanto a la medida global como a las dos componentes angostas y la componente ancha del brote A, salvo que se especifique alg'un an'alisis particular realizado sobre algunas de las componentes o medida global. 
Las l'ineas aurorales [\OIII]\,$\lambda$\,4363\,\AA, [\SII]\,$\lambda$\,4068\,\AA, [\SIII]\,$\lambda$\,6312\,\AA,  [\OII]\,$\lambda\lambda$\,7319, 7330\,\AA\ y [\NII]\,$\lambda$\,5755\,\AA\ no se han podido detectar en el espectro de la regi'on central de Haro\,15, el brote A. En este caso, utilizando la relaci'on entre temperaturas T$_e$[\SIII] vs. T$_e$[\OIII] dada por el ajuste encontrado en H06 se ha podido estimar la temperatura electr'onica del [\OIII], previa determinaci'on de T$_e$([\SIII]) con el m'etodo D07 usando el par'ametro emp'irico SO$_{23}$. Luego, siguiendo el an'alisis cl'asico, se han podido obtener de los modelos, las temperaturas T$_e$([\OII]), T$_e$([\SII]) y T$_e$([\NII]).
%

Las densidades estimadas en los cuatro casos aqu'i analizados, est'an por debajo de la densidad cr'itica por desexcitaci'on colisional. La medida global tiene una densidad de 300 part'iculas por cm$^3$;  la componente angosta 2, 200 part'iculas por cm$^3$; y tanto la componente angosta 1 como la componente ancha, 100 part'iculas por cm$^3$. A'un teniendo en cuenta que los valores de la densidad son una estimaci'on de su magnitud, dado que los errores son muy grandes, parecer'ia que la componente angosta 2 tiene una mayor densidad que las otras dos componentes. Sin embargo, y dado que la medida global es la que da una densidad mayor, no podemos afirmar nada al respecto. 

En la Tabla \ref{abundancesABCEF} se muestran los resultados tanto de las propiedades f'isicas como de las abundancias i'onica y totales, y los ICFs. En este brote no se ha medido la l'inea del [\FeIII]\,$\lambda$4658\,\AA, pero s'i se tiene una medida de las l'ineas del He. De la medida global, se pudo determinar un valor ``adoptado'' de He$^{+}$/H$^{+}$, derivado a partir de las l'ineas 4471, 5876, 6678, 7065 \AA, siendo el mismo de 0.077$\pm$0.023. Para las componentes, angosta 1, angosta 2 y ancha, se encontr'o un valor ``adoptado'' de He$^{+}$/H$^{+}$ usando s'olo la medida de la l'inea 5876\AA, dando 0.051$\pm$0.005, 0.101$\pm$0.01, 0.111$\pm$0.024 respectivamente en cada caso. El valor estimado para la componente angosta 1 es bajo, si lo comparamos con los de otras galaxias de este tipo, mientras que para las otras dos componentes los valores son t'ipicos de galaxias H {\sc ii} \citep[ver][]{2008MNRAS.383..209H}. Se debe destacar que el valor ``adoptado'' para la medida global (0.077$\pm$0.023) es aproximadamente el promedio de los otros valores pesados por la luminosidad de las componentes, y est'a dentro de los valores t'ipicos para esta clase de objetos.

En lo que respecta a la abundancia total del ox'igeno, la medida global presenta un valor de 8.17$\pm$0.05, 0.3 veces el valor solar, mientras que para las distintas componentes cinem'aticas se obtuvieron 8.04$\pm$0.05 (angosta 1), 8.11$\pm$0.08 (angosta 2), y 8.31$\pm$0.13 (ancha), que son 0.22, 0.26 y 0.42 veces la abundancia solar, respectivamente. Esto estar'ia indicando que, con errores de 1$\sigma$ (que son los dados en la tabla), la abundancia del ox'igeno discriminada por componentes cinem'aticas presenta valores diferentes entre la componente angosta 1 y la ancha, teniendo la componente ancha el valor m'as alto. La componente angosta 2 presenta el valor m'as pr'oximo a la medida global, valor medio entre la ancha y la angosta 1. Sin embargo, si se consideran los errores de 2$\sigma$ las abundancias totales de ox'igeno son las mismas para todas las componentes cinem'aticas. 
La abundancia total del ox'igeno en este brote derivada para cada una de las tres componentes y en la medida global, presenta, dentro de sus errores, los caracter'isticos valores bajos que se encuentran las galaxias H\,{\sc ii}: 12+log(O/H) entre 7.94 y 8.19 \citep{1991A&AS...91..285T,2006MNRAS.365..454H}.

La abundancia total del azufre para la medida global es 0.20 veces el valor solar, mientras que las componentes angostas 1 y 2 dan 
0.10 veces el solar y la ancha 0.32 veces el valor solar. En la abundancia total del nitr'ogeno los valores van de 0.12 a 0.19 veces el valor solar.

Las diferentes abundancias i'onicas del azufre, S$^{+}$/H$^{+}$ y S$^{2+}$/H$^{+}$, la abundancia i'onica N$^{+}$/H$^{+}$ y la abundancia i'onica Ar$^{2+}$/H$^{+}$ obtenidas para la componente ancha son m'as altas que las encontradas en las componentes angostas, entre 0.17 a 0.61 dex respecto a la angosta 1, y entre 0.28 a 0.73 dex respecto a la angosta 2. S'olo se pudieron estimar las abundancias i'onicas y total del [\NeIII] con la medida global dado que la l'inea del ne'on en 3868\,\AA\, no pudo ser resuelta por componentes cinem'aticas. Los datos est'an expuestos en la Tabla \ref{abundancesABCEF}.

Respecto a las abundancias del azufre y el nitr'ogeno relativas al ox'igeno, los valores S/O y N/O, la componente ancha tiene un valor del S/O pr'oximo al de la medida global, pero si se tienen en cuenta los errores, todos los valores son muy similares entre si (ver Figura 4.4). El valor de N/O en la componente angosta 2 es el m'as alto, pero su diferencia con las otras componentes est'a dentro de los errores como para considerarla significativa. Esto puede ser una evidencia que la evoluci'on qu'imica de las diferentes componentes es muy similar \citep[ver discusi'on en][]{Perez-Montero+11} o que directamente son distintas fases de un mismo gas. Este brote presenta un exceso muy alto de N/O (ver Figura 4.4) respecto a lo que generalmente se encuentra en las galaxias H\,{\sc ii} \citep[log(N/O)\,=\,-1.6;][]{1979MNRAS.189...95P,1979A&A....78..200A}, \cite[ver discusi'on en][]{2010ApJ...715L.128A,Perez-Montero+11}.

La extinici'on observada en el brote A es consistente con baja extinci'on. La componente angosta 2 es las que presenta mayor extinci'on y la componente angosta 1 es la que muestra menor extinci'on. La ancha tiene una extinci'on con un valor promedio de las otras dos componentes cinem'aticas, y la medida global muestra un valor similar, dentro de los errores, al de la componente ancha.

\subsubsection*{Haro\,15 B}
A diferencia del espectro obtenido en el modo de ranura larga del brote B, aqu'i s'olo fue posible derivar dos temperaturas electr'onicas con el m'etodo directo para la medida global y para las dos componentes cinem'aticas angosta y ancha, T$_e$([\OIII]) y T$_e$([\SIII]). Las precisiones alcanzadas, respectivamente para T$_e$([\OIII]) y T$_e$([\SIII]) son: para la medida global, del orden de 5\% y 13\%; 15\% y 18\% en la componente angosta y 4\% y 14\% en la ancha. La temperatura T$_e$([\OII]) se obtuvo de los modelos debido a la no detecci'on de las l'ineas aurorales [\OII]\,$\lambda\lambda$\,7319, 7330\,\AA. Y como no fue posible medir la l'inea d'ebil auroral [\SII]\,$\lambda$\,4068\,\AA, la temperatura T$_e$([\SII]) se consider'o igual a T$_e$([\OII]). La l'inea auroral [\NII]\,$\lambda$\,5755\,\AA\ tampoco en este caso se pudo medir en el espectro, y se ha hecho la consideraci'on T$_e$([\NII])\,$\approx$\,T$_e$([\OII]).

La densidad electr'onica obtenida para la medida global es de 150 part'iculas por cm$^3$ con una incerteza de 50 part'iculas por cm$^3$. Las densidades estimadas para las dos componentes cinem'aticas est'an dentro del rango de bajas densidades con valores $\sim$100 y 150 part'iculas por cm$^3$ para la componente angosta y ancha, respectivamente. En estos dos 'ultimos casos, los errores en las densidades son muy grandes, por lo cual estos valores dan una idea del orden de magnitud de la densidad. 

Los resultados de las diferentes abundancias i'onicas y totales junto con los ICFs derivados, y los errores correspondientes, se muestran en la Tabla \ref{abundancesABCEF}. Si se comparan las abundancias i'onicas derivadas para cada componente cinem'atica se puede ver que los iones pertenecientes a la zona de menor excitaci'on (O$^{+}$, S$^{+}$, N$^{+}$) muestran una mayor abundancia en la componente angosta, mientras que los iones de la zona intermendia (S$^{2+}$, Ar$^{2+}$) y los de la zona de alta excitaci'on (O$^{2+}$, Ne$^{2+}$) muestran el comportamiento opuesto, siendo mayores los valores de las abundancias determinados en la componente ancha. En este espectro no se pudo medir la l'inea del [\FeIII] en 4658\AA.

Los valores de las abundancias totales son: la del ox'igeno derivada para la medida global es 0.3 veces el valor solar, y la componente angosta y la ancha son 0.22 y 0.4 veces el valor solar, respectivamente. Estos tres valores son t'ipicos de las galaxias H\,{\sc ii}. Para las abundancias de azufre los resultados son: en el caso de la medida global 0.09 veces el solar, y 0.06 y 0.12 veces el valor solar en las respectivas componentes angosta y ancha. En el caso del nitr'ogeno la medida global es 0.08 veces el valor de la abundancia solar, y las componentes angosta y ancha son 0.06 y 0.11 veces el valor solar. La componente angosta parece tener un valor alto en abundancia de nitr'ogeno \citep[ver valores para la muestra de galaxias H\,{\sc ii} de][]{2003MNRAS.346..105P}. Haciendo una consideraci'on estad'istica con errores a 2$\sigma$ (los valores de los errores dados en la Tabla son a 1$\sigma$), los valores de las abundancias totales son iguales para las distintas componentes y para todos los elementos, dentro de esos errores, y en algunos casos incluso a 1$\sigma$. Respecto a las abundancias del azufre, nitr'ogeno, ne'on y el arg'on relativas al ox'igeno, ambas componentes presentan, pr'acticamente valores iguales, incluso comparando estos valores con los estimados para la medida global. Esto podr'ia estar indicando que ambas componentes presentan una misma evoluci'on qu'imica y/o que son dos fases distintas de un mismo gas.

Con la medida global, ha sido posible estimar la abundancia total de He a partir de la obtenci'on de la abundancia i'onica de He$^{+}$ dada por la medici'on de las l'ineas He{\sc  i}\,$\lambda\lambda$\,4471, 5876, 6678 y 7065\AA. El valor ``adoptado'' He$^{+}$/H$^{+}$ que se ha obtenido es 0.080$\pm$0.017, similar al encontrado en la medida del mismo brote en el espectro de ranura larga y en otras galaxias H\,{\sc ii} \citep{2008MNRAS.383..209H}. En el caso de la componente angosta, el valor ``adoptado'' se estim'o con la medida de las l'ineas He{\sc  i}\,$\lambda\lambda$\,5876 y 6678\,\AA\ y es de 0.043$\pm$0.006 (valor bajo para este tipo de objetos) y en la ancha 0.094$\pm$0.014, valor t'ipico para esta clase de galaxias. Este es el mismo comportamiento que se encontr'o en el brote A, entre la componente angosta 1 y las otras dos componentes.

El brote B muestra baja extinci'on. La componente angosta presenta mayor extinci'on que la componente ancha y la medida global muestra una extinci'on cuyo valor es el promedio de los valores de las otras dos componentes. La extinci'on medida en el brote B observado en el modo de dispersi'on simple, considerando los errores, tiene un valor levemente menor al que se observa en la componente angosta observada en el modo 'echelle.

\subsubsection*{Haro\,15 C, Haro\,15 E, Haro\,15 F}

A continuaci'on en esta subsecci'on se analizar'an conjuntamente los tres brotes restantes, C, E y F, analizando las medidas globales y las diferentes componentes cinem'aticas, cuando se han podido ajustar. Los espectros de estas tres regiones tienen un bajo S/N lo cual dificult'o la medici'on de no solo las l'ineas d'ebiles aurorales sino de algunas l'ineas m'as intensas necesarias para el c'alculo de temperaturas y densidades.
 
En ninguno de los tres brotes, fue posible detectar ninguna de las l'ineas aurorales, adem'as dada la baja se'nal de los espectros, en las medidas globales de los brotes C y F como en las componentes cinem'aticas de los brotes C y E, no se pudieron medir las l'ineas del [\OII]\,$\lambda\lambda$3727, 3729\,\AA\ siendo imposible aplicar el m'etodo D07 para determinar la T$_e$([\SIII]) ni el m'etodo P07 para determinar la temperatura T$_e$([\NII]), dado que estos m'etodos requieren de esta l'ineas. Por tal motivo, en estos casos se consider'o la temperatura electr'onica de [\OIII] como 10$^4$K como fue comentado en la Secci'on \ref{sec:Relaciones}. Con este valor se estim'o la temperatura de [\OII] usando la relaci'on entre ambas de P'erez-Montero \& D'iaz~(\citeyear{2003MNRAS.346..105P}), ya descripta en la misma Secci'on. Y las temperaturas T$_e$([\SII]) y T$_e$([\NII]) se consideraron iguales a la del [\OII], y T$_e$([\SIII]) se determin'o usando su relaci'on con la temperatura de [\OIII] derivada por H06. 

En cambio, en la medida global del brote E, si fue posible medir las l'ineas [\OII]\,$\lambda\lambda$3727, 3729\,\AA, con las cuales se pudo aplicar el m'etodo P07 para estimar la temperatura del [\NII] y luego con esta temperatura estimar la T$_e$([\SIII]), usando la diferencia sistem'atica de 500K entre temperaturas hallada por D'iaz et al.~(\citeyear{2007MNRAS.382..251D}), y T$_e$([\OIII]) del ajuste de H06. La temperatura T$_e$([\OII]) se obtuvo utilizando los mismos modelos y T$_e$([\SII]) considerando nuevamente su igualdad con la temperatura T$_e$([\OII]).

Las densidades electr'onicas estimadas de la medida global como de la componente angosta del brote C son de 400 part'iculas por cm$^3$. En el caso de la componente ancha se encontr'o un valor mucho m'as alto de 1400 part'iculas por cm$^3$. Estos datos dan una estimaci'on del orden de magnitud de la densidad electr'onica, dado que la incerteza en el c'alculo es muy grande. Esto en parte puede deberse a la poca se'nal en el espectro y a que la l'inea [\SII]\,$\lambda$\,6731\,\AA\ debi'o ser corregida por emisi\'on de l\'{\i}neas tel\'uricas de cielo, dificultando los ajustes de la componente ancha (ver Tabla \ref{ratiostot 4}). La densidad estimada a partir de la medida global del brote E es de 500 part'iculas por cm$^3$, mientras que la componente angosta 1 presenta una densidad electr'onica de 900 part'iculas por cm$^3$ y la componente angosta 2, 400 part'iculas por cm$^3$. En el caso de la componente angosta 1 del brote E, el alto valor de la densidad electr'onica tal vez se deba al mismo problema presentado en el brote C. En la medida global del brote F, la densidad estimada es de 100 part'iculas por cm$^3$. Debido a los errores muy altos, todos estos valores dan 'unicamente una idea del orden en magnitud de la densidad.

Las abundancias i'onicas en estas condiciones fueron dif'iciles de estimar. En las medidas globales de los brotes y en las dos componentes cinem'aticas de los brotes C y E, se pudieron obtener las abundancias i'onicas de O$^{2+}$, S$^{+}$ y N$^{+}$. En la medida global del brote E, adem'as fue posible obtener la abundancia i'onica O$^{+}$ y las abundancias totales del ox'igeno y azufre, y se pudo estimar el cociente N/O. Todos estos datos junto a las propiedades f'isicas de los brotes, est'an expuestos en la Tabla \ref{abundancesABCEF}. En estos brotes no se midi'o la l'inea  [\FeIII]\,$\lambda$4658\,\AA, por lo que no se pudo derivar una abundancia de este mismo elemento.

La abundancia de O$^{2+}$ en el brote C derivada a partir de la componente angosta tiene un valor m'as alto, $\sim$0.3 dex, que la derivada a partir de la componente ancha, pero esto podr'ia deberse a una diferencia en el grado de ionizaci'on de las componentes. Pero, para saber si hay una diferencia real entre las abundancias se debe comparar, obviamente, las abundancias totales y no las i'onicas. En la medida global, el valor de 7.83$\pm$0.03 dex es aproximadamente el promedio de dichas abundancias i'onicas derivadas para cada componente (7.85, teniendo en cuenta que la escala es logar'itmica). Esto se debe a que las intensidades de las l'ineas de emisi'on corregidas por enrojecimiento son pr'acticamente iguales. Las abundancias i'onicas del S$^{+}$ son iguales, dentro de los errores, en las dos componentes y en la medida global. Y la abundancia de N$^{+}$ es mayor en la angosta, 0.40 dex respecto de la ancha, pero esto nuevamente puede deberse a una diferencia en el grado de ionizaci'on de las componentes. Tanto en la abundancia del O$^{2+}$ como en la de N$^{+}$, la medida global es aproximadamente el valor promedio de las estimaciones para las dos componentes cinem'aticas.

Las abundancias del O$^{2+}$, S$^{+}$ y N$^{+}$ en el brote E presentan valores m'as altos, derivados a partir de la componente angosta 2 que para la derivada a partir de la angosta 1. Para la medida global nuevamente se ha encontrado, como era esperable, que es aproximadamente el promedio pesado por la luminosidad de las dos componentes cinem'aticas. La abundancia total del ox'igeno para la medida global es 0.31 veces el valor solar. En este brote se pudo estimar un valor de la abundancia de helio (0.124$\pm$0.040), que es un poco elevado pero a'un en el rango de las galaxias H\,{\sc ii} \citep{2008MNRAS.383..209H}. El mismo fue obtenido por la medida de la l'inea 5876\AA\ y fue tomado como el valor ``adoptado''.

La extinci'on observada en los brotes C y E es consistente con baja extinci'on. El brote C presenta valores de extinci'on con errores muy grandes, y solo se puede decir que las componentes cinem'aticas como la medida global presentan valores de extinci'on similares. En el brote E, la componente angosta 1 tiene un valor de extinci'on mayor a los valores de extinci'on de la componente ancha y de la medida global, ambos, como se ha comentado m'as arriba, con valores de c(H$\beta$) compatibles con cero dentro de los errores observacionales, y por esta raz'on fueron igualados a cero. En el brote F las intensidades de las l'ineas no fueron corregidas por enrojecimiento.
. 
Comparando la abundancia total de ox'igeno derivada en el brote A con el valor obtenido por L'opez-S'anchez \& Esteban~(\citeyear{2009A&A...508..615L}), 8.37$\pm$0.10, y teniendo en cuenta los errores, se puede ver que la abundancia obtenida para la medida global es menor a este valor. Esto puede deberse a las diferentes hip'otesis consideradas para determinar la temperatura electr'onica y por ende las abundancias.
En el brote B, el valor obtenido en la medida global se encuentra en muy buen acuerdo con el valor de abundancia total de ox'igeno obtenido por L'opez-S'anchez \& Esteban, en el mismo brote, 8.10$\pm$0.06. Debe destacarse que para este 'ultimo brote tanto el trabajo de L'opez-S'anchez \& Esteban como el nuestro se realiza una derivaci'on directa de la temperatura de [\OIII], y que est'an en excelente acuerdo (12000$\pm$600 K en este trabajo y 12900$\pm$700 K en el trabajo de L'opez-S'anchez \& Esteban), y una estimaci'on de T$_e$([\OII]) a partir de la de [\OIII], en el caso de  L'opez-S'anchez \& Esteban, utilizando la relaci'on de Garnett~(\citeyear{1992AJ....103.1330G}). Estas temperaturas tambi'en est'an en muy buen acuerdo, (11000$\pm$200 y 12000$\pm$500 K, en este trabajo y en el de L'opez-S'anchez \& Esteban, respectivamente).

Estudios recientes realizados en esta galaxia indican que el c'umulo estelar ionizante en el brote B ser'ia el m'as j'oven de la galaxia dado que este brote presenta un color azul y una alta emisi'on FUV, indicativo de una reciente actividad de formaci'on estelar en curso, un resultado apoyado por la detecci'on de rasgos Wolf-Rayet en el espectro \citep{2010A&A...516A.104L}. Esto est'a de acuerdo con los anchos equivalentes (EW) de la l'inea H$\beta$ (ver Tablas \ref{ratiostot 1}, \ref{ratiostot 2} y \ref{ratiostot 3}) derivados de los datos de este trabajo, que reflejan que el brote B es m'as j'oven al tener un ancho equivalente mayor al resto de los brotes \citep{2004MNRAS.348.1191T}. Por otro lado, del an'alisis realizado en este trabajo, se encuentra que la abundancia total de ox'igeno derivada en los brotes A y B es muy similar. 

Si se tienen en cuenta las abundancias i'onicas del ox'igeno derivadas para el brote B utilizando los datos de dispersi'on simple, puede notarse que la abundancia para el O$^{+}$ es menor (0.16 dex) que el valor derivado de la medida global del 'echelle. Ambos valores son compatibles si consideramos los errores a 2$\sigma$. Por otro lado, la abundancia de ox'igeno O$^{2+}$ derivada del espectro de ranura larga y de la medida global del 'echelle son compatibles dentro de los errores. La abundancia total de ox'igeno de los datos de dispersi'on simple es menor (0.12 dex) si se la compara con la estimaci'on global del 'echelle, y mucho menor (0.25 dex) si se la compara con la estimada en la componente ancha, pero es igual a la derivada a partir de la componente angosta. Sin embargo, si se tienen en cuenta los errores a 2$\sigma$, el valor derivado de los datos de ranura larga y el global del 'echelle son compatibles. Tambi'en se observa que el valor ``adoptado'' He$^{+}$/H$^{+}$ de la medida global en el brote B del 'echelle, es similar al encontrado en el espectro de ranura larga y es compatible con los valores t'ipicos para las galaxias H\,{\sc ii}.

Se ha comentado sobre el exceso de N/O en el brote A, el cual es muy alto respecto a los valores encontrados en las galaxias H\,{\sc ii}, aunque ya se han observado este tipo de excesos en otros objetos \citep[ver discusi'on en][]{Perez-Montero+11}. El brote B muestra tambi'en excesos en N/O pero el mismo no es tan marcado como en el caso del brote A.
Los valores del S/O, para los brotes A y B, aunque presentan una gran dispersi'on, est'an en el rango de este tipo de objetos \citep{2008MNRAS.383..209H} como se puede apreciar en la Figura 4.4. El brote A muestra valores mayores de este cociente, que son compatibles con el valor solar dentro de los errores, al menos para la componente ancha.
El valor del Ne/O en el brote A (medida global) y en el brote B, dentro de los errores, son consistentes con el valor solar \citep[log(Ne/O)\,$\odot$\,= 0.61 dex;][]{1998SSRv...85..161G} y se encuentran en excelente acuerdo con los valores de la literatura \citep{2006MNRAS.372..293H,2008MNRAS.383..209H, 2011arXiv1101.4140H}. El valor del cociente Ar/O para el brote A presenta una dispersi'on mayor respecto de la que presenta el brote B, esto se puede ver en la Figura 4.4. Los valores del Ar/O en el brote A son 0.08 y 0.39 dex mayores al valor solar \citep[log(Ar/O)\,$\odot$\,= -2.29 dex;][]{1998SSRv...85..161G} con la excepci'on de la componente angosta 2 que es 0.14 dex menor al valor solar. En cambio los valores obtenidos en el brote B son entre 0.1 y 0.2 dex menores al solar.
Los valores de las abundacias relativas N/O, Ne/O y Ar/O en el brote B observados en ranura larga, son consistentes con los observados en el modo 'echelle. En cambio, el valor de S/O en el modo de dispersi'on simple se aparta de los valores del 'echelle, siendo mucho mayor. 

En el caso de las abundancias totales de He, se ha encontrado un comportamiento similar para la componente angosta 1 del brote A y la angosta del brote B respecto al resto de las componentes, y es que el valor ``adoptado'' de He$^{+}$/H$^{+}$ es bajo para este tipo de objetos, pero en cambio el valor obtenido para las componentes angosta 2 y ancha del brote A y para la componente ancha del brote B son valores t'ipicos para las galaxias H {\sc ii}.

Respecto de los brotes C, E y F como es posible ver, el an'alisis comparativo sobre las abundancias i'onicas y totales entre brotes es a'un incompleto, dado que los espectros no han tenido la suficiente se'nal para realizar estas estimaciones y ser'a necesario obtener nuevos datos para un an'alisis m'as profundo. S'olo se puede decir que la abundancia total de ox'igeno del brote E para la medida global es muy similar a las abundancias obtenidas en los brotes A y B. 

La extinci'on medida en los brotes A, B, C y E, considerando los errores, presentan valores similares y todos compatibles con baja extinci'on, aunque se puede decir que el brote B tiene un valor levemente menor al del resto de los brotes.

\begin{figure*}
\begin{center}
\label{abrelativas}
\vspace{1.0cm}
\includegraphics[trim=0cm 0cm 0cm 0cm,clip,angle=0,width=7cm,height=7cm]{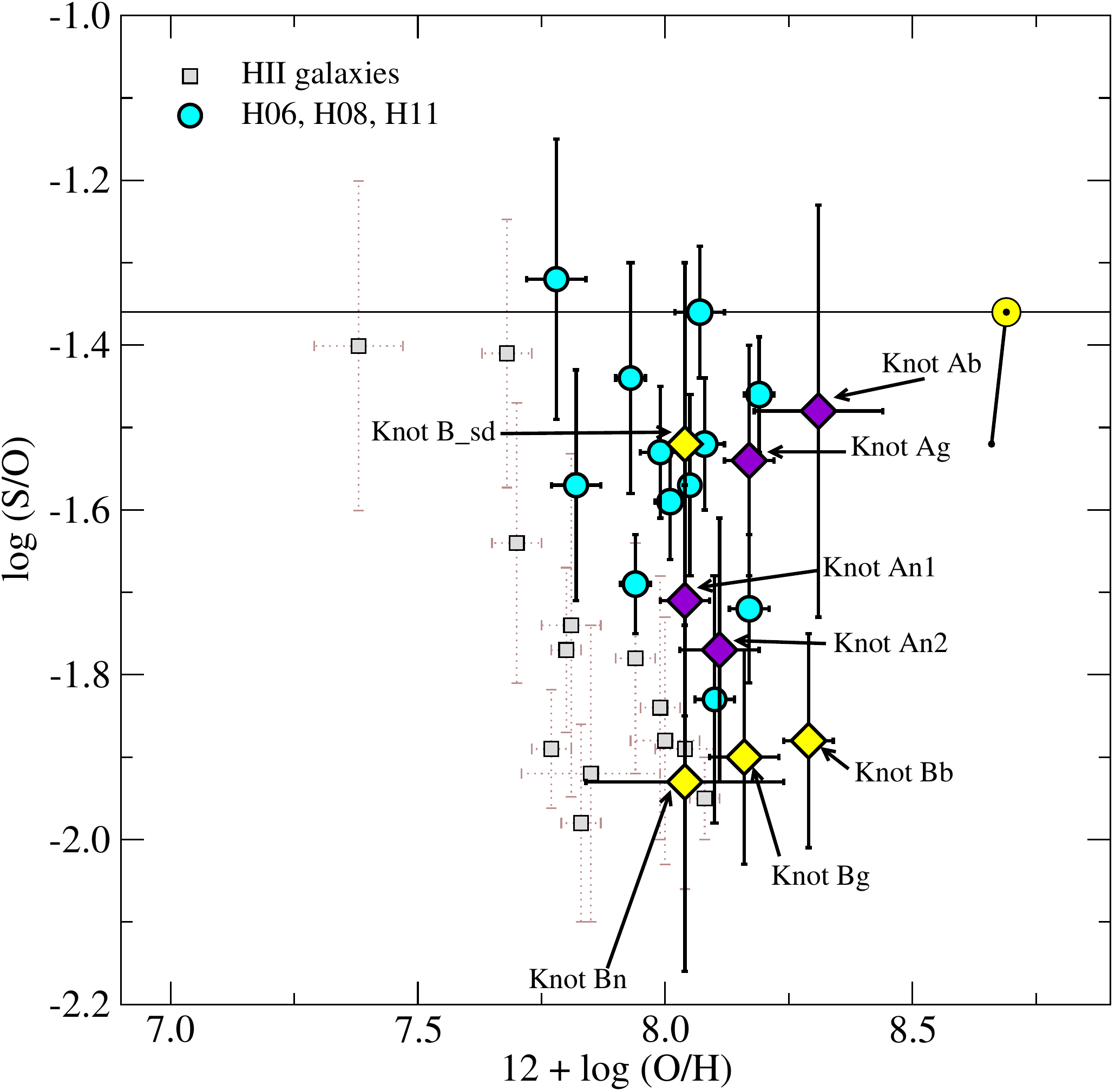}
\includegraphics[trim=0cm 0cm 0cm 0cm,clip,angle=0,width=7cm,height=7cm]{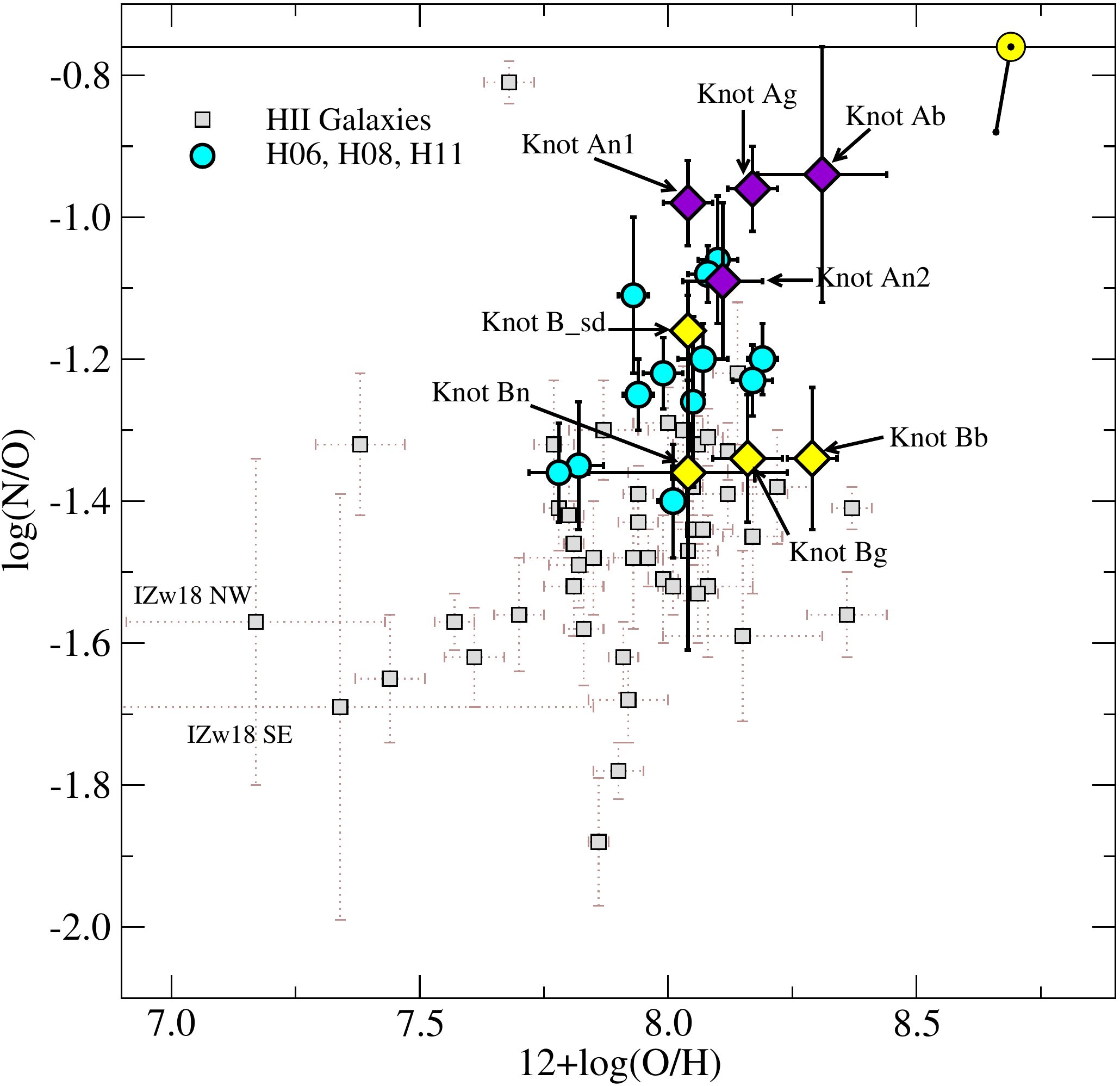}
\vspace{1cm}
\includegraphics[trim=0cm 0cm 0cm 0cm,clip,angle=0,width=7cm,height=7cm]{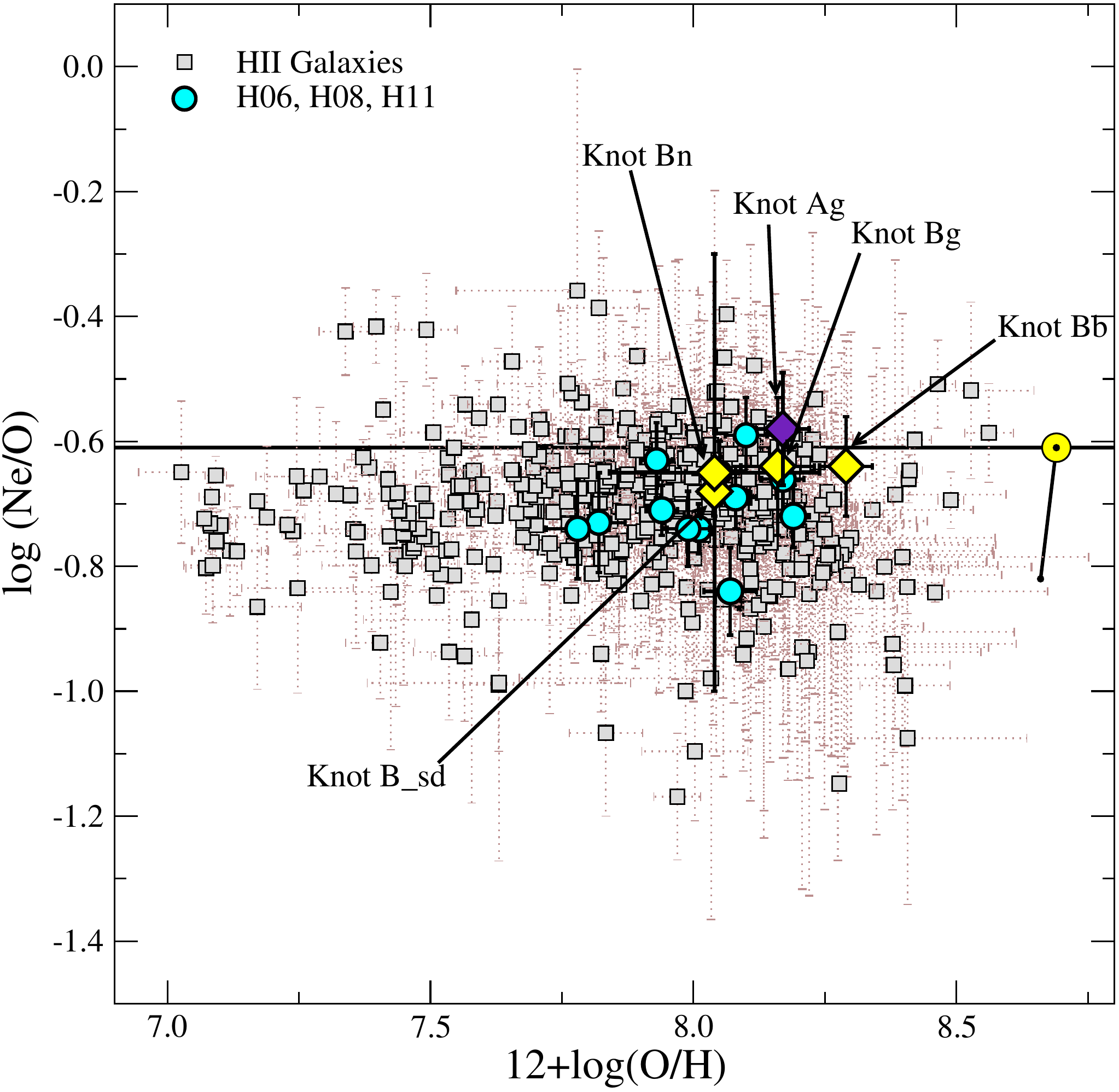}
\includegraphics[trim=0cm 0cm 0cm 0cm,clip,angle=0,width=7cm,height=7cm]{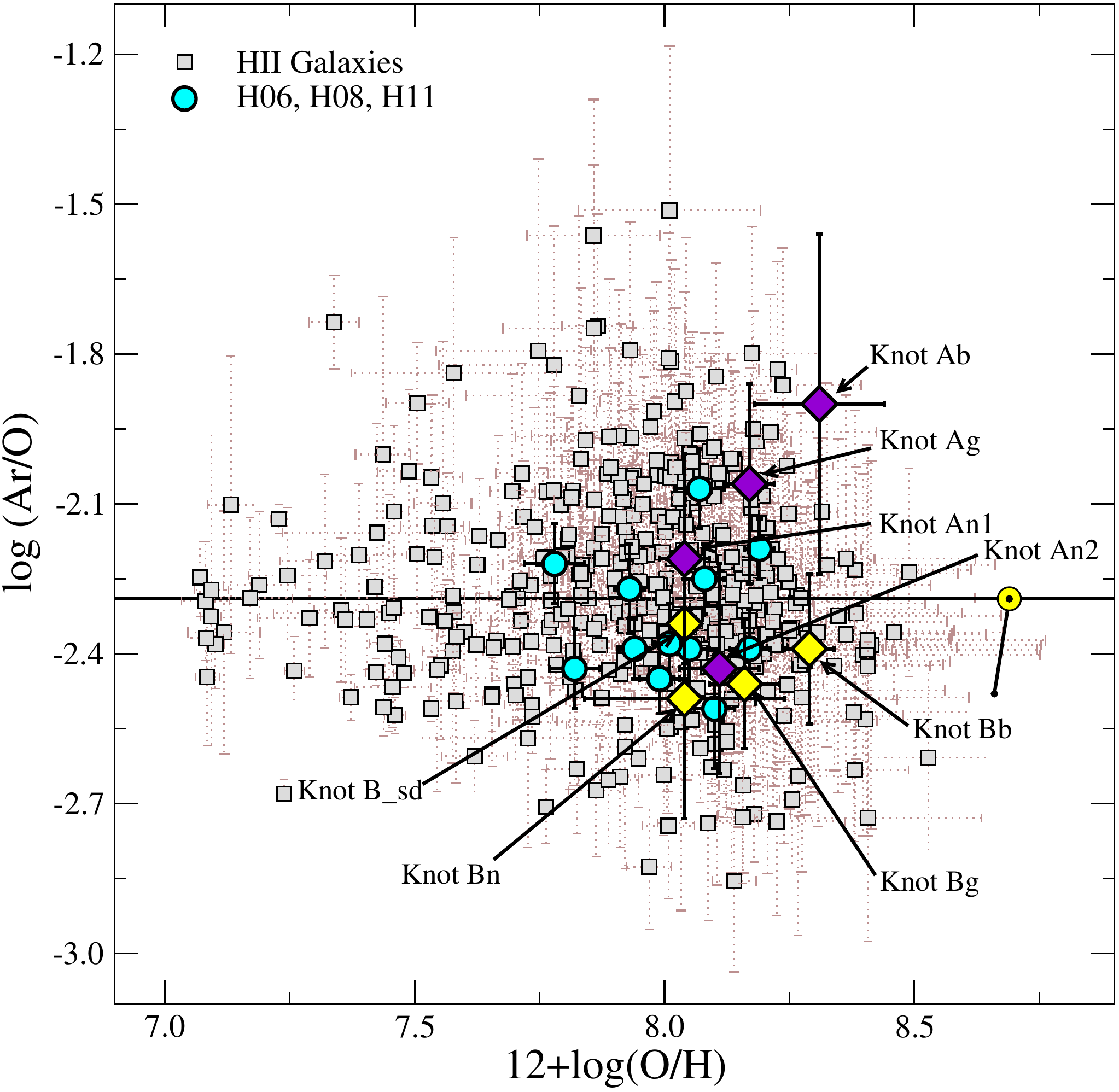}
\caption[Abundancias relativas al ox'igeno]{{\scriptsize  En los cuatro paneles se muestran los cocientes relativos al ox'igeno: S/O, N/O, Ne/O y Ar/O en funci'on de la abundancia de ox'igeno [12 + log (O/H)]. Con diamantes est'an representados los objetos observados: en amarillo, brote B, en violeta brote A. Los objetos observados en H\"agele et al.,~(\citeyear{2006MNRAS.372..293H,2008MNRAS.383..209H, 2011arXiv1101.4140H}) est'an representados por c'irculos turquesa, y las galaxias H\,{\sc ii} de la literatura recopilados por H\"agele et al.,~(\citeyear{2008MNRAS.383..209H}) para el S/O y N/O, y por P'erez-Montero et al.,~(\citeyear{2007MNRAS.381..125P}) para el Ne/O y el Ar/O con cuadrados grises. Con su habitual s'imbolo se representa los valores de las abundancias solares: del ox'igeno de Allende-Prieto et al.~(\citeyear{2001ApJ...556L..63A}),  del nitr'ogeno de Holweger~(\citeyear{2001AIPC..598...23H}) y Grevesse \& Sauval~(\citeyear{1998SSRv...85..161G}) para el azufre, el ne'on y el arg'on.}}
\end{center}
\end{figure*}



\subsection{Estructura de ionizaci'on}
\label{Estructura}
Desde hace varios a'nos, se vienen utilizado las l'ineas m'as intensas del espectro para deducir metalicidades a trav'es del uso de par'ametros emp'iricos y modelos. Encontrar una cantidad basada en la intensidad de las l'ineas de emisi'on y que se correlacione con la metalicidad no es una tarea sencilla, dado que la mayor'ia de las l'ineas de emisi'on dependen de la geometr'ia de la nebulosa a trav'es del par'ametro de ionizaci'on, o la ``dureza" de la radiaci'on ionizante. La intensidad de las l'ineas de emisi'on en el ox'igeno no tiene un comportamiento lineal con su abundancia debido a la importancia de dicho elemento en el proceso de enfriamiento de la nebulosa. De este modo, al aumentar la metalicidad, el enfriamiento se vuelve m'as eficaz, la temperatura electr'onica disminuye y con ella la intensidad de las l'ineas. Por ello las l'ineas m'as intensas tienen un comportamiento dual, de modo tal que a metalicidad baja son directamente proporcionales a la abundancia y a metalicidades altas ocurre justo lo contrario \citep{O89}.

El par'ametro de ionizaci'on, definido como el cociente entre el n'umero de fotones ionizantes y la densidad de 'atomos de hidr'ogeno, normalizado
por un factor, da una idea del grado de ionizaci'on. F'isicamente se interpreta como la velocidad m'axima que puede alcanzar el frente de ionizaci'on dentro de la nebulosa.
\[
U\,=\,\frac{Q(H)}{4\pi cn_{H}r^{2}} 
\]
donde $Q(H)$ es la tasa de producci'on de fotones de Lyman, $r$ la distancia a la estrella o c'umulo central, $n_{H}$, la densidad de 'atomos de hidr'ogeno neutros o ionizados y $c$ es la velocidad de la luz. Su valor efectivo es, esencialmente, lo que vale en el radio de Str\"omgren, $r_{S}$:
\[
\overline U\,=\,\frac{Q(H)}{4\pi cn_{H}r_{S}^{2}}\,=\,(Q(H)\epsilon n_{H})^{1/3} 
\]
donde $\epsilon$ es el llamado factor de llenado que indica las irregularidades en la densidad de la nebulosa e influye en la emisividad y en la profundidad 'optica. 
Si $\epsilon$\,$<<$\,1, como generalmente ocurre, la nebulosa es 'opticamente delgada \citep{EnriquePhDT}.

Para la determinaci'on del par'ametro de ionizaci'on, es bastante com'un usar cocientes de abundancias de especies en estados consecutivos de ionizaci'on. Sin embargo, a causa de la dificultad que tiene su determinaci'on es igual de 'util, y m'as f'acil de medir, el cociente de l'ineas de un mismo elemento en dos estados diferentes de ionizaci'on como es el caso de [\OII]/[\OIII] o de [\SII]/[\SIII]. Ambos tienen un comportamiento que tambi'en depende de la temperatura efectiva del campo de radiaci'on (Teff) de la estrella o c'umulo ionizante.
El cociente [\OII]/[\OIII] disminuye al aumentar Teff, al ir convirti'endose m'as O$^{+}$ en O$^{2+}$, y se considera v'alido en nebulosas altamente ionizadas.  El cociente [\SII]/[\SIII], por el contrario, aumenta al aumentar Teff debido a la ionizaci'on de S$^{2+}$ en S$^{3+}$ y el efecto es m'as pronunciado cuando el grado de ionizaci'on es mayor. Por tanto, se utiliza m'as en nebulosas de baja ionizaci'on (log U\,$\leq$\,2). Otros indicadores del par'ametro de ionizaci'on son los cocientes entre l'ineas colisionalmente excitadas y de recombinaci'on, como es el caso de [\SII]\,$\lambda$\,6717,6731\,\AA\,/\,H$\alpha$ o de [\OII]/H$\beta$ que son mucho m'as independientes de la temperatura efectiva y est'an menos afectados por enrojecimiento pero que, en cambio, dependen fuertemente de la metalicidad \citep{EnriquePhDT}.

V'ilchez \& Pagel~(\citeyear{1988MNRAS.231..257V}) demostraron que el cociente de las cantidades O$^{+}$/O$^{2+}$ y S$^{+}$/S$^{2+}$, llamado ``softness parameter" (cuyo significado en ingl'es es ``par'ametro de suavidad") y denotado con la letra griega $\eta$, est'a intr'insecamente relacionado con la forma del continuo ionizante y depende ligeramente de la geometr'ia. Suponiendo una geometr'ia esf'erica simple y un factor de llenado constante, el efecto geom'etrico puede ser representado por el par'ametro de ionizaci'on que, a su vez, puede estimarse a partir del cociente [\OII]/[\OIII]. Y, como se vi'o m'as arriba,  [\OII]/[\OIII] depende de Teff que, a su vez, depende de la metalicidad. 
El siguiente an'alisis s'olo es posible realizarlo en los brotes con alto S/N en sus espectros, como lo es en los brotes A y B (ver Tabla \ref{eta-etaprima}). 

La contraparte puramente observacional del par'ametro $\eta$ est'a dada por el par'ametro $\eta$'\ definido tambi'en por V'ilchez \& Pagel~(\citeyear{1988MNRAS.231..257V}) como el cociente entre las cantidades [\OII]/[\OIII] y [\SII]/[\SIII]. 
La relaci'on entre ambos par'ametros estar'a dada por el  siguiente ajuste:
\[ 
log\,\eta'\,=\,log\,\eta - 0.14 t_{e} - 0.16
\]
donde t$_{e}$ representa la temperatura electr'onica en unidades de 10$^{4}$K. La relaci'on entre $\eta$ y $\eta$'\, est'a dada a trav'es de la temperatura electr'onica, pero muy levemente, por lo que un cambio en la temperatura de 7000 a 14000 K implica un cambio en el logaritmo de un 0.1 dex, dentro de los errores de observaci'on. Siempre el log\,$\eta$'\, es inferior a log\,$\eta$.

En el panel superior de la Figura 4.5 se muestra la relaci'on entre log(O$^+$/O$^{2+}$) y log(S$^{+}$/S$^{2+}$) las cuales fueron derivadas usando m'etodos directos, modelos de fotoionizaci'on y relaciones emp'iricas, seg'un corresponda el caso, para las medidas globales y las diferentes componentes cinem'aticas de los brotes A y B (representados con diamantes violetas y amarillos, respectivamente). Por otra parte, en el panel inferior se representa el cociente entre el log\,([\OII]/[\OIII]) vs. log\,([\SII]/[\SIII]), el cual no requiere el expl'icito conocimiento de la temperaturas de l'inea involucrada en la derivaci'on de los cocientes i'onicos, y por lo tanto no depende del m'etodo para estimar estas temperaturas. Los objetos estudiados en H\"agele et al.,~(\citeyear{2006MNRAS.372..293H,2008MNRAS.383..209H,2011arXiv1101.4140H}) son representados con c'irculos turquesas, y las galaxias H{\sc ii}, cuyos datos fueron obtenidos de la literatura con cuadrados (ver descripci'on y referencias en H\"agele et al.,~\citeyear{2008MNRAS.383..209H}). Las galaxias H{\sc ii} se localizan en la regi'on donde el log\,$\eta$ est'a entre los valores -0.35 y 0.2, correspondiente a valores altos de la temperatura efectiva del campo de radiacion ionizante. 
En la Figura 4.5 las l'ineas diagonales, tanto en el diagrama superior como en el inferior, representan valores constantes de log\,$\eta$ y de log\,$\eta$'\,, respectivamente. En la Tabla \ref{eta-etaprima} se muestran los valores de los cocientes log(O$^+$/O$^{2+}$) y log(S$^{+}$/S$^{2+}$), y los cocientes log\,([\OII]/[\OIII]) y log\,([\SII]/[\SIII]) para las distintas componentes y medidas globales de los brotes A y B. Tanto en la Tabla como en la Figura 4.5, y a modo de comparaci'on, se agregaron los valores correspondientes a la medida global del brote B observado con el modo de ranura larga. Todos los objetos est'an se'nalados con sus respectivos nombres y modos de observaci'on.

\begin{figure*}
\begin{center}
\label{figH15etaSO-ion}
\vspace{1.0cm}
\includegraphics[trim=0cm 0cm 0cm 0cm,clip,angle=0,width=8cm,height=8cm]{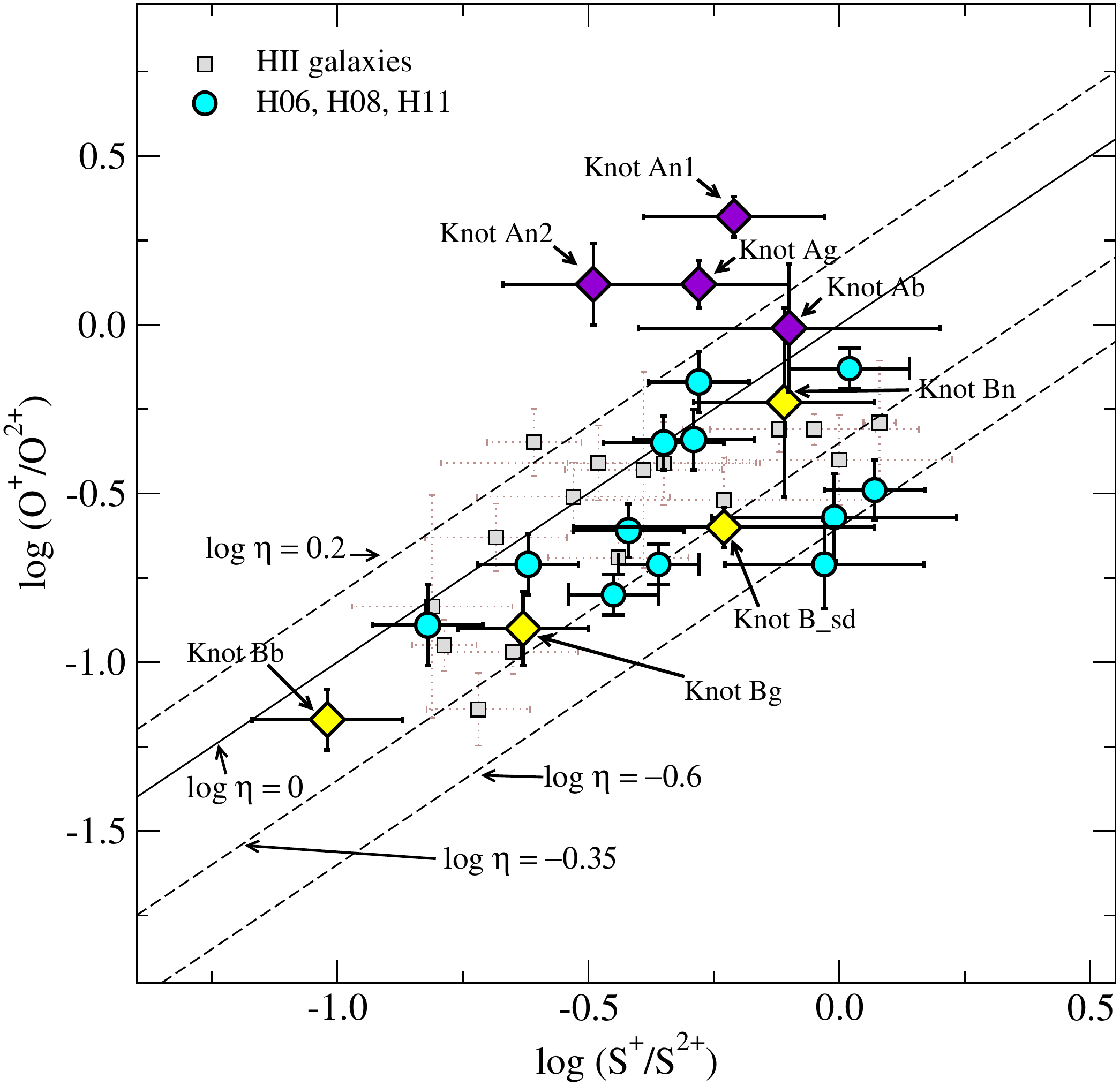}
\includegraphics[trim=0cm 0cm 0cm 0cm,clip,angle=0,width=8cm,height=8cm]{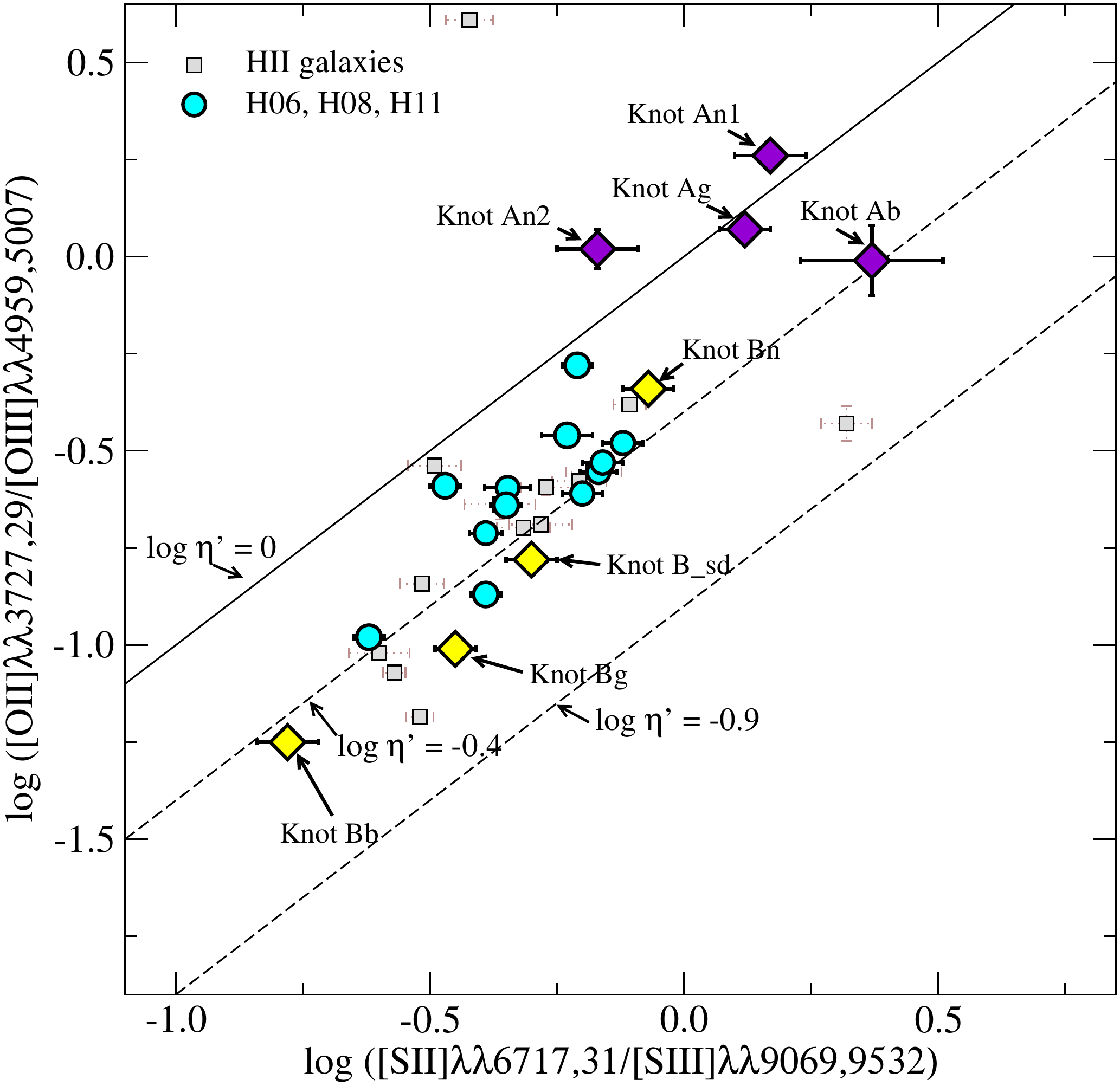}
\caption[Diagramas: log\,$\eta$ y log\,$\eta$'\, de los brotes A y B]{{\scriptsize Panel superior: log(O$^+$/O$^{2+}$) vs. log(S$^{+}$/S$^{2+}$) para las medidas globales y las diferentes componentes cinem'aticas de los brotes A y B, representados con diamantes violetas y amarillos, respectivamente. Para el brote B, se agrega el dato de dispersi'on simple. Las diagonales en este diagrama corresponden a valores constantes de log\,$\eta$. En el Panel inferior: log\,([\OII]/[\OIII]) vs. log\,([\SII]/[\SIII]), tambi'en para las mismas medidas. Las diagonales en este otro diagrama corresponden a valores constantes de log\,$\eta$'\,. En ambos diagramas, los valores est'an representados junto a objetos estudiados por H\"agele et al.,~(\citeyear{2006MNRAS.372..293H,2008MNRAS.383..209H, 2011arXiv1101.4140H}), representados con c'irculos turquesas, y a los datos de las galaxias H{\sc ii}, representados con cuadrados, y cuyos valores fueron obtenidos de la literatura \citep[ver descripci'on y referencias en][]{2008MNRAS.383..209H}. El sub'indice g se refiere a la medida global, el n se refiere a la componente angosta (1 o 2, para el brote A), el sub'indice b se refiere a la componente ancha, y el sub'indice sd se refiere al dato de dispersi'on simple}}
\end{center}
\end{figure*}

\begin{table}
{\small
\centering
\begin{center}
\caption[log\,$\eta$ y log\,$\eta$'\, de los brotes A y B]{{\scriptsize Cocientes log(S$^{+}$/S$^{2+}$) y log(O$^+$/O$^{2+}$) para obtener log\,$\eta$, y cocientes log\,([\SII]/[\SIII]) y log\,([\OII]/[\OIII]) para obtener log\,$\eta$'\,, para las distintas componentes y medidas globales de los brotes A y B. Se agrega los correspondientes valores para el brote B de dispersi'on simple (DS).}}
\vspace{0.3cm}
\label{eta-etaprima}
\begin{tabular}{@{}l@{\hspace{0.1cm}}lcccc@{}}
\hline
brote & & log(S$^{+}$/S$^{2+}$) & log(O$^+$/O$^{2+}$)&  log\,([\SII]/[\SIII]) &log\,([\OII]/[\OIII])\\
\hline
H15\,A &global& -0.28$\pm$0.18  & 0.12$\pm$0.07  &   0.12$\pm$0.05 &0.07$\pm$0.03 \\
          & angosta\,1&-0.21$\pm$0.18    &0.32$\pm$0.06    &0.17$\pm$0.07&0.26$\pm$0.03 \\
           &angosta\,2 &-0.49$\pm$0.18   & 0.12$\pm$0.12   &-0.17$\pm$0.08&0.02$\pm$0.05\\
           &ancha &-0.10$\pm$0.30  &-0.01$\pm$0.19   &0.37$\pm$0.14&-0.01$\pm$0.09 \\
\hline
H15\,B &global& -0.63$\pm$0.13  & -0.90$\pm$0.11 &  -0.45$\pm$0.04  & -1.0$\pm$0.01 \\
          & angosta &-0.11$\pm$0.18    & -0.23$\pm$ 0.28&-0.07$\pm$0.05  & -0.34$\pm$0.02\\    
          &ancha &-1.02$\pm$0.15  & -1.17$\pm$0.09& -0.78$\pm$0.06  & -1.25$\pm$0.02  \\
          &DS &-0.23$\pm$0.30 &-0.60$\pm$0.06&-0.30$\pm$0.05&   -0.78$\pm$0.01\\
\hline
\end{tabular}
\end{center}}
\end{table}

De la Figura  4.5 se desprende que el brote B tiene una estructura de ionizaci'on (o valor de $\eta$) muy similar comparando entre si la medida global y las distintas componentes cinem'aticas, e incluso si estos valores son comparados con los valores del espectro de dispersi'on simple del mismo brote, y siempre considerando los errores, lo que implica que la temperatura del campo de radiaci'on ionizante en el brote es muy parecida para las distintas componentes cinem'aticas. Esto podr'ia deberse a que el c'umulo estelar responsable de la ionizaci'on de cada componente cinem'atica ser'ia el mismo. 
En el brote A sucede algo similar, dado que tanto las componentes cinem'aticas como la medida global, presentan una estructura de ionizaci'on muy parecida y la posici'on en ambos diagramas es compatible, aunque los valores en el diagrama de $\eta$ est'an desplazados ligeramente hacia temperaturas efectivas menores del campo de radiaci'on ionizante. La 'unica componente que se distancia del resto del grupo, estando en la zona de alta ionizaci'on, es la componente ancha, pero es similar si tenemos en cuenta los errores grandes en la determinaci'on de los cocientes. Esta diferencia respecto al brote B podr'ia deberse a un estado evolutivo diferente dado que todas las componentes del brote A se ubican hacia una regi'on con temperaturas efectivas del campo de radiaci'on ionizante menores respecto a las componentes del brote B, en acuerdo con la presencia de un c'umulo estelar ionizante m'as viejo, m'as evolucionado, en el brote A, como fue sugerido por L'opez-S'anchez \& Esteban~(\citeyear{2010A&A...516A.104L}). En el brote A los valores de $\eta$ y $\eta$'\, son mayores, en casi todos los casos, a los t'ipicos valores de las galaxias H\,{\sc ii} \citep{2008MNRAS.383..209H}, solamente la componente ancha muestra valores de este tipo, sobre todo en el diagrama $\eta$'\,. En cambio para el brote B, los valores de $\eta$ y $\eta$'\, est'an en el rango encontrado para las galaxias H\,{\sc ii}.

\subsection{Abundancias qu'imicas usando par'ametros emp\'{\i}ricos}
\label{sec:Calibradores}
Las l\'{\i}neas de recombinaci\'on emitidas por los metales proporcionan las abundancias m\'as precisas gracias a su d\'ebil dependencia con la temperatura electr'onica en el interior de la nebulosa. Desafortunadamente, la gran mayor'ia de las l\'{\i}neas intensas de emisi\'on que se observan en el espectro 'optico y/o que se pueden medir con suficiente precisi'on son de naturaleza colisional y sus intensidades dependen exponencialmente de la temperatura electr'onica. En principio, esta temperatura puede ser determinada a partir de los apropiados cocientes de l\'{\i}neas. Sin embargo, estos cocientes necesitan de la detecci\'on y la medida de las l\'{\i}neas aurorales, que, como ya se ha citado anteriormente, son intr\'{\i}nsecamente d\'ebiles y no son detectables en algunas situaciones. Es este el mismo caso que sucede en las regiones con alto contenido met\'alico, donde la eficiencia del enfriamiento ejercido por los metales provoca que dichas l\'{\i}neas sensibles a la temperatura electr'onica sean indetectables, al igual que en regiones H\,{\sc ii} en galaxias distantes y objetos con bajo brillo superficial donde no se las puede detectar debido al bajo S/N de las mismas. En estos objetos, cuya derivaci'on directa de las temperaturas electr'onicas no es posible, son utilizados diferentes calibradores emp'iricos basados en las intensidades de las l\'{\i}neas de emisi'on m'as conspicuas. 

Los m'etodos emp'iricos se basan en las propiedades de enfriamiento de las nebulosas ionizadas, esto se traduce como una relaci'on entre las intensidades de l'ineas de emisi'on y la abundancia del ox'igeno. Los m'etodos emp'iricos de derivaci'on de abundancias  dependientes de las l'ineas intensas que han sido ampliamente estudiados en la literatura, se basan en la calibraci'on directa de la intensidad relativa de algunas l'ineas brillantes de emisi'on versus la abundancia de algunos iones relevantes presentes en una nebulosa \citep[ver por ejemplo,][y referencias en ellos]{2008ApJ...677..201G,2009A&A...507.1291C,2009PhDT........15G,2010MNRAS.408.2234G}. 

Entre los principales estimadores emp\'{\i}ricos utilizados en la literatura, en este apartado se analizan los siguientes par'ametros emp'iricos usados cuando fue posible disponer de las l'ineas necesarias: O$_{23}$, S$_{23}$, O$_{3}$N$_{2}$, S$_{3}$O$_{3}$, N$_{2}$ y Ar$_{3}$O$_{3}$. A continuaci'on se detallan cada uno de ellos y luego se realiza una comparaci'on entre los resultados obtenidos para los diferentes datos analizados en esta Tesis.

\subsubsection*{Par'ametro O$_{23}$}
El par'ametro O23, tambi'en conocido como R$_{23}$, propuesto por Pagel et al.~(\citeyear{1979MNRAS.189...95P}) se define como:
\[
O_{23}\,=\,\frac{I(3727,29{\textsc \AA})+I(4959{\textsc \AA})+I(5007{\textsc \AA})}{I(H\beta)}
\] \noindent 
El mayor inconveniente que existe en utilizar el O$_{23}$ es el comportamiento no lineal con la metalicidad, este par'ametro tiene una naturaleza dual.
El comportamiento dual de la calibraci'on llev'o a Skillman~(\citeyear{1989ApJ...347..883S}) a hablar en t'erminos de las ramas superior e inferior de la calibraci'on y a utilizar el valor del cociente de las l'ineas de [\OIII]\,$\lambda$\,5007\,\AA/[\NII]\,$\lambda$\,6584\,\AA\ como un criterio para decidir a qu'e rama pertenece un objeto y usar la curva de calibraci'on respectiva. Entonces es posible distinguir tres regiones diferentes: una rama m'as baja donde O$_{23}$ aumenta con una abundancia cada vez mayor; una rama superior en la que ocurre lo contrario, el par'ametro disminuye con una abundancia cada vez mayor y una regi'on de cambio de orientaci'on, que se produce en torno a log O$_{23}$ \,$\approx$\,0.8 y 12+log(O/H)\,$\approx$\,8.0.
La dependencia del O$_{23}$ con el par'ametro de ionizaci'on da lugar a incertidumbres para la metalicidad en aquellos objetos que se encuentran en la rama inferior, aunque la dispersi'on nunca es tan elevada como en la zona de cambio de orientaci'on.
 Para la rama inferior tanto Skillman como McGaugh~(\citeyear{1991ApJ...380..140M}) tuvieron en cuenta la dependencia de O$_{23}$ con el par'ametro de ionizaci'on. Pero es McGaugh quien encuentra una mejor calibraci'on basada en los modelos te'oricos en los que la ionizaci'on es proporcionada por c'umulos estelares de diferentes metalicidades, incluyendo una correcci'on para la variaci'on del par'ametro de ionizaci'on. 
En este trabajo se calculan la abundancia de ox'igeno 12+log(O/H) utilizando las calibraciones de Kobulnicky et al.~(\citeyear{1999ApJ...514..544K}) basadas en los modelos de McGaugh. Y se utiliz'o el valor del cociente entre las l'ineas [\NII]$\lambda$6584\AA\ y [\OII]$\lambda$3727\AA\ y el par'ametro N$_{2}$ (\,=\,[\NII]$\lambda$6584\AA\ / H$\beta$, abajo descripto), como un criterio para decidir a qu'e rama pertenece un objeto y as'i poder usar la curva de calibraci'on respectiva \citep{2008ApJ...681.1183K}. Si el logaritmo del primer cociente es menor a -1.2 entonces pertenece a la rama inferior, y si el logaritmo del par'ametro N$_{2}$ es menor a -1.3 (entre -1.1 y -1.3 no es posible decir nada) pertenece a la rama superior. La incerteza del ajuste da, para la rama inferior un error cuadr'atico medio de 0.13 dex y para la rama superior 0.19 dex.
\[
12 + log(O/H)_{lower}\,=\,12 - 4.944 + 0.767 x + 0.602 x{^2} - y (0.29 + 0.332 x - 0.331 x{^2})
\] \noindent
\[
12 + log(O/H)_{upper}\,=\,12 - 2.939 - 0.2 x - 0.237 x{^2} - 0.305 x{^3} - 0.0283 x{^4} - 
\] \noindent
\[
y (0.0047 - 0.0221 x - 0.102 x{^2} - 0.0817 x{^3} - 0.00717 x{^4})
\] \noindent
donde \[ x\,=\,log(O_{23})\] 
  \[
  y\,=\,log(O_{23})\,=\,log \left(\frac{[O{\textsc {iii}}]4959+5007}{[O{\textsc {ii}}]3727,29}\right)
  \] \noindent

\subsubsection*{Par'ametro S$_{23}$}

El par'ametro S$_{23}$ fue definido inicialmente por V'ilchez \& Esteban~(\citeyear{1996MNRAS.280..720V}) como:
\[
S_{23}\,=\,\frac{I(6717{\textsc \AA})+I(6731{\textsc \AA})+I(9069{\textsc \AA})+I(9532{\textsc \AA})}{I(H\beta)}
\] \noindent
y usado por primera vez como calibrador emp'irico de la abundancia de azufre por D'iaz \& P'erez-Montero~(\citeyear{2000MNRAS.312..130D}).
Una de las ventajas de este par'ametro es que las l'ineas de azufre son intensas, incluso para los objetos de mayor metalicidad. Adem'as, el m\'aximo de emisividad del azufre ocurre a temperaturas menores que en el caso del ox'igeno, con lo cual un cambio de orientaci'on en la curva del S$_{23}$ frente a la abundancia del ox'igeno se produce a metalicidad solar y no a 12+log(O/H)\,=\,8.0 como se observa en el par'ametro O$_{23}$.
Como las l'ineas de [\SIII] cumplen con la relaci'on te'orica entre ambas de: I(9532\AA)\,$\approx$\,2.44$\cdot$\,I(9069\AA), s'olo se necesita disponer de una de ellas. 

Para calcular la abundancia de ox'igeno (12+log(O/H)) se utiliz'o la calibraci'on del par'ametro emp'irico S$_{23}$ dada por P'erez-Montero \& D'iaz~(\citeyear{2005MNRAS.361.1063P}):
\[
12 + log(O/H)\,=\,8.15 + 1.85 log S_{23} + 0.58 (log S_{23}){^2}
\] \noindent
con una dispersi'on de 0.2 dex en todo el rango de abundancias, disminuyendo a 0.10 dex para las galaxias H\,II.

Tambi'en es posible calcular con este par'ametro, la abundancia de azufre 12+log(S/H) utilizando la calibraci'on del par'ametro emp'irico S$_{23}$ dada por P'erez-Montero et al.~(\citeyear{2006A&A...449..193P}):
\[
12 + log(S/H)\,=\,6.622 + 1.860 log S_{23} + 0.382 (log S_{23}){^2}
\] \noindent
con una incerteza de 0.19 dex, pero aqu'i no fue usado con este fin.

\subsubsection*{Par'ametro SO$_{23}$}

El par'ametro SO$_{23}$ es la combinaci'on entre los par'ametros S$_{23}$ y O$_{23}$ denominado y definido por D'iaz \& P'erez-Montero~(\citeyear{2000MNRAS.312..130D}).
El mismo aumenta mon'otonamente con la abundancia del ox'igeno hasta el r'egimen de abundancia sobre-solar. La calibraci'on encontrada por P'erez-Montero et al.~(\citeyear{2005MNRAS.361.1063P}) est'a dada por la relaci'on:
\[
12 + log(O/H)\,=\,9.09 + 1.03 x - 0.23 x{^2}
\] \noindent
donde 
\[
x\,=\,log\left(\frac{S_{23}}{O_{23}}\right)
\] \noindent
con una dispersi'on de 0.27 dex.

\subsubsection*{Par'ametro O$_{3}$N$_{2}$}
El par'amentro O$_{3}$N$_{2}$ est'a definido por la relaci'on:
\[
O_{3}N_{2}\,=\,log\left(\frac{I(5007{\textsc \AA})/I(H\beta)}{I(6584{\textsc \AA})/I(H\alpha)}\right)
\] \noindent
dada por Alloin et al.~(\citeyear{1979A&A....78..200A}). Este calibrador es independiente del enrojecimiento y de la calibraci'on en flujo. El uso del cociente $I(H\alpha)/I(H\beta)$ en la definici'on indica que las l'ineas fueron medidas respecto a la l'inea de recombinaci'on m'as pr'oxima para minimizar as'i la dependencia con el enrojecimiento
o la incerteza en la calibraci'on en flujo. El ajuste para el c'alculo de la abundancia de ox'igeno, 12+log(O/H), que m'as se usa es el de Pettini \& Pagel~(\citeyear{2004MNRAS.348L..59P}) v'alida solo para O$_{3}$N$_{2}$${<}$2, y con un grado de incerteza de 0.25 dex.
\[
12 + log(O/H)\,=\,8.73 - 0.32 O_{3}N_{2}
\] \noindent

\subsubsection*{Par'ametro S$_{3}$O$_{3}$}
El par'ametro S$_{3}$O$_{3}$ fue descripto por Stasinska~(\citeyear{2006A&A...454L.127S}) y est'a dado por:

\[
S_{3}O_{3}\,=\,\frac{I(9069{\textsc \AA})}{I(5007{\textsc \AA})}
\] \noindent

Su calibraci'on est'a dada por:
\[
12 + log(O/H)\,=\,8.70 + 0.28 x + 0.03 x{^2} + 0.1 x{^3}
\] \noindent
donde 
\[
x\,=\,log S_{3}O_{3}
\] \noindent
con una desviaci'on est'andar de 0.25 dex.

\subsubsection*{Par'ametro N$_{2}$}
El par'ametro N2 se define como:
\[
N_{2}\,=\,log\left(\frac{I(6584{\textsc \AA})}{I(H\alpha)}\right)
\] \noindent
definido por Storchi-Bergmann et al.~(\citeyear{1994ApJ...429..572S}). Este estimador, en contraste al O$_{23}$, tiene un car'acter univaluado en relaci'on a la metalicidad. Adem'as, es pr'acticamente independiente de la correci'on por enrojecimiento y de la calibraci'on en flujo de los espectros, dada la proximidad en longitud de onda de las dos l'ineas involucradas. Sin embargo, este par'ametro tiene una alta dispersi'on asociada a los par'ametros de la nebulosa, como por ejemplo, el par'ametro de ionizaci'on y la temperatura de la radiaci'on ionizante. Tambi'en existe un alto grado de incertidumbre debido a la variaci'on de la relaci'on de abundancias entre nitr'ogeno y ox'igeno, N/O, que a bajas metalicidades es constante al ser el nitr'ogeno de origen primario y no depender de la metalicidad. Para metalicidades mayores, en torno 12 + log(O/H)\,=\,8.0, N/O aumenta con la metalicidad debido a la aparici'on del nitr'ogeno secundario.
La calibraci'on m'as usada para calcular la abundancia de ox'igeno es la de Denicol'o, Terlevich \& Terlevich~(\citeyear{2002MNRAS.330...69D}) la cual tiene asociada una incerteza de 0.23 dex.
\[
12 + log(O/H)\,=\,(9.12 \pm 0.05) + (0.73 \pm 0.10) N_{2}
\] \noindent
\subsubsection*{Par'ametro Ar$_{3}$O$_{3}$}

Este par'ametro tambi'en fue definido por Stasinska~(\citeyear{2006A&A...454L.127S}), en busca de las mismas buenas propiedades que el calibrador S$_{3}$O$_{3}$, y se lo define como:
\[
Ar_{3}O_{3}\,=\,\frac{I(7136{\textsc \AA})}{I(5007{\textsc \AA})}
\] \noindent
y su calibraci'on est'a dada por :
\[
12 + log(O/H)\,=\,8.91 + 0.34 x + 0.27 x{^2} + 0.20 x{^3}
\] \noindent
donde 
\[
x\,=\,log Ar_{3}O_{3}
\] \noindent
con una desviaci'on est'andar de 0.23 dex.\\

A continuaci'on se analizar'an los resultados obtenidos para la determinaci'on de la abundancia de ox'igeno con los diferentes par'ametros emp'iricos en los diferentes objetos estudiados. Se comparan estas abundancias con las derivadas previamente usando las temperaturas estimadas. En la Figura 4.6 se puede ver las abundancias de ox'igeno y sus incertezas para cada brote observado en Haro\,15 calculadas usando los diferentes calibradores emp'iricos para las diferentes componentes y la medida global. De arriba a abajo: brotes A, B, C, E y F. De izquierda a derecha los diferentes par'ametros emp'iricos: S$_{23}$, O$_{23}$, SO$_{23}$, O$_{3}$N$_{2}$, S$_{3}$O$_{3}$, N$_{2}$ y Ar$_{3}$O$_{3}$. Con estrella azul se muestra la medida global en el 'echelle; con barra de error en negro los datos de dispersi'on simple; el tri'angulo turquesa hacia arriba se refiere a la componente angosta del 'echelle (para el brote A: el tri'angulo lleno representa la componente angosta\,1 y el cuadrado verde lleno representa la componente angosta\,2); el tri'angulo rosa hacia abajo es la componente ancha. Las l'ineas horizontales continuas representan el rango de la abundancia (teniendo en cuenta su error) derivada con la temperatura obtenida por el m'etodo directo (solo para el brote B en el modo de dispersi'on simple); la l'inea horizontal a trazos representa el rango de la abundancia (teniendo en cuenta su error) calculada usando la temperatura derivada de los modelos o de las relaciones emp'iricas entre temperaturas (los colores de estas l'ineas horizontales son los correspondientes a cada medida o componente, ya sea para los datos de dispersi'on simple (en color negro) o para las diferentes componentes (turquesa, verde o magenta) y para la medida global (azul) de los datos 'echelle). Por simplicidad, no se pondr'a el rango de la abundancia correspondiente a la medida global en los brotes A y B ya que su valor es el promedio pesado por luminosidad de las abundancias derivadas para las distintas componentes cinem'aticas. Luego, s'olo se grafica este valor para el brote E dado que es la 'unica estimaci'on previa de su abundancia que se pudo realizar. En el caso del brote C es posible obtener y mostrar la cantidad derivada de los datos de dispersi'on simple, y en el caso del brote F es el 'unico caso en que no se tiene una estimaci'on previa de la abundancia total de ox'igeno.

En la siguiente discusi'on de los resultados por brotes, siempre se tendr'an en cuenta los errores. 
\begin{description}
\item En el brote A existe una buena concordancia entre la abundancia total de ox'igeno de la medida global determinada usando los diferentes par'amentro emp'iricos, con la respectiva abundancia derivada usando las temperaturas obtenidas de los modelos y relaciones emp'iricas (aunque no se muestra el valor con l'ineas horizontales en la Figura). La 'unica que escapa a este resultado es la determinaci'on de la abundancia total de ox'igeno con el par'ametro emp'irico N$_{2}$, que presenta una diferencia de 0.14 dex respecto a la determinada con la temperatura obtenida con los modelos y relaciones emp'iricas si tenemos en cuenta la barra de error. En la componente angosta 1 existe una buena concordancia en la determinaci'on de la abundancia usando los par'ametros emp'iricos: S$_{23}$, O$_{23}$ y SO$_{23}$, con la abundancia determinada usando las temperaturas obtenidas de los modelos y relaciones emp'iricas (l'inea horizontal discontinua turquesa). En cambio, la abundancia determinada para la misma componente angosta 1 usando los par'ametros emp'itricos: O$_{3}$N$_{2}$, S$_{3}$O$_{3}$ y N$_{2}$ y Ar$_{3}$O$_{3}$ es mayor, aunque solamente para N$_{2}$ se encuentra una gran diferencia ($\sim$0.25 dex) y en los otros tres la diferencia entre las barras de error es del orden de 0.1 dex. La abundancia determinada para la componente angosta 2 usando los par'amentros emp'iricos presenta buena concordancia con la abundancia derivada previamente usando las temperaturas obtenidas de los modelos y relaciones emp'iricas, excepto en el par'ametro emp'irico N$_{2}$, que sobreestima el valor de la abundancia de ox'igeno en $\sim$0.15 dex. La abundancia determinada para la componente ancha usando los diferentes par'ametros emp'iricos, a excepci'on del S$_{23}$ que es levemente mayor si se tiene en cuenta la barra de error, son muy similares a la abundancia determinada usando las temperaturas obtenidas de los modelos y relaciones emp'iricas (l'inea horizontal discontinua magenta). 

 \item En el brote B, en el dato 'echelle, la abundancia de ox'igeno de la medida global determinada a partir de los par'ametros emp'iricos O$_{23}$, O$_{3}$N$_{2}$ y N$_{2}$ muestran un valor similar al derivado usando las temperaturas (aunque no se muestra este valor con l'ineas horizontales en la Figura). En cambio, la abundancia determinada a partir de los par'ametros emp'iricos S$_{23}$, SO$_{23}$, S$_{3}$O$_{3}$ y Ar$_{3}$O$_{3}$ presenta valores menores a la abundancia determinada usando las temperaturas, siendo el S$_{23}$ y el SO$_{23}$ (relacionados entre s'i) los que muestran las mayores discrepancias. La componente angosta presenta valores similares en la determinaci'on de la abundancia de ox'igeno usando los diferentes par'ametros emp'iricos y la determinada derivada usando las temperaturas (l'inea horizontal discontinua turquesa, que en este caso es discontinua porque T$_e$([\OII]) fue determinada a trav'es de modelos). La abundancia de ox'igeno obtenida para la componente ancha y determinada a partir de los par'ametros emp'iricos es siempre menor (excepto para el O$_{23}$ que da 8.17$\pm$0.13) a la determinada usando las temperaturas obtenidas (l'inea horizontal discontinua magenta). Esta diferencia es mucho m'as marcada para el par'ametro SO$_{23}$, de aproximadamente 0.75 dex (teniendo en cuenta las barras de error grandes, sobre todo en el par'ametro emp'irico), aunque se observan  diferencias muy grandes tambi'en para el S$_{23}$ y el Ar$_{3}$O$_{3}$ (0.4 dex considerando las barras de error), y una diferencia de $\sim$0.2 dex para los tres restantes (O$_{3}$N$_{2}$, S$_{3}$O$_{3}$ y N$_{2}$). En el caso del dato de dispersi'on simple, los valores de la abundancia total de ox'igeno  derivada con los par'ametros emp'iricos est'an en muy buen acuerdo con la abundancia derivada usando la temperatura obtenida por el m'etodo directo (l'inea horizontal continua negra), excepto para los par'ametros emp'iricos S$_{23}$ y SO$_{23}$ que dan valores de la abundancia de ox'igeno menores, pero marginalmente coincidentes en el primer caso, y aproximadamente 0.1 dex inferior para el segundo, considerando la barra de error.
 
\item En el brote C, los valores de las abundancias determinadas para las componentes ancha y angosta, y para la medida global del 'echelle utilizando los par'ametros emp'iricos O$_{3}$N$_{2}$ y N$_{2}$, est'an en buen acuerdo con el valor de la abundancia total del ox'igeno estimada para el mismo brote con los datos de dispersi'on simple. Adem'as, concuerdan con el valor de la abundancia de ox'igeno determinada usando las temperaturas obtenidas de modelos y relaciones emp'iricas para los datos de dispersi'on simple (l'inea horizontal discontinua negra). S'olo para los datos de ranura larga, el valor de la abundancia determinada usando el par'ametro emp'irico O$_{23}$ presenta un valor ligeramente menor al derivado usando las temperaturas.

\item En el brote E existe una buena concordancia entre las abundancias determinadas con los par'ametros emp'iricos  O$_{3}$N$_{2}$ y N$_{2}$ para las dos componentes angostas y la medida global, y coinciden con la abundancia determinada con las temperaturas estimadas de los modelos y relaciones emp'iricas (l'inea horizontal discontinua azul), excepto para la angosta 2 con el par'ametro N$_{2}$. La abundancia determinada con el par'ametro O$_{23}$ en la medida global est'a un poco por debajo de los dem'as, pero en buen acuerdo, tambi'en con la abundancia total de ox'igeno determinada a partir de las temperaturas.

\item En el brote F, la estimaci'on de la abundancia para la medida global con los par'ametros emp'iricos O$_{3}$N$_{2}$ y N$_{2}$ es similar, mostrando valores muy parecidos a los encontrados para los dem'as brotes de formaci'on estelar.
\end{description}

Se debe tener en cuenta que las relaciones emp'iricas usadas aqu'i como para los par'ametros $\eta$ y $\eta$'\ fueron derivadas a partir de medidas para regiones completas y que nunca se ha hecho un an'alisis y/o determinaci'on por componentes. Este es un trabajo que merece una discusi'on y un tratamiento m'as completo que est'a m'as all'a de lo que se puede realizar con los presentes datos. Es necesario realizar nuevas observaciones y ampliar la muestra de regiones con determinaciones de abundancias por componentes y tambi'en, dentro de lo posible, con una determinaci'on directa de la abundancia i'onia de O$^{+}$, lo que dar'ia una medida directa de O para las distintas componentes.


\begin{figure*}
\begin{center}
\label{calibradores emp'iricos}
\includegraphics[trim=0cm 0cm 0cm 0cm,clip,angle=0,width=13cm,height=18cm]{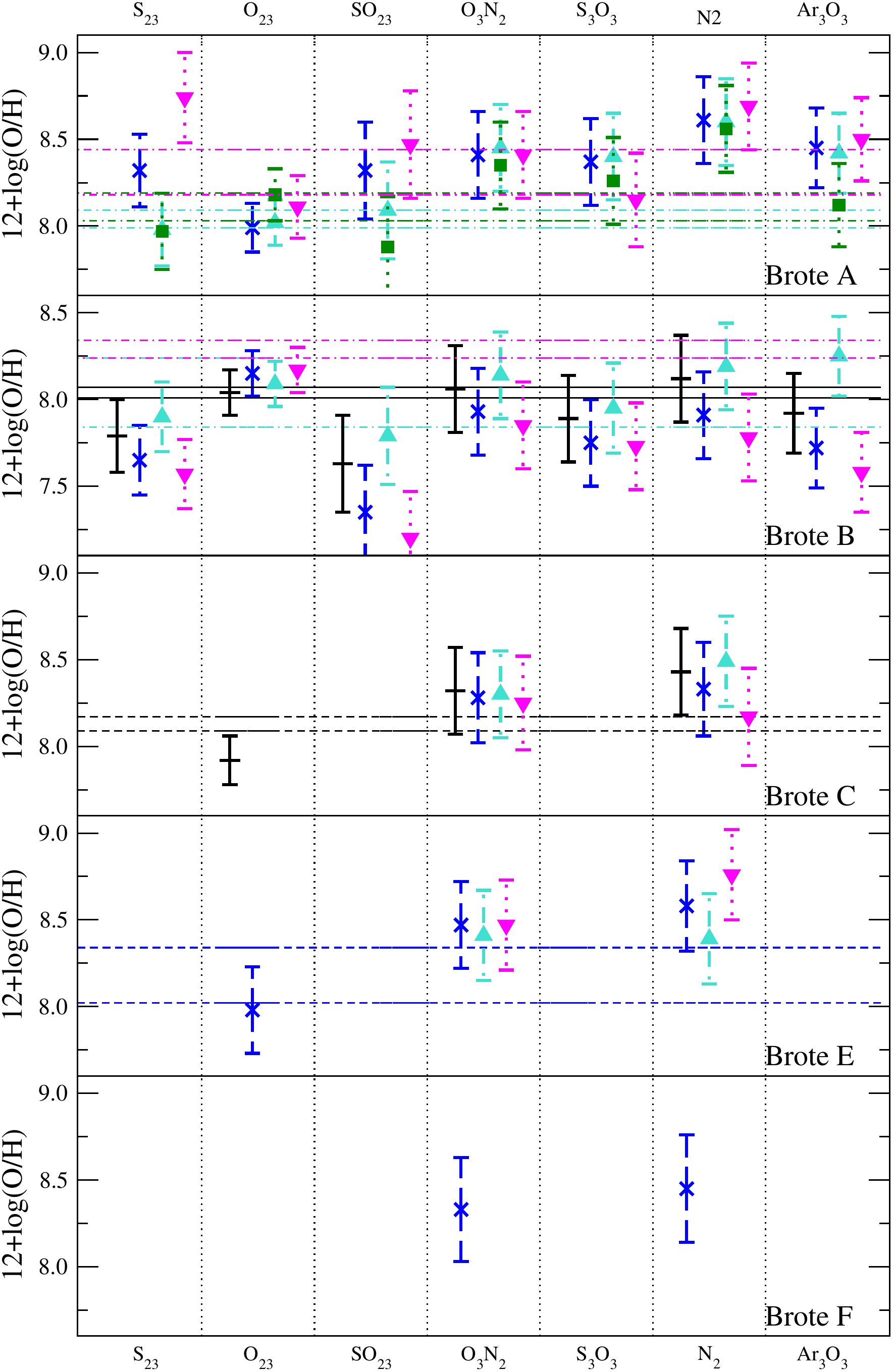}
\caption[Abundancias de ox'igeno determinadas con los par'ametros emp'iricos]{\tiny Abundancias de ox'igeno y sus incertezas para cada brote observado en Haro\,15 calculadas usando los diferentes par'ametros emp'iricos para las diferentes componentes y la medida global. De arriba a abajo: brotes A, B, C, E y F. De izquierda a derecha los diferentes par'ametros emp'iricos: S$_{23}$, O$_{23}$, SO$_{23}$, O$_{3}$N$_{2}$, S$_{3}$O$_{3}$, N$_{2}$ y Ar$_{3}$O$_{3}$. Con estrella azul se muestra la medida global en el 'echelle; con barra de error en negro los datos de dispersi'on simple; el tri'angulo turquesa hacia arriba se refiere a la componente angosta del 'echelle (para el brote A: el tri'angulo lleno representa la componente angosta\,1 y el cuadrado verde lleno representa la componente angosta\,2); el tri'angulo rosa hacia abajo es la componente ancha. Las l'ineas horizontales  representan el rango de la abundancia total de ox'igeno previamente determinada (ver texto).}
\end{center}
\end{figure*}

\clearpage
\section{Resumen y conclusiones}
\label{sec:resumen y conclusiones}

En este Cap'itulo se ha realizado un an\'alisis de las caracter\'{\i}sticas f'isicas del gas ionizado discriminado para cada componente y para el flujo global de las l'ineas de emisi'on del gas en los brotes m'as brillantes de formaci'on estelar de la galaxia BCD Haro\,15. Para esto se estimaron temperaturas y densidades electr'onicas, y se obtuvieron abundancias i'onicas y totales de las diferentes especies.

Como en las relaciones entre temperaturas electr\'onicas no fue posible obtener una estimaci\'on realista de las incertidumbres asociadas a los modelos dado que los errores de los modelos no se pueden cuantificar, los errores que se han dado en las abundancias son cotas inferiores.

De los espectros de los brotes B y C obtenidos en modo dispersi'on simple tomados con la c\'amara WFCCD del telescopio de 100"\ du Pont  de LCO, Chile, se puede resumir los siguiente resultados:\\

\begin{itemize}
 \item en el brote B, cuatro temperaturas electr'onicas, T$_e$([\OIII]), T$_e$([\OII]), T$_e$([\SIII]) y T$_e$([\SII]) fueron derivadas de mediciones directas con precisiones del orden de 2\%, 3\%, 10\% y 14\% respectivamente, y se estim'o la T$_e$([\NII]) considerado la aproximaci'on T$_e$([\NII])\,$\approx$\,\\T$_e$([\OII]). Mientras que en el brote C, para determinar las temperaturas electr'onicas T$_e$([\OIII]), T$_e$([\OII]), T$_e$([\SIII]), T$_e$([\SII]) y T$_e$([\NII]) se recurri'o tanto a los modelos como a las relaciones emp'iricas. Es de notar que en el brote B, T$_e$([\OII]) es 4100 K m'as alta que T$_e$([\SII]). En cuanto a la temperatura de [\NII] en el brote B, el valor obtenido es mayor que la del brote C, a'un estim'andola usando el m'etodo P07. 
 \item la densidad electr'onica calculada a partir del cociente de l'ineas del [\SII] para ambos brotes est'a por debajo de la densidad electr'onica cr'itica por desexcitaci'on colisional, tal como sucede en los procesos de formaci'on estelar en las galaxias H {\sc ii}.
\item para el brote B, se dedujeron las abundancias i'onicas del O$^{+}$, O$^{2+}$, S$^{+}$, S$^{2+}$, N$^{+}$, Ne$^{2+}$, Ar$^{2+}$, Ar$^{3+}$, Fe$^{2+}$, He$^{+}$ y He$^{2+}$. Tambi'en en este brote se calcularon las abundancias totales del O, S, N, Ne, Ar, Fe y He, y los cocientes logar'itmicos N/O, S/O, Ne/O y Ar/O. En el brote C, se estimaron las abundancias i'onicas O$^{+}$, O$^{2+}$, S$^{+}$ y N$^{+}$, y se calcularon las abundancias totales del O y N, y la abundancia relativa N/O. El brote B tiene una abundancia de ox'igeno 12 + log (O/H)\,=\,8.04$\pm$0.03 (0.22 veces el valor solar), similar, dentro de los errores, al encontrado por L'opez-S'anchez \& Esteban~(\citeyear{2009A&A...508..615L}) en el mismo brote. La abundancia de azufre en este brote es 6.52$\pm$0.22 (0.15 veces el valor solar) y la abundancia de nitr'ogeno es 6.88$\pm$0.20 (0.09 veces el valor solar). El valor estimado para N/O es un poco mayor en el brote C que en el B, aunque igual dentro de los errores, mostrando en ambos casos una sobre abundancia respecto a lo encontrado, generalmente, para las galaxias H {\sc ii}.
\item se ha encontrado que es mayor el grado de excitaci'on en el brote B que en el C.
\item la abundancia total del He estimada en el brote B tiene un valor t'ipico, como los que se encuentran para las galaxias H {\sc ii}.
\end{itemize}

De los espectros de alta resoluci'on tomados con el espectr'ografo 'echelle del telescopio de 100"\ du Pont  de LCO, Chile, de los brotes A, B, C, E y F se puede resumir los siguiente resultados:\\

\begin{itemize}
 \item en el brote A, para la medida global y para las dos componentes angostas y la componente ancha, se utiliz'o la relaci'on encontrada por H06 entre temperaturas T$_e$[\SIII] vs. T$_e$[\OIII] para estimar la temperatura electr'onica del [\OIII], previa determinaci'on de T$_e$([\SIII]) con el m'etodo D07. Luego, siguiendo el an'alisis cl'asico, se obtuvieron de los modelos las temperaturas T$_e$([\OII]), T$_e$([\SII]) y T$_e$([\NII]). 
 
 En el brote B, se derivaron T$_e$([\OIII]) y T$_e$([\SIII]) con el m'etodo directo para la medida global y para las dos componentes cinem'aticas,  con precisiones del orden de 5\% y 13\% para la medida global, 15\% y 18\% en la componente angosta y 4\% y 14\% en la ancha. La temperatura T$_e$([\OII]) se obtuvo de los modelos y las temperaturas T$_e$([\SII]) y T$_e$([\NII]) se consideraron iguales a T$_e$([\OII]).
 
  En las medidas globales de los brotes C y F como en las componentes cinem'aticas de los brotes C y E, se consider'o la temperatura electr'onica de [\OIII] como 10$^4$K. Luego, con este valor se estim'o la temperatura de [\OII] usando la relaci'on entre ambas de P'erez-Montero \& D'iaz~(\citeyear{2003MNRAS.346..105P}), las temperaturas T$_e$([\SII]) y T$_e$([\NII]) se consideraron iguales a la temperatura de [\OII], y finalmente T$_e$([\SIII]) se determin'o usando su relaci'on con la temperatura de [\OIII] derivada por H06. En la medida global del brote E, se pudo aplicar el m'etodo P07 para estimar la temperatura del [\NII], y con esta temperatura se estim'o T$_e$([\SIII]) haciendo uso de la diferencia sistem'atica de 500K entre temperaturas hallada por D07. Luego, T$_e$([\OIII]) se estim'o usando el ajuste de H06. La temperatura T$_e$([\OII]) se obtuvo utilizando los modelos y se consider'o que T$_e$([\SII]) es igual a la temperatura de [\OIII].
  
 \item las densidades electr'onicas estimadas tanto para las medidas globales de los cinco brotes como para las diferentes componentes cinem'aticas de los brotes A, B, C y E fueron calculadas a partir del cociente de l'ineas del [\SII] y est'an por debajo de la densidad electr'onica cr'itica por desexcitaci'on colisional. Estos valores dan una estimaci'on del orden de magnitud de la densidad, dado que la incerteza en el c'alculo es muy grande, salvo en el caso del brote B, donde la densidad electr'onica obtenida para la medida global es de 150 part'iculas por cm$^3$ con una incerteza de 50 part'iculas por cm$^3$.
 
 \item  diferentes abundancias i'onicas y totales junto con los ICFs fueron derivados tanto para las medidas globales de los cinco brotes como para las diferentes componentes cinem'aticas de los brotes A, B, C y E. En el brote B, los iones pertenecientes a la zona de menor excitaci'on (O$^{+}$, S$^{+}$, N$^{+}$) muestran una mayor abundancia en la componente angosta, mientras que los iones de la zona intermendia (S$^{2+}$, Ar$^{2+}$) y los de la zona de alta excitaci'on (O$^{2+}$, Ne$^{2+}$) muestran el comportamiento opuesto. 
 
 Las abundancias totales de O, S, N, Ne y Ar, donde fue posible obtenerlas, se encuentran en el mismo rango de las abundancias totales medidas para las galaxias H {\sc ii}. La abundancia total de ox'igeno derivada en el brote B est'a en muy buen acuerdo con el valor obtenido por L'opez-S'anchez \& Esteban~(\citeyear{2009A&A...508..615L}), pero en el brote A las abundancias aqu'i derivadas son menores ($\sim$0.2 dex) al valor encontrado por L'opez-S'anchez \& Esteban, aunque si se tienen en cuenta las barras de error la diferencia se reduce a $\sim$0.05 dex. Del an'alisis realizado se encuentra que, dentro de los errores, la abundancia total de ox'igeno derivada en los brotes A, B y la medida global del brote E son muy similares. 
 
 Si se comparan los valores derivados en el brote B para los datos 'echelle con los estimados a partir de los datos de ranura larga se puede apreciar que la abundancia total de ox'igeno es menor para los datos de dispersi'on simple compar'andola con la estimaci'on global (aunque muy parecido se considera los errores), y m'as a'un si se la compara con la componente ancha, pero es igual al derivado para la angosta. 
 
 Se encontr'o un exceso de N/O en el brote A y B, siendo en A muy alto respecto a los valores encontrados en las galaxias H\,{\sc ii}, aunque este tipo de excesos ya fue observado en otros casos \citep[ver discusi'on en][]{Perez-Montero+11}. Los valores del S/O, para los brotes A y B, aunque presentan una gran dispersi'on, est'an en el rango en el que suelen encontrarse este tipo de objetos. Y los valores del Ne/O en el brote A (medida global) y en el brote B, dentro de los errores, son consistentes con los valores de la literatura. Finalmente, el valor del cociente Ar/O para el brote A presenta una dispersi'on mayor respecto de la que presenta el brote B, y en ambos casos dentro de los valores t'ipicos para esta clase de objetos. La similitud, dentro de los errores, en las abundancias relativas S/O y N/O en las tres componentes del brote A podr'ia ser una evidencia que la evoluci'on qu'imica de las diferentes componentes es muy similar o que directamente son distintas fases de un mismo gas. Una situaci'on similar ocurre con las abundancias relativas S/O, N/O, Ne/O y Ar/O de las dos componentes del brote B. 
Los valores de las abundacias relativas N/O, Ne/O y Ar/O en el brote B observados en ranura larga, son consistentes con los observados en el modo 'echelle, salvo el valor S/O que es mayor en dispersi'on simple que en 'echelle.

 \item se encontr'o que las abundancias totales de He tienen un comportamiento similar para la componente angosta 1 del brote A y la angosta del brote B respecto al resto de las componentes, y es que el valor ``adoptado'' de He$^{+}$/H$^{+}$ es bajo para este tipo de objetos, pero en cambio el valor obtenido para las componentes angosta 2 y ancha del brote A, y para la componente ancha del brote B son valores t'ipicos de los encontrados para galaxias H {\sc ii}. Sin embargo, se necesitan datos mejores para estudiar este comportamiento y tratar de reducir los errores de estas estimaciones.
 
  \item la extinci'on medida en los brotes A, B, C y E, considerando los errores, presentan valores similares y todos compatibles con baja extinci'on, aunque se puede decir que el brote B tiene un valor levemente menor al del resto de los brotes.
  
 \item ser'a necesario obtener nuevos datos de los brotes C, E y F para un an'alisis m'as profundo de las abundancias i'onicas y totales.

\item analizando el brote A y el brote B por separado, y considerando los errores, las diferentes componentes cinem'aticas de cada brote presentan una estructura de ionizaci'on similar entre ellas, lo cual implicar'ia que la temperatura del campo de radiaci'on ionizante dentro de cada brote es muy parecida para las distintas componentes cinem'aticas. Esto podr'ia deberse a que el c'umulo estelar responsable de la ionizaci'on de cada una de las componentes ser'ia el mismo. Para el brote B, este resultado es v'alido tanto para los datos 'echelle como para los datos de dispersi'on simple.
  
\item las abundancias totales de ox'igeno estimadas con los diferentes par'ametros emp'iricos, en general, muestran valores consistentes con las abundancias derivadas usando las temperaturas electr'onicas obtenidas ya sea de los modelos y relaciones emp'iricas o por m'etodos directos, excepto en determinados casos donde los valores de las abundancias determinadas con algunos par'ametros dan valores que sobreestiman o subestiman la abundancia total de ox'igeno obtenida previamente. 

Cabe destacar que las relaciones emp'iricas usadas aqu'i como las usadas para los par'ametros $\eta$ y $\eta$'\ fueron derivadas a partir de medidas para regiones completas y que nunca se ha hecho un an'alisis y/o determinaci'on por componentes. Este es un trabajo que merece una discusi'on y un tratamiento m'as completo que est'a m'as all'a de lo que se puede realizar con los presentes datos.
 \end{itemize}
 
 \clearemptydoublepage

\newpage
\clearemptydoublepage



\clearemptydoublepage
\newpage

\chapter*{{\color{night}Conclusiones y Trabajo a futuro}}
\label{capitulo5}
\addcontentsline{toc}{chapter}{\numberline{}{Conclusiones y Trabajo a futuro}}

\pagestyle{plain}

En esta tesis he identificado y analizado las caracter'isticas f'isicas y cinem'atica de una muestra de Regiones H{\sc ii} Gigantes en galaxias observables con telescopios del Hemisferio Sur. Tambi'en he definido la naturaleza y las propiedades f'isicas de las componentes gaseosas en algunos brotes de formaci'on estelar masiva de la galaxia BCD Haro\,15, permitiendo analizar las abundancias qu'imicas i'onicas y totales de las mismas.
Para llevar a cabo este trabajo se han utilizado t'ecnicas de espectroscop'ia de alta resoluci'on como lo es la espectroscop'ia 'echelle, y tambi'en he utilizado, para una galaxia, espectros de dispersi'on simple de baja resoluci'on.

Analizando 11 regiones H{\sc ii} luminosas en galaxias espirales y en una galaxia BCD, he obtenido los siguientes resultados generales importantes:\\
\begin{dinglist}{52}
\item he encontrado comportamientos cinem'aticos complejos en seis regiones de formaci'on estelar de las galaxias espirales NGC\,6070, NGC\,7479 y en cinco brotes de formaci'on estelar violenta en la galaxia compacta azul Haro\,15. En cuatro de ellas se hallaron indicios de la presencia de componentes con velocidades radiales distintas, pr'acticamente imposibles de detectar espacialmente.

\item he realizado un estudio detallado de las propiedades f'isicas de los principales brotes de formaci'on estelar de la galaxia BCD Haro\,15 utilizando espectroscop'ia de ranura larga y 'echelle obtenida en LCO.
\end{dinglist}

Dentro de esos resultados puedo decir que:\\
\begin{itemize}
\item Fue posible determinar la dispersi'on de velocidades dada por el ensanchamiento del perfil de l'ineas de emisi'on nebular para cada regi'on. La alta relaci'on S/N, junto a la resoluci'on del espectro 'echelle permiti'o resolver el perfil de las l'ineas de emisi'on y calcular la dispersi'on de velocidades del gas ionizado. Este an'alisis fue hecho midiendo el ancho observado del perfil de las l'ineas de recombinaci'on de hidr'ogeno despu'es de corregir el mismo por el perfil instrumental y por la contribuci'on t'ermica. Fue posible confirmar la naturaleza gigante de las regiones candidatas.

\item De la muestra de seis GH{\sc ii}R en las galaxias NGC\,6070 y NGC\,7479, se ha encontrado que todas ellas muestran evidencia de un residuo presente en los perfiles de las l'ineas de emisi'on. En estas regiones se ha intentado ajustar una componente ancha o dos componentes sim'etricas separadas en velocidad respecto a la componente principal. Todas las regiones estudiadas presentan componentes cinem'aticas supers'onicas.

\item Los cinco brotes de la galaxia Haro\,15, a excepci'on del brote F, muestran m'ultiples componentes supers'onicas. En el brote F el ensanchamiento del perfil integrado solamente pudo ser detectado en la l'inea H$\alpha$. En esta l'inea de emisi'on se pudo ajustar una componente ancha supers'onica de 22 \kms\ y una angosta subs'onica de 8 \kms, dispersi'on de velocidades t'ipica de regiones H\,{\sc ii} cl'asicas, aunque la presencia de una componente ancha sea casi exclusiva de las Regiones H\,{\sc ii} Gigantes.

\item Si bien las caracter'isticas de estas Regiones H{\sc ii} Gigantes presentan una correlaci'on en el plano $\log(L) - \log(\sigma)$, la presencia de m'ultiples componentes cinem'aticas influye fuertemente en la ubicaci'on final en dicho plano. Esto representa un alerta importante ante el uso indiscriminado de dicha regresi'on cuando no se tienen datos completos sobre el comportamiento din'amico de los brotes de formaci'on estelar. Las componentes angostas identificadas en los perfiles complejos siguen la relaci'on de sistemas virializados, en cambio los perfiles ajustados con una componente Gaussiana simple muestran una pendiente m'as plana.

\item El an'alisis de las observaciones realizadas por H\"agele y colaboradores~(\citeyear{2006MNRAS.372..293H,2008MNRAS.383..209H,2011arXiv1101.4140H}) y la metodolog'ia definida por ellos para obtener abundancias elementales precisas de O, S, N, Ne, Ar y Fe en el gas ionizado, me ha permitido estimar en los brotes de la galaxia Haro\,15 por lo menos cuatro temperaturas de l'inea: T$_e$([\OIII]), T$_e$([\SIII]), T$_e$([\OII]) y T$_e$([\SII]), y una densidad electr'onica, N$_e$([\SII]) para los datos de ranura larga. Para los datos 'echelle, cada brote fue analizado por separado, estudiando las temperaturas, densidades, abundancias i'onicas y totales estimadas para la medida global y para cada componente cinem'atica separadamente. En el brote B fue el 'unico brote para el que se derivaron T$_e$([\OIII]) y T$_e$([\SIII]) con el m'etodo directo para la medida global y para las dos componentes cinem'aticas, con precisiones del orden de 5\% y 13\% para la medida global, 15\% y 18\% en la componente angosta y 4\% y 14\% en la ancha. En el resto de los brotes (A, C, E y F) las temperaturas electr'onicas fueron estimadas usando, seg'un el caso, los modelos, relaciones entre temperaturas, m'etodos emp'iricos o incluso en algunos casos cuando no fue posible derivar ninguna temperatura, se consider'o la temperatura electr'onica de [\OIII] como 10$^4$K y de ella, usando relaciones entre temperaturas, se obtuvo el resto. Las densidades electr'onicas N$_e$([\SII]) fueron calculadas a partir del cociente de l'ineas del [\SII] y todas est'an por debajo de la densidad electr'onica cr'itica por desexcitaci'on colisional. 

\item Con estas medidas y un tratamiento cuidadoso y realista de los errores observacionales, se obtuvieron estimaciones de las abundancias qu'imicas i'onicas y totales del O, S, N, Ne, Ar, Fe y He, seg'un fuere el caso. Este an'alisis detallado de las abundancias fue realizado en cada brote de formaci'on estelar con m'as de una componente y en cada una de las componentes individuales, ya sea en los datos de dispersi'on simple como en los datos espectrosc'opicos 'echelle. Las abundancias totales de O, S, N, Ne y Ar, donde fue posible obtenerlas, se encuentran en el mismo rango de las abundancias totales medidas para las galaxias H {\sc ii}. La abundancia total de ox'igeno derivada en el brote B est'a en muy buen acuerdo con el valor obtenido por L'opez-S'anchez \& Esteban~(\citeyear{2009A&A...508..615L}), pero en el brote A las abundancias aqu'i derivadas son menores ($\sim$0.2 dex) al valor encontrado por L'opez-S'anchez \& Esteban, aunque si se tienen en cuenta las barras de error la diferencia se reduce a $\sim$0.05 dex. Del an'alisis realizado se encuentra que, dentro de los errores, la abundancia total de ox'igeno derivada en lo brotes A, B y la medida global del brote E son muy similares. Como en las relaciones entre temperaturas electr\'onicas no fue posible obtener una estimaci\'on realista de las incertidumbres asociadas a los modelos dado que los errores de los modelos no se pueden cuantificar, los errores que se han dado en las abundancias son cotas inferiores.
\item Se encontr'o un exceso de N/O en el brote A y B, siendo en A muy alto respecto a los valores encontrados en las galaxias H\,{\sc ii}, aunque este tipo de excesos ya fue observado en otros casos \citep[ver discusi'on en][]{Perez-Montero+11}. Los valores del S/O, para los brotes A y B, aunque presentan una gran dispersi'on, est'an en el rango en el que suelen encontrarse este tipo de objetos. Y los valores del Ne/O en el brote A (medida global) y en el brote B, dentro de los errores, son consistentes con los valores de la literatura. Finalmente, el valor del cociente Ar/O para el brote A presenta una dispersi'on mayor respecto que la del brote B, y en ambos casos el valor del cociente est'a dentro de los valores t'ipicos para esta clase de objetos. La similitud, dentro de los errores, en las abundancias relativas S/O y N/O en las tres componentes del brote A podr'ia ser una evidencia que la evoluci'on qu'imica de las diferentes componentes es muy similar o que directamente son distintas fases de un mismo gas. Una situaci'on similar ocurre con las abundancias relativas S/O, N/O, Ne/O y Ar/O de las dos componentes del brote B. 
Los valores de las abundacias relativas N/O, Ne/O y Ar/O en el brote B observados en los datos espectrosc'opicos obtenidos con ranura larga, son consistentes con los observados en el modo 'echelle, salvo el valor S/O que es mayor en dispersi'on simple que en modo 'echelle.

\item Se encontr'o que las abundancias totales de He tienen un comportamiento similar para la componente angosta 1 del brote A y la angosta del brote B respecto al resto de las componentes. El valor ``adoptado'' de He$^{+}$/H$^{+}$ que se obtuvo es bajo para este tipo de objetos. En cambio, el valor obtenido para las componentes angosta 2 y ancha del brote A, y para la componente ancha del brote B son valores t'ipicos a los encontrados para las galaxias H {\sc ii}. Es necesario obtener nuevos y mejores datos para estudiar este comportamiento, adem'as de tratar de reducir los errores de las estimaciones de las abundancias i'onicas y totales de helio. 

  \item La extinci'on medida en los brotes A, B, C y E, considerando los errores, presentan valores similares y todos compatibles con baja extinci'on, aunque se puede decir que el brote B tiene un valor levemente menor al del resto de los brotes.
  
\item En los brotes A y B de Haro\,15, la estructura de ionizaci'on y, por ende la temperatura del campo de radiaci'on ionizante, es muy similar para cada componente del respectivo brote estudiado. Esto implicar'ia que el c'umulo ionizante de cada una de las componentes del brote en cuesti'on ser'ia el mismo. La excepci'on estar'ia en la componente ancha del brote A que difiere levemente de su grupo pero es similar si consideramos los errores. El brote B tiene una ionizaci'on m'as alta que el brote A.

\item Para cada brote de Haro\,15, y cuando fue posible disponer de las l'ineas m'as intensas, se determin'o la abundancia de ox'igeno a partir del an'alisis de los diferentes par'ametros emp'iricos: O$_{23}$, S$_{23}$, SO$_{23}$, O$_{3}$N$_{2}$, S$_{3}$O$_{3}$, N$_{2}$ y Ar$_{3}$O$_{3}$. Las abundancias totales de ox'igeno estimadas con los diferentes par'ametros emp'iricos, en general, mostraron valores consistentes con las abundancias derivadas usando las temperaturas obtenidas ya sea de los modelos y relaciones emp'iricas o por m'etodos directos, excepto en determinados casos donde los valores de las abundancias determinadas con algunos par'ametros dan valores que sobreestiman o subestiman la abundancia total de ox'igeno obtenida previamente. 
Cabe destacar que las relaciones emp'iricas usadas tanto para los par'ametros emp'iricos como para los par'ametros $\eta$ y $\eta$'\ fueron derivadas a partir de medidas para regiones completas y que nunca se ha hecho un an'alisis y/o determinaci'on por componentes. Este es un trabajo que merece una discusi'on y un tratamiento m'as completo que est'a m'as all'a de lo que se puede realizar con los presentes datos.
\end{itemize}
 
Los resultados de esta Tesis amplian de distintas formas el estudio de las Regiones H{\sc ii} Gigantes. Si bien no es la primera vez que se combina la informaci'on cinem'atica - que permite separar componentes indistinguibles en espectroscop'ia de baja resoluci'on - junto con un an'alisis detallado de las propiedades f'isicas de cada componente, este trabajo presenta una mayor completitud frente a otros trabajos publicados dada la gran cantidad de l'ineas analizadas, entre las que se encuentran las d'ebiles l'ineas aurorales.  Adem'as, a esto se suma la posibilidad de descomponer sus perfiles en m'as de una componente angosta y una componente ancha. Para ello, he debido avanzar sobre las t'ecnicas convencionales de an'alisis espectral, desarrollando una metodolog'ia de trabajo original, necesaria para llevar esta investigaci'on a un resultado convincente.

\section*{Trabajo futuro}

La influencia de las estrellas masivas deja su impronta en la fase gaseosa, principalmente excitando, ionizando y modificando estructuralmente el ISM a partir de su radiaci'on UV y vientos. La intensidad de las l'ineas de emisi'on del gas ionizado permite analizar las propiedades f'isicas de dichas regiones, su din'amica interna, y caracterizar, utiliz'andolas como sondas, la cinem'atica de las galaxias que las albergan.
Siguiendo esta l'inea y teniendo en cuenta los progresos m'as recientes en este campo, especialmente la nueva informaci'on que ofrecen los grandes telescopios, el plan de trabajo a futuro se estructura en dos objetivos principales:\\

\begin{dinglist}{42}
\item Continuar la identificaci'on y consiguiente an'alisis de Regiones H{\sc ii} Gigantes en galaxias cercanas. La identificaci'on es posible
mediante espectroscop'ia de alta resoluci'on que permite verificar la naturaleza supers'onica de la cinem'atica del gas ionizado. Por otro
lado, estudiar la estructura cinem'atica de dichas regiones con espectroscop'ia 3D que permita combinar informaci'on cinem'atica y espacial, con una un'ivoca determinaci'on de los nodos individuales que contribuyen al perfil global observado.

\item Definir la naturaleza y propiedades f'isicas de los brotes de formaci'on estelar masiva, como as'i tambi'en entender los mecanismos
que permiten su retroalimentaci'on y la formaci'on de nuevas generaciones de estrellas que producen a la vez, el enriquecimiento qu'imico del Universo Local. El an'alisis de los par'ametros f'isicos tales como densidad electr'onica, temperatura electr'onica, y composici'on qu'imica permitir'a caracterizar los brotes de formaci'on estelar m'as extremos analizando su entorno. 
\end{dinglist}

Se dispone de espectros de alta resoluci'on obtenidos con los telescopios de 2.5 y 6.5m del Observatorio de Las Campanas (LCO) para la detecci'on de Regiones H{\sc ii} Gigantes en m'as galaxias espirales y BCDs. Recientemente, se re-observaron dos de los brotes de la galaxia Haro\,15 (A y B) con un mejor cociente S/N, con lo que se espera poder detectar las d'ebiles l'ineas aurorales y poder realizar ajustes en estas l'ineas que sean independientes de las soluciones de las l'ineas m'as intensas.

Para estudiar comportamientos cinem'aticos complejos en las regiones de formaci'on estelar har'e uso de las facilidades instrumentales
de espectroscopia 3D de los Telescopios Gemini, tanto en el rango 'optico (GMOS), como en el infrarrojo cercano (NIFS). Cuento con
observaciones asignadas con alta prioridad (Banda 1) para obtener espectroscop'ia 3D de campo integrado en una regi'on de formaci'on
estelar que presenta m'as de una componente en el perfil de sus l'ineas de emisi'on. La propuesta aceptada plantea hacer uso de la alta
calidad de im'agenes de los telescopios Gemini en su modo IFU (sigla en ingl\'es de Integral Field Unit) para mapear el campo de velocidades. De esta manera espero dilucidar si el perfil, aparentemente supers'onico de algunas regiones HII gigantes puede resolverse en varias
componentes, cada una de ellas con distintas velocidades. Tambi'en, junto a Guillermo Bosch y Guillermo H\"agele, obtuve tiempo de observaci'on durante el semestre 2009A, con el cual planteamos mejorar la resoluci'on estudiando regiones visibles con Gemini
Norte utilizando la resoluci'on espacial al l'imite de difracci'on en el infrarrojo cercano, utilizando 'optica Adaptable con estrella gu'ia
laser en el instrumento GNIRS. El uso de 'optica Adaptable permite obtener im'agenes de la misma calidad que la del Telescopio
Espacial (FWHM $\approx$0.08 segundos de arco) en un telescopio de 8 metros de di'ametro. Esto nos permitir'a estudiar la estructura
cinem'atica de dichas regiones con una resoluci'on sin precedentes necesaria para distinguir la presencia de distintas componentes,
con velocidades diferentes, que puedan contribuir al ancho observado en el perfil de la emisi'on global.

Se pretende ampliar la muestra tanto de Regiones H{\sc ii} Gigantes como de galaxias H{\sc ii} para las que se pueda obtener una abundancia precisa
utilizando lo que se conoce como el m'etodo directo. Usaremos los datos estudiados en \cite{Firpo05,2010MNRAS.406.1094F} y nuevas observaciones de regiones H{\sc ii} en otras galaxias BCDs. Esto permitir'a producir y calibrar relaciones emp'iricas entre las diferentes temperaturas
de l'inea que reemplazar'ian a las que se utilizan actualmente y que est'an basadas en secuencias de modelos de fotoionizaci'on muy
simples, y pobremente contrastados, debido a la falta de datos id'oneos. Para ello, se requiere el uso de espectr'ografos de amplia cobertura espectral y
resoluci'on espectral intermedia. La muestra debe ser dise'nada para abarcar objetos en los extremos bajo y alto de temperaturas
electr'onicas. Esta cobertura total en temperaturas electr'onicas proveer'a una limitaci'on muy necesaria de la estructura en temperaturas de las
galaxias H{\sc ii}. 
\\
   \begin{center}
     \includegraphics[width=.15\textwidth]{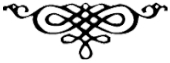}
   \end{center}

\clearpage
\clearemptydoublepage
\newpage

\clearemptydoublepage
\newpage

\renewcommand{\baselinestretch}{1} {\small
\renewcommand{\bibname}{Bibliograf'ia}
\bibliographystyle{astron}
\bibliography{biblos}
\addcontentsline{toc}{chapter}{Bibliograf'ia}}
\clearemptydoublepage

{\color{night}\section*{\Large\ Agradecimientos}}
\addcontentsline{toc}{chapter}{\numberline{}Agradecimientos}
\thispagestyle{empty}
\vskip 0.1cm
Quiero expresar mi agradecimiento:\\

A mi director de Tesis, Guillermo Bosch por su generosidad al brindarme la oportunidad de ser parte del Grupo de Estrellas Masivas, y de compartir su capacidad y experiencia cient\'{\i}fica en un marco de confianza, afecto y amistad, fundamentales para la concreci\'on de este trabajo. Tambi\'en le agradezco haberme dado la libertad y todos los medios adecuados para la realizaci\'on de esta Tesis. Lo humano prevalece por sobre todas las cosas. Muchas Gracias Guille!!

A mi co-director de Tesis, Guillermo H\"agele quien supo compartir, con una disposici\'on sin igual, toda su experiencia de trabajo adquirida en el grupo del Departamento de F\'{\i}sica Te\'orica de la Universidad de Madrid, y quien me brind\'o la oportunidad de acercarme a conocer a \'Angeles y a su grupo de trabajo. Y como nombrar a Guiye sin nombrar a su familia, Moni y Vale, quienes me dieron todo su cari\~no y afecto en las estad\'{\i}as en Madrid, en particular a Moni siempre presente ayudando con scripts y trucos computacionales. Gracias ``familia''!!  

A Nidia Morrell, por ense\~narme las t\'ecnicas de observaci\'on, invitarme a participar de sus proyectos de observaci\'on y brindarme la oportunidad de obtener los excelentes datos observacionales que sin ellos, toda esta Tesis no se hubiera podido concretar. Por todas esas incre\'ibles noches estrelladas en Las Campanas ... Gracias Nidia!

A \'Angeles D'iaz, por su generosidad cient\'{\i}fica y valiosas cr\'{\i}ticas al discutir  el an\'alisis y los resultados de este trabajo y por invitarme a participar del Proyecto ``Estallidos y su huella en la evoluci\'on c\'osmica de las galaxias'' con el cual pude viajar en m\'as de una oportunidad a la ciudad de Madrid a realizar las estad\'ias de investigaci\'on. Gracias \'Angeles!

A mi ``compa\~nera de Tesis'', Cecilia Fari\~na, por su presencia incondicional, su permanente disposici\'on y desinteresada ayuda y por sobre todo su amistad.

A Elena y Roberto Terlevich por su calidez y por sus valiosas sugerencias durante el desarrollo de este trabajo.

A Virpi Niemela, por abrirme las puertas del grupo y aceptar ser mi aval en los comienzos de la beca de la UNLP. Se te extra\~na mucho Virpi!

A Roberto Venero, Amalia Meza y Pablo Cincotta por su permanente disposici\'on y desinteresada ayuda. 

A mis amigos del Observatorio (incluyendo algunas parejas de amigas/os) por su continuo y afectuoso aliento. Si tuviera que especificar a todos me llevar\'{\i}a un cap\'{\i}tulo entero de esta Tesis, pero cada uno de ustedes sabe reconocerse en estas palabras.

A mis amigas Mar'ia Paula Natali, Mar'ia Florencia Pan y a mis amigas de Dom\'{\i}nico por su cari\~no, comprensi\'on y constante est\'{\i}mulo.

A mis amigos del Ministerio por acompa\~narme en todos los momentos importantes.

A mis padres por ense\~narme que la perseverancia y el esfuerzo son el camino para lograr objetivos. Y a mi familia por ense\~narme a enfrentar los obst\'aculos con alegr\'{\i}a y afecto.

A la Facultad de Ciencias Astron\'omicas y Geof\'{\i}sicas por brindarme el espacio para trabajar y la calidad acad\'emica necesaria para alcanzar este t\'{\i}tulo. Y un agradecimiento al personal no docente de la Facultad por la ayuda brindada durante todo este tiempo. 

A la Universidad Nacional de La Plata por haberme dado la posibilidad de tener una beca de investigaci\'on para realizar todo el trabajo de Tesis.

Al Director y personal t\'ecnico del Observatorio de Las Campanas (LCO), Chile, por su generosa hospitalidad y soporte t\'ecnico durante los turnos de observaci\'on.

Y un especial agradecimiento al jurado evaluador de la Tesis, Cristina Cappa, Enrique P\'erez-Montero y Lydia Cidale, por sus valiosas sugerencias y correcciones para mejorar esta Tesis.

A todos y cada uno ... Muchas Gracias!!

\begin{picture}(50,200)
\put (210,160){\em \small  Tarda en llegar}\\
\put (210,150){\em \small  Y al final, al final}\\
\put (210,140){\em \small  Hay recompensa}\\
\put (210,130){\em \small  En la zona de promesas}\\
\put (210,120){\em \small  (Zona de promesas - G.Ceratti)}\\
\put (230,90){\includegraphics[width=.15\textwidth]{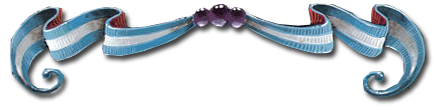}}
\end{picture}


\thispagestyle{empty}

\begin{figure*}
   \begin{center}
     \leavevmode
     \vskip 3cm
     \includegraphics[width=.95\textwidth]{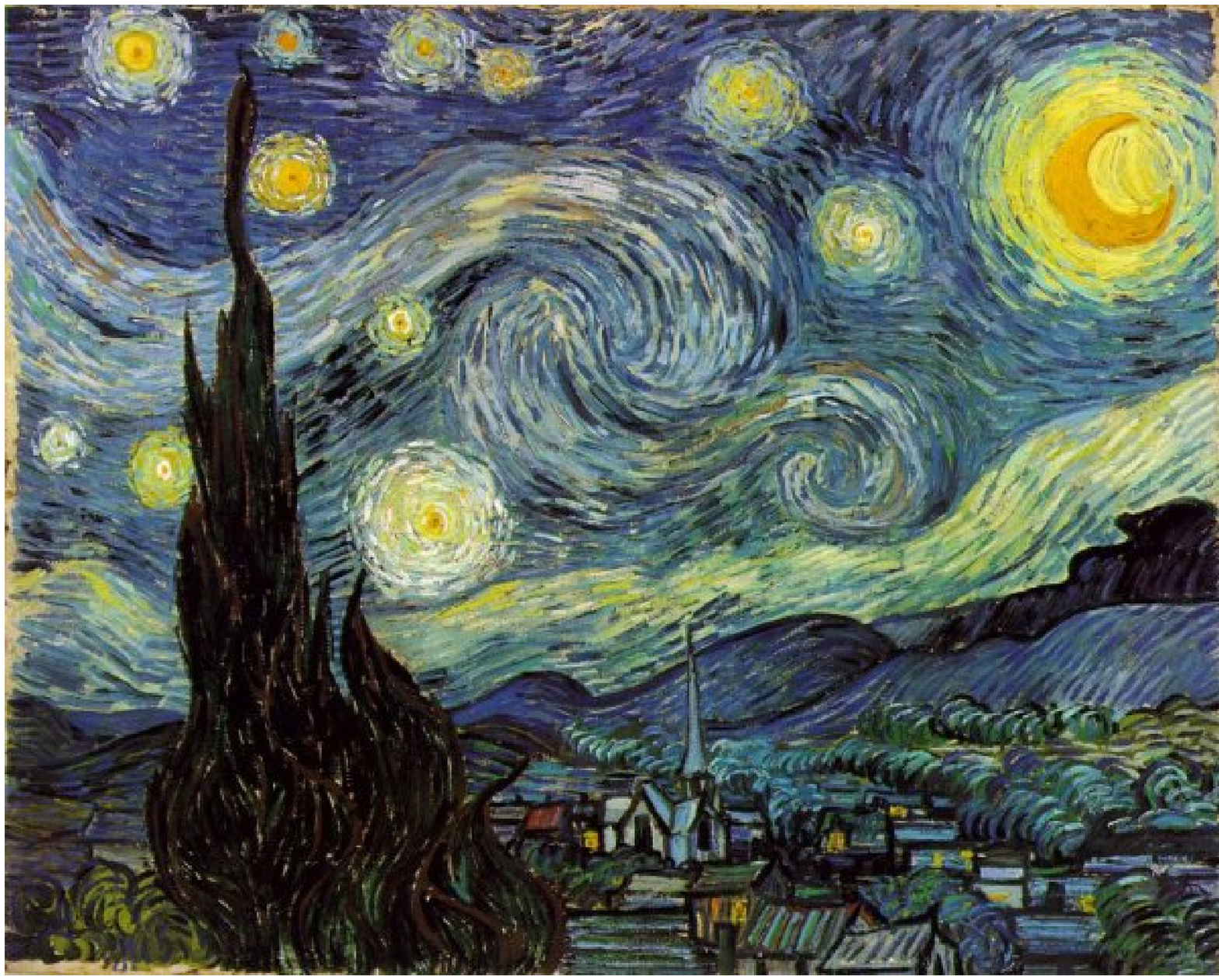}
   \end{center}
 \end{figure*}
 \noindent
 \scriptsize
 {\em \Large ``Cuando siento una necesidad de religi'on\\
 salgo de noche para pintar las estrellas"\\
 Vincent Van Gogh}
 \normalsize

\clearemptydoublepage
\newpage

\clearemptydoublepage
\newpage

\end{spacing}

\end{document}